\documentclass{elsarticle}
\usepackage[utf8]{inputenc}
\usepackage{graphicx}
\usepackage{amsfonts}
\usepackage{amssymb}
\usepackage{mathtools}
\usepackage{subcaption}
\usepackage{xcolor}
\usepackage{url}
\journal{ }
\newcommand{\bsym}[1]{\boldsymbol{#1}}

\title{Bridging scales with Machine Learning: From first principles statistical mechanics to continuum phase field computations to study order-disorder transitions in Li$_x$CoO$_2$}

\author[umme]{G.H. Teichert}
\author[umme]{S. Das}
\author[umme]{M. Faghih Shojaei}
\author[umap]{J. Holber}
\author[jhu,tri]{T. Mueller}
\author[tri]{L. Hung}
\author[umme,umms,umap,micde]{V. Gavini}
\author[umme,umm,umap,micde]{K. Garikipati\corref{mycorrespondingauthor}}
\ead{krishna@umich.edu}
\cortext[mycorrespondingauthor]{Corresponding Author}

\address[umme]{Department of Mechanical Engineering, University of Michigan}
\address[tri]{Toyota Research Institute, Los Altos, CA}
\address[umms]{Department of Materials Science \& Engineering, University of Michigan}
\address[umm]{Department of Mathematics, University of Michigan}
\address[micde]{Michigan Institute for Computational Discovery \& Engineering, University of Michigan}
\address[umap]{Applied Physics Program, University of Michigan}
\address[jhu]{Materials Science \& Engineering, Johns Hopkins University}

\begin{document}

\begin{abstract}
Li$_x$\textit{TM}O$_2$ (TM={Ni, Co, Mn}) forms an important family of cathode materials for Li-ion batteries, whose performance is strongly governed by Li composition-dependent crystal structure and phase stability. Here, we use Li$_x$CoO$_2$ (LCO) as a model system to benchmark a machine learning-enabled framework for bridging scales in materials physics. We focus on two scales: (a) assemblies of thousands of atoms described by density functional theory-informed statistical mechanics, and (b) continuum phase field models to study the dynamics of order-disorder transitions in LCO. Central to the scale bridging is the rigorous, quantitatively accurate, representation of the free energy density and chemical potentials of this material system by coarse-graining formation energies for specific atomic configurations. We develop active learning workflows to train recently developed integrable deep neural networks for such high-dimensional free energy density  and chemical potential functions. The resulting, first principles-informed, machine learning-enabled, phase-field computations allow us to study LCO cathodes' phase evolution  in terms of temperature, morphology, charge cycling and particle size.
\end{abstract}

\maketitle

\section{Introduction}
Layered oxides of type Li$_x$\textit{TM}O$_2$ ($x \in [0,1]$, TM={Ni, Mn, Co}, hence called NMC materials) are commonly used cathodes in Li-ion batteries because of the tunability of the chemistry specific to an application, as well as their high energy density and charging rate. Of these, LiCoO$_2$ (LCO) has been the choice for consumer electronics, while Ni-rich compositions are attractive for high energy density batteries, important for electrification of transportation. NMCs undergo phase transitions during cycling, which can, in some cases, affect their performance. 
LCO suffers deleterious stacking sequence phase transitions from the O3 to H1-3 structure below $x=1/3$ \cite{VanderVen1998}.
Quantitative control of cycling and structural stability of layered NMC oxides requires a first principles-based understanding of the thermodynamics of Li intercalation, its impact on the crystal structure's stability and long-term electrochemical properties, and a modeling framework that spans scales from atoms to the continuum. However, scale-bridging is a hard problem and prior studies, as we outline below, have treated the different length scales in isolation. Here, we present a key advance in further developing a scale bridging framework \cite{Teichert2019,Teichert2020} by rigorously and systematically tying first principles energetics with the free energy---in a high dimensional space accounting for all symmetries---describing the phase evolution. This is demonstrated for the O3 phase of LCO with insights to the free energy underpinnings of the thermodynamic phase evolution and order-disorder transitions under Li intercalation.

The rich phase-behavior of LCO has been extensively studied by experiments. A first order phase transition has been identified for compositions $0.75 \leq x \leq 0.94$, caused by a metal-insulator transition \cite{Reimers1992,Menetrier1999}. Reimers and Dahn \cite{Reimers1992} reported an ordering at $x = 1/2$, confirmed as a row ordering \cite{ShaoHorn2003,Takahashi2007}.
Shao-Horn et al. \cite{ShaoHorn2003} found evidence of ordering at $x = 1/3$ at $\sim$100 K.
Additionally, charge ordering Co$^{3+}$ and Co$^{4+}$ atoms at $x = 1/2$ and $x = 2/3$ was observed at 175 K, while these charge states' occupation was random at room temperature \cite{Motohashi2009,Takahashi2007}. The transition from the O3 to the H1-3 structure also has been identified for $x < 0.33$ \cite{Chen2004,Chang2013}.

There also have been numerous computational studies of LCO. \emph{Ab initio} thermodynamic approaches combining density functional theory (DFT)  and statistical mechanics \cite{VanderVen1998, Wolverton1998} have been used to systematically predict phase diagrams, free energy data, and voltage curves. The first-principles methods have been combined with experiment and CALPHAD models \cite{Kaufman1970} to define Gibbs free energy functions \cite{Abe2011,Chang2013} using traditional Redlich-Kister polynomials.  DFT predictions have been combined with experimental data, such as in phase field modeling of an observed metal-insulator phase transition \cite{Nadkarni2019}. However, while the  link between \emph{ab initio} and statistical mechanics computations is well established, first-principles methods have not been included in a comprehensive scale bridging framework that connects to continuum treatments such as phase field models to study the thermodynamics and kinetics of phase transitions in these systems.

\begin{figure}[t]
    \includegraphics[width=\textwidth]{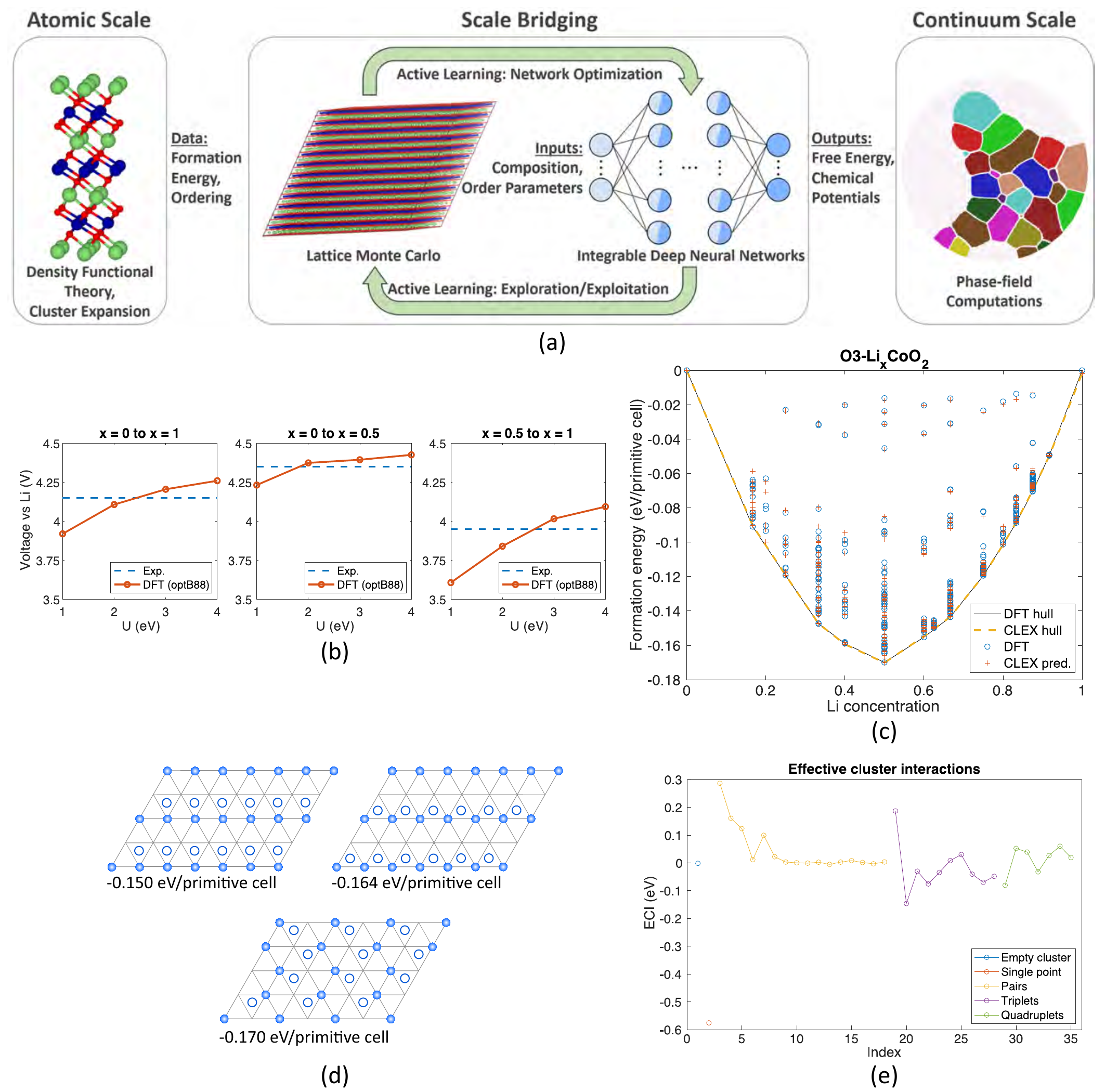}
    \caption{(a) Flowchart outlining the data, computational methods and machine learning-enabled linkages that bridge from the atomic up to the continuum scale. The Li atoms are green, Co are blue, and O are red (visuals generated using VESTA \cite{Momma2011}. (b) Calculated voltage across various compositions as a function of U, compared with experimental voltages \cite{amatucci1996coo2,xia2007phase}. (b) Formation energies and convex hull predicted by the cluster expansion (CLEX) are compared with those from the DFT calculations. The configurations predicted by the cluster expansion to lie on the convex hull are the same as those on the DFT convex hull. (d) Three configurations with Li concentration $x = 0.5$ and their corresponding calculated formation energies. The filled circles represent Li atoms within a Li layer, and the empty circles represent Li atoms in the adjacent Li layer. (e) ECI values associated with each basis function in the cluster expansion.}
    \label{fig:Fig1}
\end{figure}

The current work demonstrates a rigorous framework combining advanced first-principles methods with our recent work in machine learning to bridge across DFT-informed statistical mechanics  and continuum scale simulations (Figure \ref{fig:Fig1}a). Our use of O3 layered LCO to benchmark the approach is motivated by the extensive studies of this chemistry. To our knowledge, the systematic scale bridging demonstrated here has not been presented for this cathode material.

We perform DFT$+U$ calculations ~\cite{Cococcioni2005} to capture  electronic correlation in transition metal oxides with localized $d$ orbitals. Combined with van der Waals functionals, this validates the DFT approaches by close agreement with experimental voltages in layered LCO~\cite{Aykol2015}. A cluster expansion Hamiltonian is parameterized by the DFT computations and used in Semi-Grand Canonical Monte Carlo sampling to compute Li intercalation thermodynamics as a function of temperature and chemical potential, and to construct temperature - composition phase diagrams \cite{PhysRevB.86.134117, y2002first}. 

The crystallographic structure of a material determines the ordered arrangements that inserted atoms can adopt on the lattice at certain compositions. The free energy is a function of the composition, parameters describing these orderings, as well as other variables such as strain and temperature. For the regimes of LCO's phase stability considered here, six parameters determined from symmetry group considerations fully describe its ordering (\S \ref{sec:order parameters}, \ref{sec:order params methods}).
The statistical mechanics simulations yield free energy density derivatives; i.e., chemical potentials, with respect to the composition and order parameters. The machine learning framework that is central to this work adaptively learns a seven-dimensional representation of the free energy density as a function of composition and six order parameters from free energy derivative data. Strain and vibrational entropy have not been included; therefore mechanics and heat transport are not accounted for. Integrable deep neural networks (IDNNs) previously introduced by the authors \cite{Teichert2019,Teichert2020}, can be trained to derivative data, e.g. chemical potentials  obtained through Monte Carlo sampling, and analytically integrated to recover the antiderivative function, e.g. the free energy density function needed in phase field models. This requires adaptive sampling of non-convex regions, extrema, boundaries of admissible regions and high error points of the seven-dimensional free energy density. 

These machine learnt thermodynamic representations have a quantifiable precision controlled by convergence tolerances. They are used to confirm a match with extensively reported phase diagrams for LCO, as well as experimental voltage measurements. We also perform large-scale DFT calculations to compute the anti-phase boundary energies and interface energies between the different phases. Finally, using the obtained free energy densities and anti-phase boundary/interface energies, we carry out phase field studies on the dynamics of an order-disorder transition in isolated particles as well as during charge-discharge cycles on single LCO particles at temperatures of practical interest. 




\subsection{Formation energies and configurations from DFT}

The  mean voltages predicted from DFT$+U$ for Li composition $x =\{0,1\}, \{0,0.5\}, \{0.5,1\}$ and increasing values of $U$ are plotted in Figure \ref{fig:Fig1}b. We found $U = 2.5$ eV for calculation of LCO formation energies to provide a good match with the experimental voltages reported in Ref. \cite{Aykol2015} for $x = \{0,1\}$ and $x = \{0.5,1\}$. DFT$+U$ computations were carried out on 333 configurations identified by the CASM (Clusters' Approach to Statistical Mechanics) software \cite{casm}. The calculated formation energies and associated convex hull with the predicted ground states are plotted in Figure \ref{fig:Fig1}c. The unit cell volume for LiCoO$_2$ was calculated to be 32.502 \AA$^3$.

Figure \ref{fig:Fig1}d shows Li in three ordered configurations encountered in this work at $x=1/2$. We found the zig-zag ordering to be the ground state with a formation energy $E_\text{f}=-0.170$ eV rather than the two row configurations ($E_\text{f}=-0.150$ and $-0.164$ eV). Charge splitting appears in our DFT results only for the row configuration on the left in Figure \ref{fig:Fig1}d.

\subsection{Statistical mechanics}

\subsubsection{Cluster expansions for $E_\text{f}$}

A cluster expansion was developed for the formation energy, $E_\text{f}$, within the CASM code using a genetic algorithm that selected 35 out of an initial 221 basis functions. The RMS error for this cluster expansion prediction was 2.49 meV per primitive cell, obtained under 10-fold cross-validation. The corresponding effective cluster interactions (ECIs) are shown in Figure \ref{fig:Fig1}e. The cluster expansion predicted  $E_\text{f}$ values are plotted with those from DFT$+U$ in Figure \ref{fig:Fig1}c. Note that the respective convex hulls  match in formation energies and configurations.

\subsubsection{Monte Carlo simulations of ordered/disordered structures}

\begin{figure}[t]
    \includegraphics[width=\textwidth]{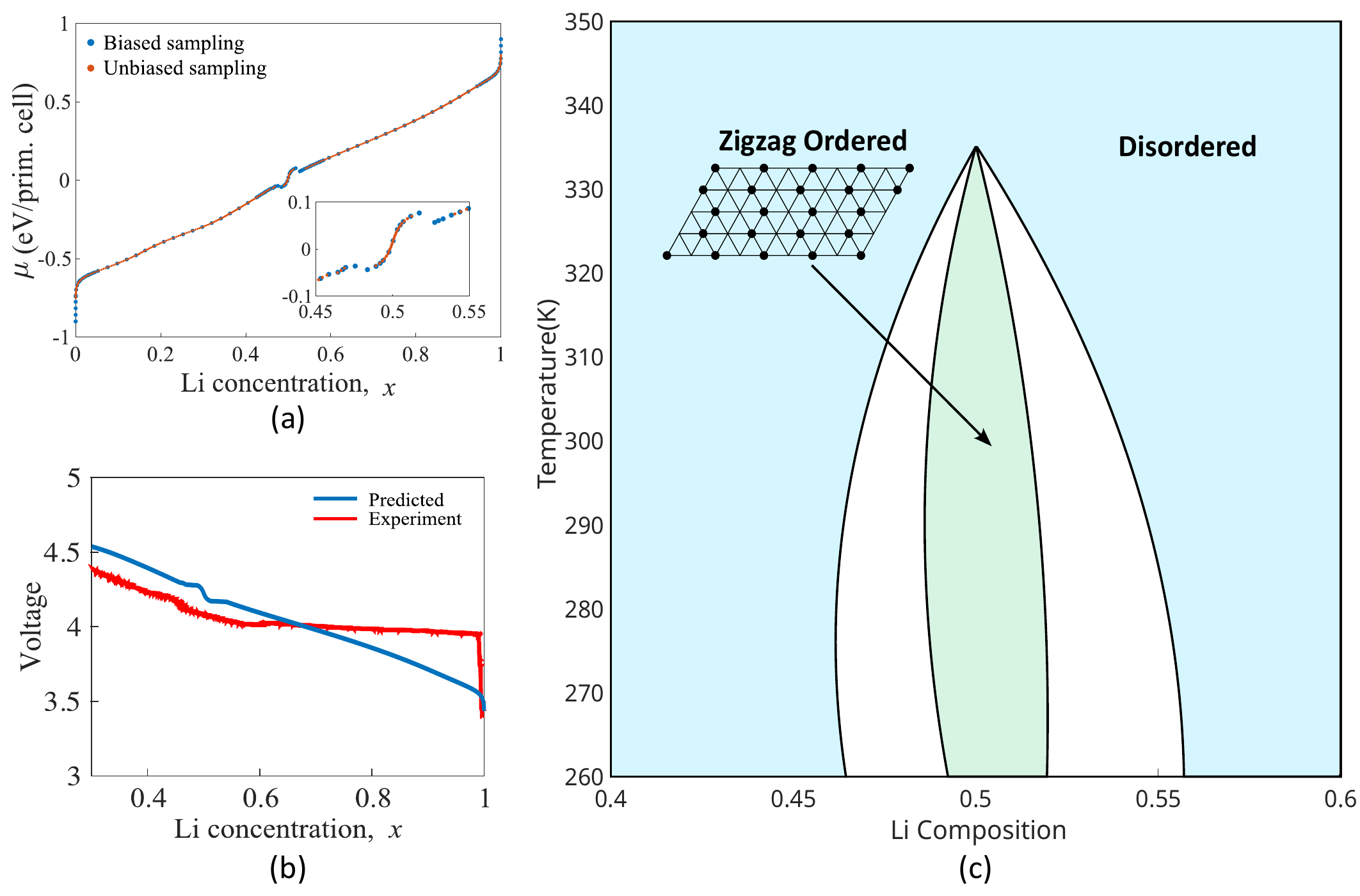}
    \caption{(a) Chemical potential values at 300 K, predicted using Monte Carlo simulations with and without bias potentials. (b) Comparison of the predicted voltage at 300 K (blue) with the experimental voltage (red) from Ref. \cite{Amatucci1996}. (c) Phase diagram for the O3 structure for LCO with the ordering of interest at $x=1/2$, based on Monte Carlo results. The gaps in unbiased sample points in (a) correspond to order-disorder phase transitions (white region) at 300 K.}
    \label{fig:Fig2}
\end{figure}


Li content as a function of chemical potential and temperature was computed through Semi-Grand Canonical Monte Carlo for temperatures ranging from 200 K to 400 K. Unbiased and biased (umbrella) sampling were used \cite{Torrie1977,Sadigh2012a,Sadigh2012b}. Two-phase regions are easily seen as gaps in the composition values of the unbiased data (Figure \ref{fig:Fig2}a). These data were used to construct a phase diagram of O3 LCO, for Li compositions $x \ge 1/3$ that is shown in Figure \ref{fig:Fig2}c.

From the Monte Carlo results, the order-disorder transition temperature for the ordering at $x\approx 1/2$ is slightly greater than 330 K, consistent with experiment \cite{Reimers1992}. A fairly good agreement also exists between the computed voltage profile at 300 K and experimental measurements \cite{Amatucci1996} ( Figure \ref{fig:Fig2}b). The order-disorder transition is visible in the fluctuation around $x \approx 1/2$.  Similar to past first-principles studies on LCO \cite{VanderVen1998,y2002first}, however, the calculations lack the metal-insulator transition that leads to the two-phase region observed over $x\sim 0.7$-$0.9$. Therefore, the corresponding voltage plateau is not captured in the simulations. The Monte Carlo results in this work, and the predicted voltage, are only for the O3 structure and, therefore, valid for compositions $x \ge 1/3$.

\subsubsection{Symmetry-adapted order parameters}
\label{sec:order parameters}
\begin{figure}[t]
    \includegraphics[width=\textwidth]{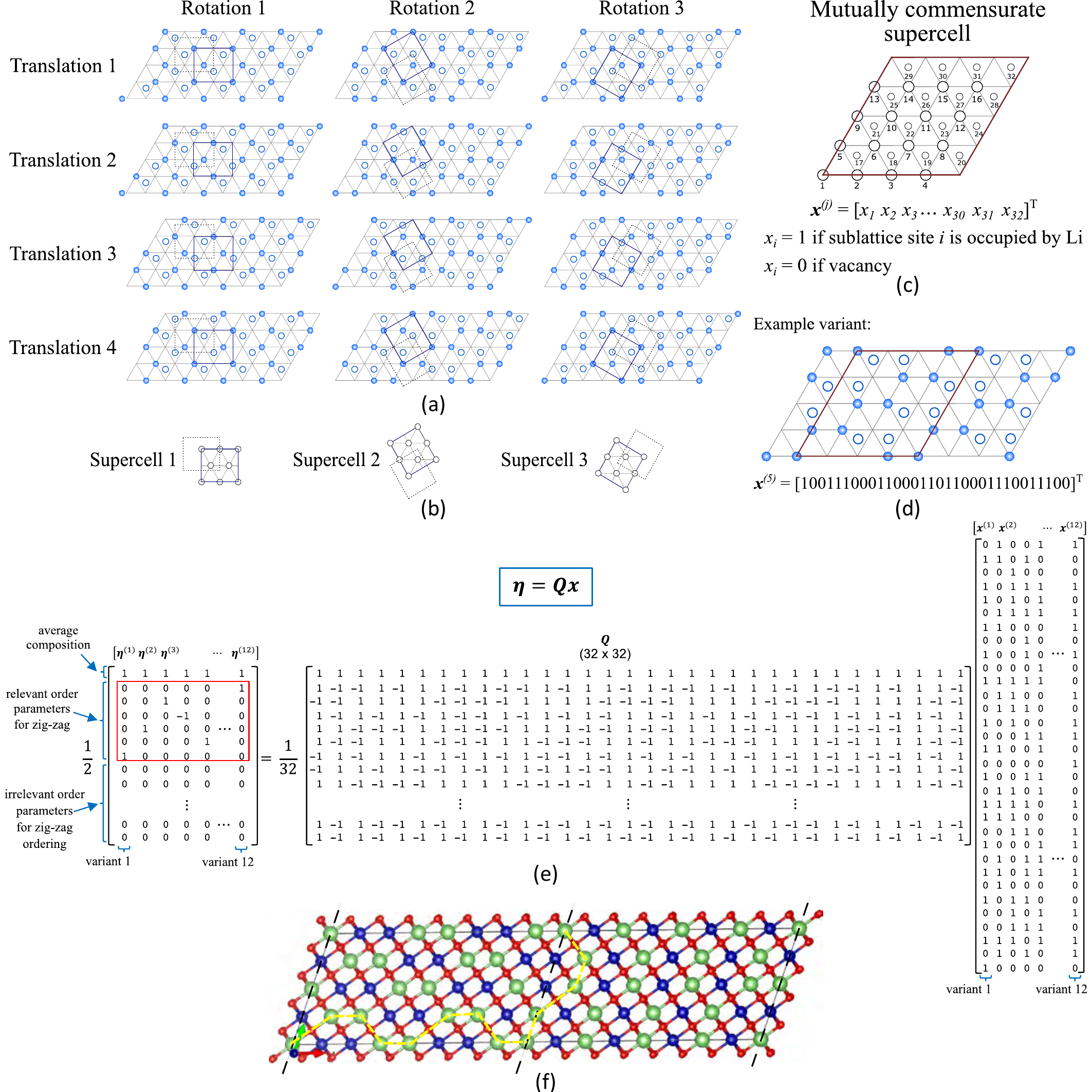}
    \caption{(a) The zig-zag ordering has 12 variants, resulting from combinations of 3 rotations and 4 translations. (b) The supercells corresponding to each variant are also shown. Note that these can remain invariant under translations, but not under rotations. The filled circles represent Li atoms occupying the same layer; empty circles are Li atoms in the  layer below (or above). (c) The mutually commensurate supercell that includes all the supercells is also shown with its 32 sublattice sites. (d) One example of the 12  variants of zig-zag ordering is shown in its sublattice representation. (e) The transformation $\boldsymbol{\eta} = \boldsymbol{Qx}$ from sublattice to order parameter space. The 12 zig-zag variants in the $\boldsymbol{\eta}$ space have zeroes in rows 8-32. Therefore seven order parameters are needed to describe the composition and ordering. (f) Atomic configurations for anti-phase boundary (dashed lines) energy computations between ordered variants $\eta_1 = 1/2$ (left) and $\eta_3 = 1/2$ (right). The yellow path traces the change in zig-zag ordering.}
    \label{fig:orderparams}
\end{figure}

Symmetry-adapted order parameters precisely represent the distinct variants of the zig-zag ordering that was identified by DFT as having the lowest energy of all possible decorations of the triangular Li sub-lattice at the composition of $x = 1/2$. The zig-zag ordering has 12 variants that are shown in  Figure \ref{fig:orderparams}a. They arise from the 3 rotations $\times$ 4 translations that belong to the symmetry group of the triangular lattice. The mutually commensurate supercell that includes the supercells of all the rotations and translations was computed and includes 32 unique sublattice sites on two successive Li layers (Figures \ref{fig:orderparams}b-\ref{fig:orderparams}c). This yields a 32-dimensional basis on which 12 vectors $\{\bsym{x}^{(1)},\ldots,\bsym{x}^{(12)}\}$ uniquely describe the 12 variants (Figures \ref{fig:orderparams}c-\ref{fig:orderparams}d).

 A more efficient representation of the 12 variants is obtained by seven order parameters, $\eta_0,\dots,\eta_6$ corresponding to the first seven rows of the matrix $\boldsymbol{\eta}$ in Figure \ref{fig:orderparams}e, with its remaining rows being zeros. Of these, $\eta_0 = x$ corresponds to the composition averaged over all 32 sublattice sites. It is associated with a single distinct eigenvalue of an orthogonal matrix $\boldsymbol{Q} \in \mathbb{R}^{32\times 32}$ that is invariant to the symmetry group of the zig-zag ordering (Figure \ref{fig:orderparams}e and \S \ref{sec:order params methods}). The other six order parameters are associated with eigenvectors of one of the degenerate eigenvalues and represent the 12 variants by $\eta_1,\dots \eta_6 = \pm 1/2$ in Figure \ref{fig:orderparams}e. 

\subsection{Large-scale DFT calculations for the anti-phase boundary energies between ordered variants}
The coherent anti-phase boundary energy between the different LCO variants at the ordering composition $x=1/2$ is another key input required for the  phase-field simulations.  Using the variants defined by $\eta_1 = 1/2$ and $\eta_3 = 1/2$  (Figure \ref{fig:orderparams}f) we have carried out large-scale DFT computations that predict an anti-phase boundary energy of $\gamma = 30.9$ mJ/m$^2$ (see \S \ref{sec:DFTintfc}). Guided by the estimate of a factor of $1/2$ relating the order-disorder interface energy to the anti-phase boundary energy~\cite{wang1998}, we use $\sim 15.45$ mJ/m$^2$ for the order-disorder interface energy.

\subsection{Active learning an integrable deep neural network (IDNN) for the seven-dimensional free energy surface}

\begin{figure}[t]
    \includegraphics[width=\textwidth]{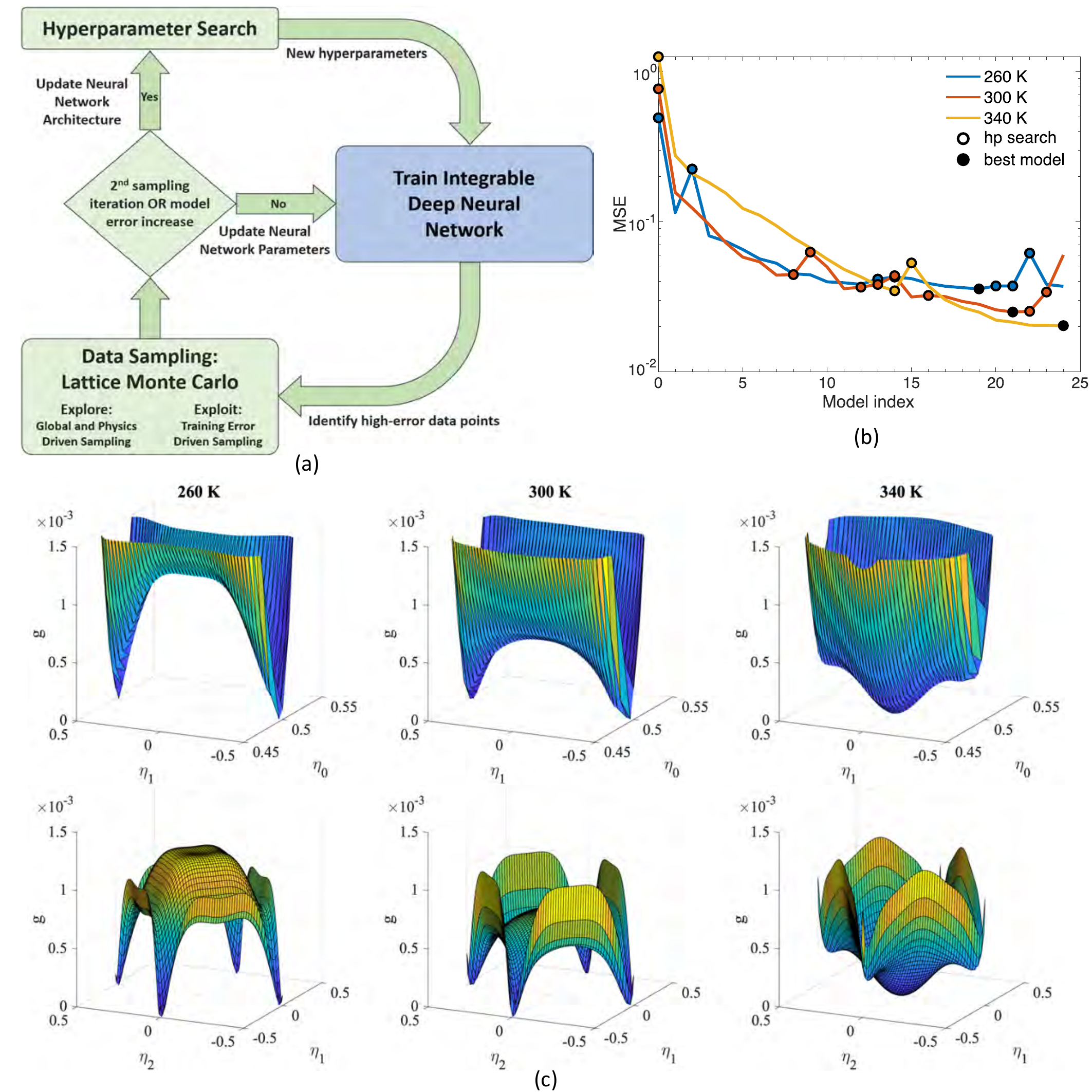}
    \caption{(a) A schematic of the active learning workflow with exploration, training and exploitation to guide additional sampling. (b) Active learning curves for the IDNN with respect to model index for 260, 300, and 340 K. Each model corresponds to progressively larger datasets. Note the logarithmic scale on the y-axis. Open circles indicate new hyperparameter searches to update the IDNN. Losses are MSEs for the final ($25^\text{th}$) dataset. Solid circles mark the best IDNN model at each temperature. (c) The free energy density surfaces plotted as $\eta_0-\eta_1$ slices ($\eta_2,\dots,\eta_6 = 0$) and $\eta_1-\eta_2$ slices ($\eta_0 = 1/2$, $\eta_3,\dots\eta_6 = 0$). The ordered variants correspond to the wells located at $\eta_0 = x \sim 1/2$ and $\eta_1,\dots,\eta_6 \sim \pm 1/2$ at $260$ and $300$ K. At $340$ K the only well is at $\eta_1,\dots,\eta_6 = 0$ for all values of $\eta_0$; i.e., only the disordered form exists.}
    \label{fig:Fig4}
\end{figure}


Figures \ref{fig:Fig4}a, b, show respectively, a schematic and the active learning curves with respect to IDNN models indexed by the training dataset number. See \S \ref{sec:IDNN} for details on IDNNs. The MSE is reported for each IDNN model using the the last ($25^\text{th}$) dataset. The best model (lowest MSE) at each temperature is marked by a solid circle. Models with hyperparameter searches performed are marked by open circles. Figure \ref{fig:Fig4}c shows the free energy density surfaces as $\eta_0-\eta_1$ and $\eta_1-\eta_2$ slices for the best model at each temperature. The active learning resulted in IDNNs with two hidden layers and 173 neurons per layer at 260 K, three hidden layers and 163 neurons per layer at 300 K and two hidden layers with 193 neurons per layer at 340 K. 

\subsubsection{Symmetry-respecting functions as IDNN features}
\label{sec:IDNNsymm}
The features for IDNN representations of the  free energy density are the following polynomials, up to sixth order in $\eta_0,\dots \eta_6$. They are invariant under transformations of the triangular Li sublattice that map between the ordered variants. Therefore, IDNN representations in terms of these polynomials inherit symmetry with respect to the free energy. These symmetry-respecting IDNN features are determined by applying the Reynolds operator to monomials in $\eta_0,\dots,\eta_6$ \cite{dresselhaus2007group}:

{\begin{minipage}[h]{0.45\textwidth}
\small
	\begin{align*}
    h_1 &= \eta_0\\
    h_2 &= \frac{2}{3}\sum_{i=1}^6\eta_i^2\\
    h_3 &= \frac{8}{3}\sum_{i=1}^6\eta_i^4\\
    \begin{split}
    h_4 &= \frac{4}{3}\big[\left(\eta_1^2+ \eta_2^2\right)\left(\eta_3^2 + \eta_4^2 + \eta_5^2 + \eta_6^2\right) + \\
    &\phantom{=\left(\eta_1^2+ \eta_2^2\right)}\left(\eta_3^2 +\eta_6^2\right)\left(\eta_4^2 + \eta_5^2\right)\big]
    \end{split}\\
    h_5 &= \frac{16}{3}\left(\eta_1^2\eta_2^2 + \eta_3^2\eta_6^2 + \eta_4^2\eta_5^2\right)\\
    h_6 &= \frac{32}{3}\sum_{i=1}^6\eta_i^6
    \end{align*}
\end{minipage}
\begin{minipage}[h]{0.45\textwidth}    
\small
    \begin{align}
    \begin{split}
        h_7 &= \frac{8}{3}\big[\left(\eta_1^4+ \eta_2^4\right)\left(\eta_3^2 + \eta_4^2 + \eta_5^2 + \eta_6^2\right) + \left(\eta_3^4 +\eta_6^4\right)\left(\eta_4^2 + \eta_5^2\right) + \nonumber\\
        &\phantom{= \frac{\sqrt{10}}{4}\big[}\left(\eta_1^2+ \eta_2^2\right)\left(\eta_3^4 + \eta_4^4 + \eta_5^4 + \eta_6^4\right) + \left(\eta_3^2 +\eta_6^2\right)\left(\eta_4^4 + \eta_5^4\right)\big]
    \end{split}\nonumber\\
    \begin{split}
        h_8 &= \frac{16}{3}\big[\eta_1^2\eta_2^2(\eta_3^2 + \eta_4^2 + \eta_5^2 + \eta_6^2) + \eta_3^2\eta_6^2(\eta_1^2 + \eta_2^2 + \eta_4^2 + \eta_5^2) +\nonumber\\
        &\phantom{=\frac{\sqrt{30}}{2}\big(}\eta_4^2\eta_5^2(\eta_1^2 + \eta_2^2 + \eta_3^2 + \eta_6^2)\big]
    \end{split}\nonumber\\
    h_9 &= \frac{32}{3}\left(\eta_1^4\eta_2^2 + \eta_1^2\eta_2^4 + \eta_3^4\eta_6^2 + \eta_3^2\eta_6^4 + \eta_4^4\eta_5^2 + \eta_4^2\eta_5^4\right)\nonumber\\
    h_{10} &= 8(\eta_1^2 + \eta_2^2)(\eta_3^2 + \eta_6^2)(\eta_4^2 + \eta_5^2)\nonumber\\
    \begin{split}
    h_{11} &= \frac{64}{5}\big[\eta_1\eta_2(\eta_3^2 - \eta_6^2)(\eta_4^2 - \eta_5^2) + \eta_3\eta_6(\eta_1^2 - \eta_2^2)(\eta_4^2 - \eta_5^2) + \nonumber\\
    &\phantom{= \frac{64}{5}\big[}\eta_4\eta_5(\eta_1^2 - \eta_2^2)(\eta_3^2 - \eta_6^2)\big]
    \end{split}\nonumber\\
    h_{12} &= 64\sqrt{5}\eta_1\eta_2\eta_3\eta_4\eta_5\eta_6
    \label{eq:symmfunc}
\end{align}
\end{minipage}
}

\subsection{Phase field simulations}
\label{sec:phasefieldresults}

\begin{figure}[tb]
    \centering
    \begin{minipage}[b]{0.75\textwidth}
        \includegraphics[width=\textwidth]{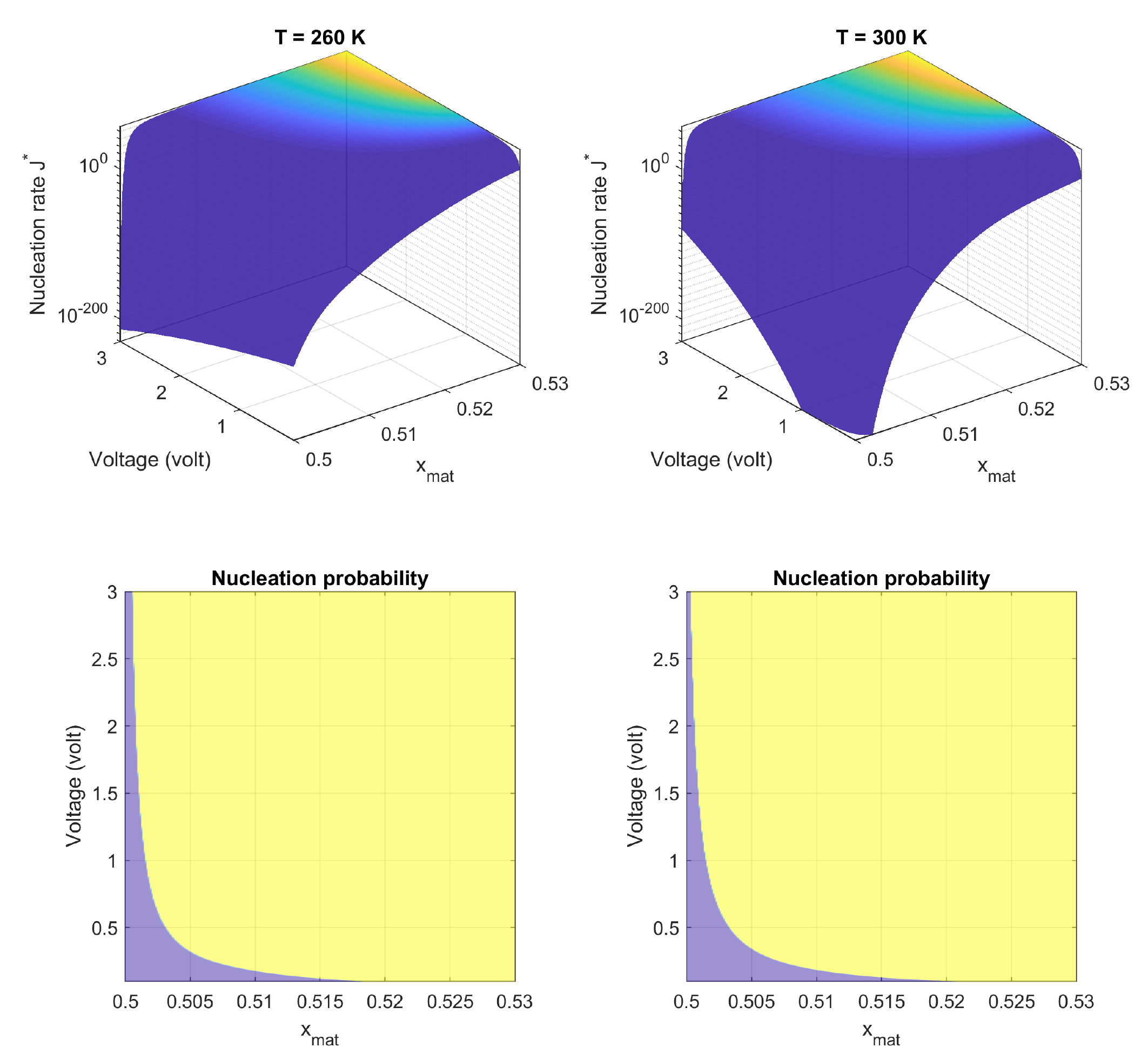}
        \subcaption{Nucleation rates and probabilities}
    \end{minipage}
    \begin{minipage}[t]{1\textwidth}
        \includegraphics[width=\textwidth]{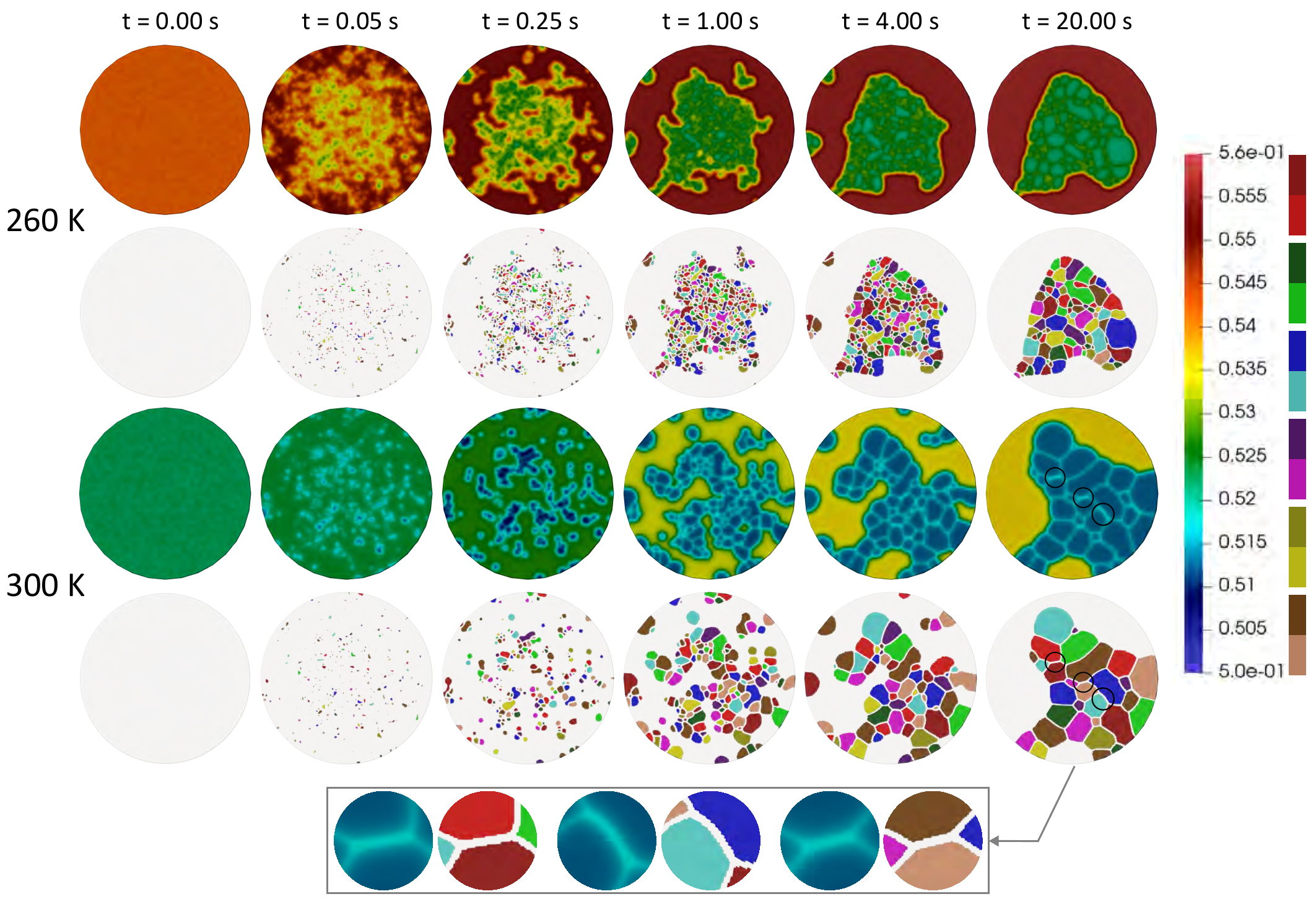}
        \subcaption{Phase field simulations of growing ordered regions}
    \end{minipage}
\caption{(a)  Nucleation rates at 260 K and 300 K as functions of the voltage and the composition of a potential nucleation site in the disordered region $x_\mathrm{mat}$, and the corresponding nucleation probabilities $P_n$ for $\Delta t=1\mathrm{e}-4$ sec. The dark and light colors show $P_n=0$ and $P_n=1$, respectively (b) 2D phase field simulations at $260$ K ($300$ K) showing the Li composition in a 1 $\mu$m diameter particle. The initial Li composition wass randomly perturbed about $x = 0.5425$ ($x = 0.5225$) and no boundary flux. The inset shows some of the widest order-order interfaces at 300 K formed by $\eta_1 = \pm \frac{1}{2}$ (left),  $\eta_3 = \pm \frac{1}{2}$ (middle), and $\eta_6 = \pm \frac{1}{2}$ (right).}
\label{fig:Fig5}
\end{figure}

The representations learnt by the IDNN show that the decrease in free energy density for heterogeneous nucleation of an ordered phase at $260$ K is $1$ meV/unit cell, while at $300$ K it is $0.5$ meV/unit cell (Figure S1).  Using the DFT results for anti-phase boundary energy $\gamma = 30.9$ mJ/m$^2$ and order-disorder interface energy $\sim 15.45$ mJ/m$^2$, we obtain a far higher nucleation rate of the ordered phase at $260$ K than at  $300$ K. Figure \ref{fig:Fig5}a shows the nucleation rates \emph{versus} applied voltage and the associated composition of the disordered matrix ($x_\text{mat}$) obtained  at $260$ and $300$ K  (see \S S3 for further details). At $340$ K only the disordered phase exists, as seen in the phase diagram (Figure \ref{fig:Fig2}c) and the learnt free energy density surfaces (Figure \ref{fig:Fig4}c). 

The scale bridging framework is applied to two  phase field studies of LCO cathodes involving the order-disorder transition around $x = 1/2$. The gradient parameters in the Cahn-Hilliard and Allen-Cahn phase field equations (\ref{eq:continuouschempots}-\ref{eqn:CH-AC}) were found to be $\chi_0 = 1\times 10^{-4}\; \text{mJ/m}$ ($1.88\times 10^{-4}\; \text{mJ/m}$) and $\chi_1,\dots,\chi_6 = 2.12\times 10^{-8}\; \text{mJ/m}$ ($4.91\times 10^{-8}\; \text{mJ/m}$) from the anti-phase boundary/interface energies and the IDNN representations of the free energy surfaces at $260$ K ($300$ K) temperature (see \S \ref{sec:phasefield}). 

The first phase field study considers the dynamics of the order-disorder transition in two-dimensional particles without external Li fluxes. Figure \ref{fig:Fig5}b shows a dynamics closely resembling spinodal decomposition followed by Ostwald ripening of ordered phase domains in a particle of diameter $1\;\mu$m, which has an average composition  $x = 0.525$ at $260$ K and $x = 0.53$ at $300$ K. These compositions lie in the order-disorder region of the phase diagram. At each temperature, the first row shows composition, $x$ and the second shows the orderings $\eta_1,\dots,\eta_6$. Investigation of the seven-dimensional free energy surface shows that a single one out of twelve variants forms within each ordered region (\S S7 and Fig S7). These dynamics were not run until equilibrium microstructures were obtained (at $260$ and $300$ K) because of the long computation time needed. However, the mass fractions of the ordered and disordered phases predicted by the coupled phase field equations at each temperature are in good agreement with the lever rule applied to the phase diagram after accounting for the slight shift in order-disorder boundaries as predicted by the IDNN learnt free energy density (\S S7, \S S8, and  Fig S8). This suggests that near-equilibrium states have been attained in the computations. We call attention to the greater numbers of ordered regions and the narrower anti-phase boundaries/interfaces at $260$  K relative to $300$ K due to the correspondingly lower boundary/interface energies.  
Note the higher nucleation rates but the slower motion of anti-phase boundaries at $260$ K over $300$ K. The nucleation rate calculations appear in \S S3(see Figures S1-S4).

The interface widths are larger between the ordered variants corresponding to $\eta_i = \pm 1/2$ than a pair of variants represented by $\eta_i = 1/2$ or $-1/2$ and $\eta_j = 1/2$ or $-1/2$, where $i\neq j$, instances of which are shown as an inset on the fifth row of Figure \ref{fig:Fig5}b. This results from symmetries learnt by the IDNN that are additional to those imposed as features above via the Reynolds operator (\S \ref{sec:IDNNsymm}), and are discussed further below.

\begin{figure}[tb]
	\includegraphics[width=1.0\textwidth]{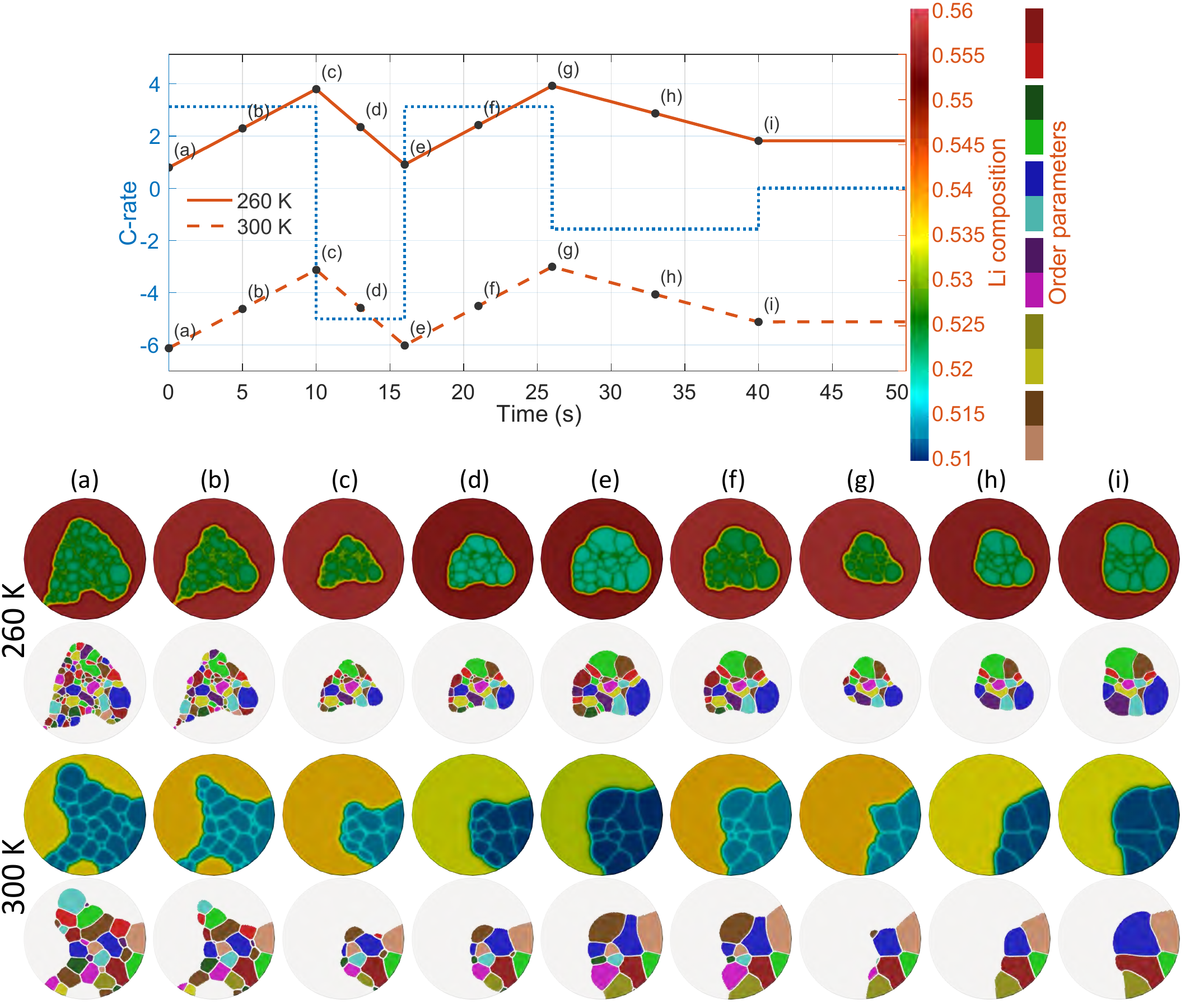}
    \caption{Phase field results showing the Li composition field resulting from applying a cycling current density for 260 and 300 K, for a 1 $\mu$m diameter particle. The dashed blue line shows the applied C-rate, where a positive sign denotes discharging and negative is charging, and the red lines show the corresponding average Li composition of the particle.}	
    \label{fig:Fig6}
\end{figure}

 The second phase field study is also on two-dimensional particles at 260 and 300 K, and is focused on the effect of charge cycling of an LCO particle, by specifying time-varying externally applied Li fluxes \cite{Jiang2016}. See (Figure \ref{fig:Fig6}). At each temperature, the final configuration of ordered variants in a disordered matrix from the study in Figure \ref{fig:Fig5} was taken as the initial condition. An initial current density of 0.6677 A m$^{-2}$ was applied. The applied current density was adjusted to -1.0683, 0.6677, -0.3338, and 0 A m$^{-2}$ at 10, 16, 26, and 40 s, respectively. 
This cycling corresponds to C-rates of 3.125, -5, 3.125, -1.5625, and 0 C, where a positive sign denotes discharging and negative is charging.
Over the composition range, discharging, a-c and e-g in Figure \ref{fig:Fig6}, injects Li, driving the particle further into the disordered regime and shrinking the ordered domains; conversely, charging, c-e and g-i, extracts Li, making the ordered domains grow. Note the continued growth/shrinkage of different ordered domains as the anti-phase boundaries migrate and decrease in curvature.
This cycle at 260 (300) K results in a net discharging, starting from the average Li composition of 0.5425 (0.5225) and increasing that to an average final composition of 0.5455 (0.5254), over the voltage plateau of $\sim 4.15$ V (see Figure \mbox{\ref{fig:Fig2}b}).

\section{Discussion}
\label{sec:discussion}

Our study has been an extensive one, beginning with hundreds of DFT computations to inform tens of thousands of semi-grand canonical Monte Carlo calculations. The active learning framework guided Monte Carlo sampling for data to train IDNN representations of continuum, homogeneous free energy density functions. With gradient free energies informed by other large-scale DFT calculations of anti-phase boundary energies, this continuum free energy description enabled detailed first principles-informed phase field studies. To our knowledge, a systematic scale bridging framework of this nature has not been presented for a materials system. Guided by machine learning, it endows our predictions with quantifiable precision at every stage. In order to place our work in context, we address its scope, open questions and future developments starting with the DFT calculations, and proceeding through first principles statistical mechanics and machine learning to the continuum computations.

We have carried out DFT$+U$ calculations on the O3-layering structure, which is the stable form of LCO for $x \ge 1/3$. The $U$ parameter was  obtained by matching with experimental  measurements of the average voltage between $x =0$ and $x = 1$, $x =0$ and $x = 1/2$, and $x = 1/2$ and $x = 1$ (see Figure \ref{fig:Fig1}b). A different approach could be to choose $U$ for an optimal fit across the entire composition range in Figure \ref{fig:Fig2}a by iteratively updating the cluster expansion for $E_\text{f}$ and making voltage predictions at intermediate $x$. Empirical approaches to determine the $U$ parameter can be completely circumvented by using the self-consistent Hubbard method~\cite{timrov2021}.



Moving on to the configurations predicted by our DFT study, we note the experimental evidence for the row orderings at $x = 1/2$ in Figure \ref{fig:Fig1}d \cite{Reimers1992,ShaoHorn2003,Takahashi2007}. Takahashi, et al. \cite{Takahashi2007} presented the row ordering on the left, and Shao-Horn, et al. \cite{ShaoHorn2003} suggested a different row ordering in the center. In their DFT study, Van der Ven, et al. \cite{VanderVen1998} found the row ordering in the center to have the lowest formation energy, followed closely by the zig-zag ordering on the right. They make the observation, however, that the energy difference, of order 1 meV in their work, is small enough that it is difficult to determine the ground state with certainty solely from DFT calculations. The differences in formation energies between our results and those of earlier DFT work are likely due to our using DFT$+U$, vdW-DF, and generalized gradient approximations (GGA), as opposed to the local-density approximation (LDA) used in the prior study. The scale bridging framework remains applicable to the other orderings, although presented here for the zig-zag case.

Given this zig-zag structure, the symmetry-adapted order parameters $\eta_0,\dots, \eta_6$ are important to the scale-bridging framework by enabling a more efficient representation of the 12 variants in $\mathbb{R}^7$ compared to the sub-lattice vectors $\boldsymbol{x} \in \mathbb{R}^{32}$, since $\eta_i = \pm 1/2$, $i = 1,\dots,6$ represent perfect orderings (Figure \ref{fig:orderparams}e). Monte Carlo sampling in $\mathbb{R}^7$ over regions of higher error, in the free energy wells, over the order-disorder transitions and near the bounds in $\boldsymbol{\eta}$-space, however, presents a challenge, which we have approached with the active learning workflow (Figure \ref{fig:Fig4}a), which, as Figure \ref{fig:Fig4}b shows, improves the IDNN training over datasets, but not monotonically. As the Monte-Carlo sampled data grow with active learning-guided exploitation and exploration, an increasing complexity of representation is needed for the IDNN homogeneous free energy density model. New hyperparameter searches were imposed if the two most recent IDNN model updates  showed increasing MSE against the most recent dataset.  This IDNN model plotted over two-dimensional slices of $\mathbb{R}^7$ (Figure \ref{fig:Fig4}c) is more complex at $300$ K with three hidden layers of 163 neurons than at $260$ K (two hidden layers, 173 neurons each) and $340$ K (two hidden layers, 193 neurons each).

In addition to the homogeneous free energy density, the energy of the anti-phase boundary between any two of the 12 variants also is needed for phase field computations. The large-scale DFT computations yielded anti-phase boundary energy $\gamma = 30.9$ mJm$^{-2}$ between $\eta_1 = 1/2$ and $\eta_3 = 1/2$ variants (Figure \ref{fig:orderparams}f). However, order-disorder interface energy computation presents difficulties for systematic convergence studies since randomly sampled disordered configurations change with size. We have therefore followed the approach suggested in the literature \cite{wang1998} that the order-disorder interface energy is approximately half the order-order anti-phase boundary energy. While the state of order/disorder causes local atomic relaxation, the lattice remains coherent. Therefore, the elastic stresses at anti-phase boundaries and order-disorder interfaces are not expected to be large, and these transitions, while important to the phase dynamics, may not contribute to the degradation as discussed in the experimental literature \cite{choi2006particle,leng2015effect,pender2020electrode}. 

This machine learning-enabled scale bridging from first-principles statistical mechanics to continuum free energy functions allows quantitatively rigorous phase field computations. These reveal complexity that is linkable to the lower scales: Thicker anti-phase boundaries form between (A) ordered variants represented by extreme values of the same parameter, $\eta_i = \pm 1/2$ than between (B) variants represented by $\eta_i = 1/2$ or $-1/2$ and $\eta_j = 1/2$ or $-1/2$, $i\neq j$ (\S \ref{sec:phasefieldresults}, Figure \ref{fig:Fig5}b bottom row). Type A must include a disordered region $(\eta_1,\dots,\eta_6 =0)$, which does not appear in type B. This is explained by the two-dimensional homogeneous free energy density slices in Figure \ref{fig:Fig4}c. Consider an example of type B with $x = \eta_0 = 1/2$ and $\eta_3,\dots, \eta_6 = 0$. The homogeneous free energy density surface has paths between domains with $\eta_1 = 1/2$ or $-1/2$ and $\eta_2 = 1/2$ or $-1/2$ that do not pass through $\eta_1, \eta_2 = 0$. Consequently, the transition between regions with orderings of type B does not include a fully disordered interface, and appears as a sharp transition (Figure \ref{fig:Fig5}b, bottom inset). In contrast, as also seen in Figure \ref{fig:Fig4}c, for $x = \eta_0 = 1/2$ and $\eta_3,\dots, \eta_6 = 0$ the path between type A variants: $(\eta_1,\eta_2) = (-1/2,0)$ and $(\eta_1,\eta_2) = (1/2,0)$ has to pass through the point $x = \eta_0 =1/2$ and $(\eta_1,\dots,\eta_6) = (0,\dots,0)$--a fully disordered region. This anti-phase boundary appears with a larger width.  The Allen-Cahn equation (\ref{eqn:CH-AC}) shows that in a physical particle at equilibrium, the order parameter profiles along paths that are perpendicular to type A anti-phase boundaries are well approximated by the $\tanh$ function of position. For  type B variants, the IDNN learns scale bridging-informed, physics-constrained symmetries of the form $\mu_i = \mu_j$ along paths $\eta_i = \pm\eta_j \pm 1/2$ in $\boldsymbol{\eta} \in \mathbb{R}^7$. These symmetries are distinct from those imposed on the IDNN features in \S \ref{sec:IDNNsymm}, Equation (\ref{eq:symmfunc}).

In addition to equilibrium structures, the phase dynamics of order-disorder transition are of interest. They require nucleation of an ordered variant at composition $x_\text{nuc}$ from a disordered matrix at $x_\text{mat}$, which depends on the associated free energy density decrease. For homogeneous nucleation, our scale-bridging results in $\Delta g_\text{hom}^{260} \approx 2\Delta g_\text{hom}^{300}$. A significantly higher nucleation rate is thus attained at $260$ K than at $300$ K. Under an external field $V$ we have a critical free energy decrease for heterogeneous nucleation $\Delta g^*_\mathrm{het} = f(\theta)(16\pi\gamma^2/3(\Delta g_\mathrm{hom}^3 \pm V(x_\text{mat}-x_\text{nuc}))$ depending on the sign of $V$, where $f(\theta)$ is a function of the contact angle of the nucleus. We find $x_\text{mat} - x_\text{nuc} \sim 0.01$. See \S S3 and Figures S1b, S1c \cite{legoues1984,simmons2000,simmons2004,sear2007}. With $x_\text{nuc} = 1/2$, Figure \ref{fig:Fig5}a shows the variation of nucleation rate and probability with voltage and $x_\text{mat}$. The very steep, almost discontinuous, transition from zero to exponentially high rates and probability from $0$ to $1$ guarantees that the formation of the ordered phases is essentially instantaneous at either $260$ K or $300$ K once the combination of voltage and $x_\text{mat}$ is favorable.

More insight to the order-disorder transition at $x=1/2$ can be gained from the IDNN homogeneous free energy density function model itself. An examination of the lowest free energy paths followed on this seven-dimensional manifold, in combination with the phase field dynamics of the Cahn-Hilliard equation for $\eta_0$ and the Allen-Cahn equation for $\eta_1$, can provide detailed insight to these transitions. If ordering were ignored, we obtain a one-dimensional homogeneous free energy density parameterized by only the composition, $x = \eta_0$. In the neighborhood of $x = 1/2$ non-convexities appear that suggest spinodal decomposition as a phase separation mechanism (\S S1.5, \S S4, Figs S4, and S8a). However, an analysis of the seven-dimensional IDNN free energy density model shows that, in the ordered region five of six order parameters remain at zero, while one changes nearly discontinuously (with respect to $\eta_0$) to $\sim 1/2$ (see \S S7, Figure S9). Two-dimensional slices (Fig \ref{fig:Fig4}c and Fig S7) viewed in the $g-\eta_0$ plane show  a discontinuity in slope; i.e., chemical potential $\mu_0$. With a one-dimensional free energy density learnt by the IDNN using only $\eta_0$ as input this $\mu_0$-discontinuity gets smoothed to a non-convexity of the $g-\eta_0$ curve (\S S7 and Fig S8a) leading to the suggestion of spinodal decomposition noted above. The  full seven-dimensional IDNN model, however, reveals a more complex trajectory: As $\eta_0$ enters the region of ordering, a non-convexity is encountered in the $g-\eta_0-\eta_i$ space for some $i\in \{1,\dots,6\}$. Fig S7 shows this for the $g-\eta_0-\eta_1$ space exploiting the symmetry with respect to $\eta_1,\dots,\eta_6$. However, it is not a non-convexity of $g$ with respect to $\eta_0$ alone, but one that is aligned with eigenvectors of the Hessian $\partial^2 g/\partial\eta_0\partial \eta_1$ for $(\eta_2,\dots,\eta_6) = (0,\dots,0)$. This is a spinodal decomposition in the $g-(\eta_0,\dots,\eta_6)$ space, and not fully described by considering only the $g-\eta_0$ space. The seven-dimensional perspective resolves  the phase transition that determines the chemical potentials, and thus the voltage, at the lower lithiation limit of LCO. Thereby, it guides a complete understanding of battery performance and ultimately, its control. Figure S8b locates the points related to the above analysis on the phase diagram. Enabled by machine learning, the scale-bridging reveals details of the dynamics under quantifiable precision that is directly connected to first principles statistical mechanics and DFT calculations. 

As seen in Figure \ref{fig:Fig5}b, notable variations can arise in the order-disorder morphology of a cathode particle undergoing rapid charge-discharge cycles in a temperature range of $260-300$ K (and up to $330$ K) that falls within the window of relevance for actual batteries. Order-disorder transitions do not create lattice mismatches, and therefore are not expected to cause stresses-induced damage \cite{wang1999tem,pender2020electrode}. The absence of order-disorder transitions above $340$ K for the composition range studied (Figures \ref{fig:Fig2}c and \ref{fig:Fig4}c) eliminates even the low stresses that may otherwise form at anti-phase boundaries. This could motivate doping with elements that stabilize the disordered structure. Such a mechanism has relevance to the Al and La doped stabilization of the order-disorder transition as observed recently \cite{pender2020electrode}.

Smaller particles have a more uniform morphology. This is well-understood: The length scale of the transient microstructure in the simulations is defined by the gradient coefficient, which is related to the anti-phase boundary and interfacial energies, and is independent of the particle size. As the particle size decreases from $1\;\mu$m to $50$ nm, the result is a less complex microstructure. See \S S4 Figure S6 and Ref \cite{teichert2021li}. Current cycling $50$ nm particles at the rates considered here showed no differences with temperature. The particles are small enough that the relative values of the mobility, particle size, and microstructure length scale lead the Li atoms to diffuse uniformly and extremely rapidly in comparison with the time scale of the simulation such that any temperature effects are not seen (Figure S6).

We note that for compositions $x < 1/3$, DFT$+U$ studies will need to be completed for the H1-3 structure, which is the stable form over the lower compositional range, and subjected to our scale-bridging framework, thus extending it to the entire composition range. A systematic incorporation of additional effects remains. These include elasticity, changes in lattice structure and vibrational entropy, which can be scaled up to continuum thermomechanics. An extension to the equations of electrochemistry at the active particle-electrolyte scale will allow the study of phenomena such as the mosaic instability observed in nanoparticulate batteries \cite{orvananos2014particle}. Our predictions rest upon the fidelity of the DFT methods, which also are far from \emph{ab initio}. Therefore, we do not claim to present quantitatively accurate computational predictions. Yet, we have furnished a scale-bridging framework to systematically infuse continuum phase field computations with quantitative information from first principles statistical mechanics with a precision that can be tightened. This will allow us to study how morphology evolves under various microstructure-specific, environmental, kinetic and cycling conditions at a computational fidelity not previously accessed with regard to the length scales. From such simulations we can make connections to observations in the experimental literature on LCO \cite{wang1999tem,choi2006particle,pender2020electrode,leng2015effect,merryweather2021operando}.

\section{Methods}
\label{sec:methods}
\subsection{DFT calculations for energies}
\label{sec:DFTmethods}

The formation energy $E$ of a configuration with Li composition $x$ can be calculated using total energy values computed using density functional theory (DFT) according to the following equation \cite{VanderVen1998}:
\begin{align}
    E_{\mathrm{Li}_{x}\mathrm{CoO}_2} &= E^\mathrm{tot} - xE^\mathrm{tot}_{\mathrm{LiCoO}_2} - (1 - x)E^\mathrm{tot}_{\mathrm{CoO}_2}
    \label{eqn:form_energy}
\end{align}
where $E^\mathrm{tot}$, $E^\mathrm{tot}_{\mathrm{LiCoO}_2}$, and $E^\mathrm{tot}_{\mathrm{CoO}_2}$ are the total energies for the given configuration, LiCoO$_2$, and CoO$_2$, respectively. 

We use a simplified rotational-invariant formulation of DFT+$U$~\cite{Cococcioni2005} and a vdW-DF exchange correlation functional, namely the optB88 exchange correlation functional \cite{Thonhauser2007,Klime2009,Langreth2009}, to calculate the formation energy on a chosen subset of LCO configurations using \texttt{Quantum Espresso} \cite{QE-2009,QE-2017}. The \texttt{CASM} (Clusters' Approach to Statistical Mechanics) software suite was used to identify a set of configurations with the O3 crystal structure for parametrizing the cluster expansion. 

We calibrate the Hubbard $U$ parameter to match experimental average lithiation voltages over various ranges of Li composition. The average voltage from $x_1$ to $x_2$ is calculated using the following equation \cite{Aydinol1997,Meredig2010,Aykol2015}:
\begin{align}
    V = -\frac{E_{\mathrm{Li}_{x_2}\mathrm{CoO}_2} - E_{\mathrm{Li}_{x_1}\mathrm{CoO}_2} - (x_2 - x_1)E_{\mathrm{Li}}}{(x_2 - x_1)e}
\end{align}
where $E$ is the calculated total energy from DFT and $e$ is the charge on an electron. We compute the average voltage for Li composition: $\{x_1,x_2\} = \{0,1\}, \{0,1/2\}$ and $\{1/2,1\}$. By comparing with experimental voltages (Figure \ref{fig:Fig1}b), we select an appropriate value of $U = 2.5$ eV. Additional details on the DFT calculations for formation energy are provided in \S S1.1.


\subsubsection{Interface energy}
\label{sec:DFTintfc}
Large-scale DFT computations were performed for the coherent interface energy of the anti-phase boundary between the $\eta_1=1/2$ and $\eta_3=1/2$ ordered LCO rotational variants. These computations were performed using the DFT-FE software~\cite{das2022dft,MOTAMARRI2020106853,Motamarri2018,MOTAMARRI2013308}, a recently developed massively parallel open-source code for large-scale real-space Kohn-Sham DFT studies based on a finite-element discretization. We employ the PBE exchange-correlation functional and the optimized norm-conserving Vanderbilt pseudopotentials (ONCV)~\cite{oncv2013} from the Pseudo Dojo library~\cite{van2018pseudodojo}. 

Table S1 shows the energies and system sizes used to determine the anti-phase boundary energy $\gamma$, which is estimated to be  30.9 mJ/$\textrm{m}^2$. Details of the periodic simulation cells, treatment of anti-phase boundaries, accounting of elastic misfit strain energy, and convergence tolerances are provided in \S S1.1.1.

\subsection{Statistical mechanics}
\label{sec:statmech}


\subsubsection{Cluster expansion for formation energy}
We adopt cluster expansions to access the large numbers of configurations needed in the statistical mechanics studies. The formation energy is written as $E_\text{f}(\bsym{\sigma})$, where  $\bsym{\sigma}$ is the configuration vector with the occupancy variable $\sigma_i = 1$ if Li occupies the site $i$ and $\sigma_i = 0$ if the site contains a vacancy. We define $E_\text{f}(\bsym{\sigma})$ by computing the formation energy with DFT for a subset of configurations and use these values to parameterize a cluster expansion \cite{Sanchez1984,deFontaine1994} as a rapidly queryable surrogate for the formation energy. We use the \texttt{CASM} software\cite{casm}, which facilitates the construction and parameterization of cluster expansion Hamiltonians and their use in Monte Carlo simulations, to select configurations for the DFT computations and perform the statistical mechanics calculations in this work \cite{VanderVen2010,thomas2013,puchala2013}. 
Details on the definition of cluster basis functions, effective cluster interaction coefficients, regression techniques and algorithms to choose among the candidate basis functions appear in \S S1.2.1.

\subsubsection{Symmetry-adapted order parameters}
\label{sec:order params methods}
We follow Natarajan, et al \cite{Natarajan2017} for identifying symmetry-adapted order parameters representing the variants in Figure \ref{fig:orderparams}a. 
The symmetry group, $P$, of the zig-zag ordering consists of 384 unique linear transformations between the 12 variants  each represented by a matrix in $\mathbb{R}^{32\times 32}$. Following the algorithm in Ref. \cite{thomas2017}, we constructed a $P$-invariant matrix and performed its eigenvalue decomposition resulting in eight nonzero eigenvalues: two distinct, two repeated three times, and four repeated six times, and eight corresponding sets of eigenvectors, for a total of 32 eigenvectors forming the rows of the orthogonal transformation matrix $\bsym{Q}\in\mathbb{R}^{32\times 32}$, that maps a sublattice composition vector $\bsym{x} \in \mathbb{R}^{32}$ to order parameter space $\bsym{\eta} \in \mathbb{R}^{32}$.

Using zero indexing, components $\eta^{(i)}_7$ through $\eta^{(i)}_{31}$ of $\bsym{\eta} \in \mathbb{R}^{32}$ are zero for all variants $i = 1,\dots, 12$, and are irrelevant for describing the zig-zag ordering (Figure \ref{fig:orderparams}e). Of the seven relevant order parameters, $\eta_0$ is associated with one of the distinct eigenvalues and corresponds to the composition averaged over all 32 sublattice sites; i.e., $\eta_0 = x$. The other six order parameters, $\eta_1,\dots,\eta_6$ are associated with one of the degenerate eigenvalues that has six corresponding eigenvectors.


Since the  free energy density is invariant under transformations of the triangular Li sublattice that map between the ordered variants, the IDNN representation is presented with features that are symmetric functions of $\eta_0,\dots \eta_6$ under these transformations. Monomials of up to sixth order were chosen and subjected to the Reynolds operator, by summing: 
\begin{equation}
 h(\bsym{\eta}) = \sum_{\bsym{M}^{(\eta)}\in\mathcal{M}}f(\bsym{M}^{(\eta)}\bsym{\eta}),
\end{equation}
where for each $\bsym{M} \in P$ we have $\bsym{M}^\eta = \bsym{QM}$. This operation yields the P-invariant polynomial functions in Eq. (\ref{eq:symmfunc}) as IDNN features. Further details on the symmetry-respecting functions of the order parameters in $\boldsymbol{\eta}$ space are available in \S S5

\subsubsection{Monte Carlo sampling}
Given $E_\text{f}(\bsym{\sigma})$, we sample within the semi-grand canonical ensemble, in which the chemical potential is specified and the corresponding composition and/or order parameters are determined through ensemble averaging. 
This approach, however, does not produce data within the unstable regions associated with phase separation. However, phase field simulations require free energy information within these two-phase regions in order to consistently resolve phase interfaces. Additional Monte Carlo calculations were performed for temperatures of 260 K, 300 K, and 340 K using bias potentials for umbrella sampling within the unstable regions of the order-disorder transition \cite{Natarajan2017,Torrie1977,mishin2004,Sadigh2012a,Sadigh2012b}. 
The partition function is:
\begin{align}
    \Theta &= \sum_{\bsym{\sigma}} \exp{\left(-\frac{E(\bsym{\sigma})  + \sum_{i=0}^6\phi_i(\eta_i(\bsym{\sigma}) - \kappa_i)^2}{k_B T}\right)} \label{eqn:part3}
\end{align}
where $\phi_i$ and $\kappa_i$ determine the curvature and center of the bias potential, respectively, and the inner sum is over the composition and six order parameters. 
The ensemble average of the composition $\langle \eta_0\rangle$ and each order parameter $\langle\eta_i\rangle$, $i = 1,\dots,6$ is related to its corresponding chemical potential through the bias parameters:
\begin{align}
    \frac{1}{M}\mu_i\Big|_{\langle\bsym{\eta}\rangle} &= -2\phi_i(\langle\eta_i\rangle - \kappa_i), \qquad i=0,\ldots,6
    \label{eq:mu-eta-kappa}
\end{align}



The Monte Carlo calculations are run with an allowed variance in ensemble averages $\langle\eta_i\rangle=3 \times 10^{-4}$ as a convergence criterion, from which the precision of the order parameter is computed as Var$(\langle\eta_i\rangle - \kappa_i)$. The precision in $\langle\mu_i\rangle$ follows using (\ref{eq:mu-eta-kappa}). 
Additional details on the Monte Carlo simulations, are provided in \S S1.2.3.

\subsection{Integrable deep neural networks (IDNNs) for free energy representations}
\label{sec:IDNN}

The IDNN representation is obtained for the free energy density function by training on derivative data: chemical potentials as labels and the corresponding symmetry-invariant functions of composition or order parameter as features (\S \ref{sec:order params methods}) \cite{Teichert2019,Teichert2020}. The integrability of the IDNN follows from the fundamental theorem of calculus since it is the derivative of a fully-connected deep neural network (DNN). 
A DNN is a function $Y(\bsym{X},\bsym{W},\bsym{b})$ representing the ensemble averaged chemical potentials $\langle \mu_i\rangle$, with arguments or inputs $\bsym{X}$ representing the symmmetry-invariant functions of composition and order parameters $\langle\eta_i\rangle$, $i = 0,\dots6$, weights $\bsym{W}$, and biases $\bsym{b}$. DNN training is an optimization problem for the weights and biases, given the dataset $\{(\bsym{\widehat{X}}_\theta,\widehat{Y}_{\theta})\}$. Here, however, the dataset $\{(\bsym{\widehat{X}}_\theta,\widehat{Y}_{\theta})\}$ is not available. Instead, we have the derivative dataset $\{(\bsym{\widehat{X}}_\theta,\widehat{y}_{\theta_k})\}$, where $\widehat{y}_{\theta_k}$ corresponds to the partial derivative of $\widehat{Y}_{\theta}$ with respect to the $k^\text{th}$ component of $\widehat{\bsym{X}}_\theta$. By defining the IDNN as the gradient of $Y$ with respect to its inputs $\bsym{X}$, i.e. $\partial Y(\bsym{X},\bsym{W},\bsym{b})/\partial X_k$, the training/optimization problem is:
\begin{align}
    \bsym{\widehat{W}},\bsym{\widehat{b}} = \underset{\bsym{W},\bsym{b}}{\mathrm{arg\,min}}\,\sum_{k=1}^n\mathrm{MSE}\left(\frac{\partial Y(\bsym{X},\bsym{W},\bsym{b})}{\partial X_k}\Big |_{\bsym{\widehat{X}}_\theta},\widehat{y}_{\theta_k}\right)
\end{align}
The optimized weights $\bsym{\widehat{W}}$ and biases $\bsym{\widehat{b}}$ are used with the IDNN function $\partial Y(\bsym{X},\bsym{\widehat{W}},\bsym{\widehat{b}})/\partial X_k$ to predict the chemical potential. The same weights and biases  are used in its antiderivative DNN function $Y(\bsym{X},\bsym{\widehat{W}},\bsym{\widehat{b}})$ to predict the hommogeneous free energy density.

\subsection{Sampling and active learning workflow}
\label{sec:activelearning}
IDNN training to represent the chemical potential for the zig-zag ordering requires sampling data in the seven-dimensional 
$\boldsymbol{\eta}$ space. Uniform, dense sampling in this space would require a prohibitive number of Monte Carlo simulations. Instead, we focus on physically significant regions with difficult-to-capture features. These  include the energy wells related to the variants of the zig-zag ordering and the divergent behavior of the chemical potential at the boundaries of the order parameter space, including the composition end members at $\eta_0 = x=\{0,1\}$. Some general, unguided \emph{exploration} sampling of the order parameter space is also performed to capture overall trends.

We improve the partially trained IDNN by combining exploration with \emph{exploitation} sampling in areas with high point-wise error. The active learning workflow iterates over cycles of exploration sampling, IDNN training, and exploitation sampling  until a stopping criterion is met. This sampling must be carried out within the boundaries of the admissible domain in $\boldsymbol{\eta}$-space. We use the Billiard Walk \cite{Polyak2014} random sampling algorithm in this space. More detail is available in \S S1.4.

The workflow (Figure \ref{fig:Fig4}a,b) forced a new  search for the IDNN hyperparameters on (a) the second workflow iteration, and (b) if the mean square error (MSE) calculated for the two previous IDNN models using the most recent dataset increased from one to the other. If the MSE decreased, then the workflow allowed training to continue with the previous IDNN on the most recent data (Figure \ref{fig:Fig4}b).

\subsection{Phase field theory and associated computational framework}
\label{sec:phasefield}

  Neglecting elastic effects, the total free energy of the system is:
\begin{align}\label{eqn:free_energy}
    \Pi[x,\widehat{\bsym{\eta}}] = \int\limits_\Omega \left(f(x,\widehat{\bsym{\eta}}) + \frac{1}{2}\chi_0|\nabla x|^2 + \sum_{i=1}^6\frac{1}{2}\chi_i|\nabla\widehat{\eta}_i|^2\right)\,\mathrm{d}V
\end{align}
where $\chi_i$ are the gradient parameters, and $f(x,\widehat{\bsym{\eta}})$ is the free energy density, represented by the analytically integrated DNN in this work. The chemical potentials $\widetilde{\mu}_i$ used in the phase field equations are variational derivatives of the total free energy: $\widetilde{\mu}_0 := \delta\Pi/\delta x$ and $\widetilde{\mu}_i := \delta\Pi/\delta\widehat{\eta}_i$, $i = 1,\ldots,6$ (the $\widehat{\bullet}$ notation is used for clarity; however, $\widehat{\eta}_i = \eta_i$ from previous sections):
\begin{equation}
    \widetilde{\mu}_0 = \frac{\partial f}{\partial x} - \chi_0 \nabla^2 x,\qquad
    \widetilde{\mu}_i = \frac{\partial f}{\partial \widehat{\eta}_i} - \chi_i \nabla^2 \widehat{\eta}_i, \; i=1,\ldots,6
    \label{eq:continuouschempots}
\end{equation}
The Cahn-Hilliard phase field  equation \cite{CahnHilliard1958} models the dynamics of the conserved composition, and the Allen-Cahn equation \cite{Allen1979} models nonconserved order parameter fields, respectively:
\begin{equation}
    \frac{\partial x}{\partial t} = \nabla\cdot(\widetilde{M}\nabla\widetilde{\mu}_0),\quad\text{and}\quad
    \frac{\partial \widehat{\eta}_i}{\partial t} = -L\widetilde{\mu}_i, \qquad i=1,\ldots,6
    \label{eqn:CH-AC}
\end{equation}
where $\widetilde{M}$ is the mobility and $L$ is a kinetic coefficient. We substitute in the chemical potentials and write the governing equations in weak form to be solved using a mixed finite element method:
\begin{align}
    0 &= \int_\Omega \left(w_x\frac{\partial x}{\partial t} + \widetilde{M}\nabla w_x\cdot\nabla\tilde{\mu}_0\right)\mathrm{d}V - \int_{\partial\Omega}w_xj_n\mathrm{d}S\\
    0 &= \int_\Omega \left[w_{\tilde{\mu}_0}\left(\tilde{\mu}_0 - \frac{\partial f}{\partial x}\right) - \chi_0\nabla w_{\tilde{\mu}_0}\cdot\nabla x\right]\mathrm{d}V\\
    0 &= \int_\Omega \left[w_i\frac{\partial \widehat{\eta}_i}{\partial t} + L\left(w_i\frac{\partial f}{\partial \widehat{\eta}_i} + \chi_i\nabla w_i\cdot\nabla\widehat{\eta}_i\right)\right]\mathrm{d}V, \qquad i=1,\ldots,6
\end{align}
where $w_x$, $w_{\tilde{\mu}_0}$, and $w_i$ are weighting functions. For the equations written in this mixed formulation, the following boundary conditions have been applied to $x$, $\tilde{\mu}_0$, and $\widehat{\eta}_i$, $i = 1,\ldots,n$,  on $\partial\Omega$, where $\bsym{n}$ is the outward unit normal and $j_n$ is an influx:$\nabla x\cdot\bsym{n} = 0,\; \widetilde{M}\nabla \tilde{\mu}_0\cdot\bsym{n} = j_n,\;
    \nabla \widehat{\eta}_i\cdot\bsym{n} = 0, \; i=1,\ldots,6$

We define a composition-dependent surrogate function for the diffusivity $D$ at $300$ K, using the predicted values from Ref. \cite{VanderVen2000}: $D = 0.01\exp(-274(1.05-x)(0.47-x)(1.-x))$, and approximate $D$ at other temperatures as outlined in \S S1.5. The diffusivity surrogate function and the predicted data appear in Figure S5, where the effective vibrational frequency $\nu^*$ is reported to be on the order of $10^{13}$ s$^{-1}$ \cite{VanderVen2000}. 
The mobility $\widetilde{M}$ is related to $D$ by the equation $\widetilde{M} = Dx/(k_BT)$ \cite{Jiang2016}. We solve an inverse problem for the gradient parameters $\chi_i$ in Eq. (\ref{eqn:free_energy}) by constraining the phase field calculation of the interface energy to agree with the DFT results in \S \ref{sec:DFTintfc}, see \S S1.5

The simulations are performed using the finite element method with the \texttt{mechanoChemFEM} code (available at \url{github.com/mechanoChem/mechanoChemFEM}), which is based on the \texttt{deal.II} \cite{dealii2017} library. Adaptive meshing with hanging nodes and adaptive time stepping are used. Further details on the phase field methods are provided in \S S1.5.

\section*{Acknowledgements}
\label{sec:acknowledgements}
We thank Brian Puchala and Anirudh Natarajan for their insight regarding the \texttt{CASM} software and related methods, as well as Chirranjeevi Gopal and Muratahan Aykol for their suggestions on DFT methods for layered oxides. We gratefully acknowledge the support of Toyota Research Institute, Award \#849910, ``Computational framework for data-driven, predictive, multi-scale and multi-physics modeling of battery materials''.
This work has also been supported in part by National Science Foundation DMREF grant \#1729166, ``Integrated Framework for Design of Alloy-Oxide Structures''.
Additional support was provided by Defense Advanced Research Projects Agency (DARPA) under Agreement No. HR0011199002, ``Artificial Intelligence guided multi-scale multi-physics framework for discovering complex emergent materials phenomena''. We acknowledge the support of the U.S. Army Research Office through the DURIP grant W911NF1810242, which provided computational resources for this work.
Simulations in this work were run on the Great Lakes HPC cluster at the University of Michigan. Additional simulations were performed using resources provided by the Extreme Science and Engineering Discovery Environment (XSEDE) Comet at the San Diego Supercomputer Center and Stampede2 at the Texas Advanced Computing Center through allocations TG-DMR180072 and TG-MCH200011. XSEDE is supported by National Science Foundation grant number ACI-1548562.
Some simulations were performed using resources provided by the NSF via grant 1531752 MRI: Acquisition of Conflux, A Novel Platform for Data-Driven Computational Physics (Tech. Monitor: Ed Walker), with additional support by the University of Michigan.

\section*{Author contributions}
\label{sec:contributions}
The study's scale bridging methodology was developed by KG and GHT; electronic structure studies were planned by GHT, SD and VG;  electronic structure calculations were performed by GHT and SD; statistical mechanics and phase field studies were carried out by GHT, JH and MFS; the results were analyzed by all the authors; the paper was written by all the authors.

\section*{Competing interests}
The authors declare no competing interests.

\clearpage
\bibliographystyle{unsrt}
\bibliography{references}

\end{document}


\maketitle

\section{Detailed Methods}
\label{sec:methods}
\subsection{Density functional theory methods}
\label{sec:DFTmethods}

The formation energy $E$ of a configuration with Li composition $x$ can be calculated using total energy values computed using density functional theory (DFT) according to the following equation \cite{VanderVen1998}:
\begin{align}
    E_{\mathrm{Li}_{x}\mathrm{CoO}_2} &= E^\mathrm{tot} - xE^\mathrm{tot}_{\mathrm{LiCoO}_2} - (1 - x)E^\mathrm{tot}_{\mathrm{CoO}_2}
    \label{eqn:form_energy}
\end{align}
where $E^\mathrm{tot}$, $E^\mathrm{tot}_{\mathrm{LiCoO}_2}$, and $E^\mathrm{tot}_{\mathrm{CoO}_2}$ are the total energies for the given configuration, LiCoO$_2$, and CoO$_2$, respectively. In performing DFT calculations for Li$_x$CoO$_2$, Aykol and co-authors found that it is important to include not only the Hubbard correction (DFT+$U$), which reduces self-interaction errors \cite{Anisimov1991,Zhou2004,Aykol2014}), but to also include the van der Waals (vdW) interactions \cite{Aykol2015}. The effect of the vdW interactions allows the voltage predicted by DFT to match the experimental voltage when using an appropriately tuned value of $U$. While both vdW-corrections and vdW-density functionals (vdW-DF) improve the DFT results, predictions are most consistent and accurate with vdW-DF. 

In this workflow, we use a simplified rotational-invariant formulation of DFT+$U$~\cite{Cococcioni2005} and a vdW-DF exchange correlation functional, namely the optB88 exchange correlation functional \cite{Thonhauser2007,Klime2009,Langreth2009,Sabatini2012,Aykol2015,Thonhauser2015,Berland2015}, to calculate the formation energy on a chosen subset of LCO configurations. The \texttt{CASM} (Clusters' Approach to Statistical Mechanics) software suite was used to identify a set of configurations with the O3 crystal structure for parametrizing the cluster expansion. Fully-relaxed DFT calculations were completed for 333 of these configurations, with \texttt{CASM} automatically adjusting the size of the k-points grid according to atomic configuration. We perform the ground-state DFT calculations with geometry optimization in \texttt{Quantum Espresso} \cite{QE-2009,QE-2017}, using projector augmented-wave (PAW) pseudopotentials calculated with the Perdew-Burke-Ernzerhof (PBE) functional from \texttt{PSlibrary 1.0.0} \cite{dalcorso2014,pslibrary}. The values for the wave function and charge density cutoffs are chosen in two steps. Starting with the cutoff values suggested in the pseudopotential file for Co, we first increase only the charge density cutoff until the total energy converges to $<$1 meV/atom. Next, we increase both the wave function and the charge density cutoffs, maintaining the ratio of the two values, until the total energy again converges to $<$1 meV/atom, giving a wave function cutoff of 55 Ha and a charge density cutoff of 301.5 Ha. A k-point grid of $6 \times 6 \times 3$ is also used to ensure total energy convergence within $<$1 meV/atom. Structural optimization is performed until cell stress and ionic forces are under 0.5 kbar and 0.00005 Ha/Bohr, respectively. The crystal structure for O3-LiCoO$_2$ is shown in Figure 1a.

In this work, we calibrate the Hubbard $U$ parameter to match experimental average lithiation voltages over various ranges of Li composition. To do this, we compute the voltage at increasing $U$ for given composition values of $x_1$ and $x_2$ \cite{Meredig2010}, using crystal structures previously reported from experiments. The average voltage from $x_1$ to $x_2$ is calculated using the following equation \cite{Aydinol1997,Aykol2015}:
\begin{align}
    V = -\frac{E_{\mathrm{Li}_{x_2}\mathrm{CoO}_2} - E_{\mathrm{Li}_{x_1}\mathrm{CoO}_2} - (x_2 - x_1)E_{\mathrm{Li}}}{(x_2 - x_1)e}
\end{align}
where $E$ is the calculated total energy from DFT and $e$ is the charge on an electron. We compute the average voltage for Li composition: $\{x_1,x_2\} = \{0,1\}, \{0,1/2\}$ and $\{1/2,1\}$. By comparing with experimental voltages (Figure 1b), we select an appropriate value of $U = 2.5$ eV. 

The development of charge ordering was monitored through the magnetic moments in the DFT calculations \cite{Aykol2015}. Evidence of charge splitting appears in the DFT results in this work for the row ordering on the left in Figure 1d, but not in the other row ordering or in the zig-zag ordering. Charge ordering of Co$^{3+}$ and Co$^{4+}$ atoms has been experimentally observed at low temperatures and Li compositions of $x = 1/2$ and $x = 2/3$ \cite{Motohashi2009}.

\subsubsection{DFT calculations for interface energy}
\label{sec:DFTintfc}
Large-scale DFT computations were performed for the coherent interface energy of the anti-phase boundary between the $\eta_1=1/2$ and $\eta_3=1/2$ ordered LCO rotational variants. These computations were performed using the DFT-FE software~\cite{das2022dft,MOTAMARRI2020106853,Motamarri2018,MOTAMARRI2013308}, a recently developed massively parallel open-source code for large-scale real-space Kohn-Sham DFT studies based on a finite-element discretization. We employ the PBE exchange-correlation functional and the optimized norm-conserving Vanderbilt pseudopotentials (ONCV)~\cite{oncv2013} from the Pseudo Dojo library~\cite{van2018pseudodojo}. All numerical parameters in DFT-FE were chosen such that the ground-state energies  converged to an accuracy of 1.5 meV/atom. Additionally, ionic forces and cell stresses were relaxed to under 5 meV/Å and 0.5 Kbar respectively, and Fermi-Dirac smearing with a temperature of 500 K was used for all simulations. 

In order to compute the interface energy, we considered two types of simulation cells. The first was a periodic simulation cell representing the interface system, constructed with equal number of mutually commensurate (MC) supercells (see \S \ref{sec:order params methods}) for each ordered variant resulting in two interfaces upon accounting for the periodicity of the simulation domain. The MC supercells were arranged along the lattice vector that is not part of the plane forming the interface, as shown in Figure 3f for a two MC supercell case. Additionally, we consider another periodic simulation cell to represent the bulk, constructed using one MC supercell corresponding to either of the ordered variants. Since the corresponding supercells are related through rotations, the bulk energies are equal for both the ordered variants. Next, we performed ground-state DFT calculations on the interface system and the bulk system with full structural relaxation of ionic forces and cell stress. The ionic forces and cell stresses are relaxed self-consistently where each cell stress relaxation update involves a full ionic forces relaxation keeping cell vectors fixed. The total interface energy per unit area of interface, considered to be the average of the two interfaces and denoted by $\gamma^{\prime}$, is given by the following energy difference 
\begin{equation}
    \gamma^{\prime}=\frac{E_{(\textrm{Var1}+\textrm{Var2})}-2N_{\textrm{cell}}E_{\textrm{bulk}}}{2 A}\,,
\end{equation}
where $N_{\textrm{cell}}$ denotes the number of MC supercells for each variant, $A$ denotes the area of the interface after structural relaxation, and the factor of 2 in the denominator accounts for the averaging over the two interfaces in the periodic system. The interface energy $\gamma^{\prime}$ is composed of two contributions. The first is a short-ranged term due to the local deviation of the atomic arrangement across the interface, which is denoted by $\gamma$. The second is a long-ranged contribution due to the elastic misfit strain to maintain the coherency of the interface. Although the misfit elastic strain decays away from the interface, it is not computationally feasible to resolve it by explicit DFT calculations with periodic supercells. Due to the use of periodic boundary conditions in the interface calculations, the misfit elastic strain contributes an energy to $\gamma^{\prime}$ that scales linearly with number of MC supercells:
\begin{equation}
    \gamma^{\prime}=\gamma+(k_1+k_2)\times N_{\textrm{cell}}\,,
    \label{eq:intfcengy}
\end{equation}
where $k_1$ and $k_2$ denote the misfit elastic energy per MC supercell for the two ordered variants. Next, the linear scaling elastic misfit strain energy contribution to $\gamma^{\prime}$ is eliminated by finding the intercept of Eq. (\ref{eq:intfcengy}) at $N_{\textrm{cell}}=0$  to obtain $\gamma$. A similar approach has been employed in previous atomistic calculations of coherent interface energies ~\cite{FARKAS1994367,MISHIN20041451}.  Table S1 shows the energies and system sizes used to determine $\gamma$ in the LCO system, where we consider $N_{\textrm{cell}}$ ranging from 1--3 and obtain the expected close to a linear relationship between $\gamma^{\prime}$  and $N_{\textrm{cell}}$. Finally, $\gamma$ is estimated to be  30.9 mJ/$\textrm{m}^2$.  We additionally remark that the above calculations required simulation domains consisting of $\sim$ 500--1400 atoms (4,000--12,000 electrons) combined with around 200 geometry updates for each simulation domain size, indicating the large-scale nature of these calculations.

\subsection{Statistical mechanics}
\label{sec:statmech}

In this work, we follow the statistical mechanics approach outlined by Van der Ven, et al. \cite{VanderVen1998} in their first-principles study of Li$_x$CoO$_2$ (LCO), and of Natarajan, et al. \cite{Natarajan2017}, among others.

\subsubsection{Cluster expansion for formation energy}
Given the expense of DFT calculations, we adopt cluster expansions to access the large numbers of configurations needed in the statistical mechanics studies. The formation energy is written as $E_\text{f}(\bsym{\sigma})$, where  $\bsym{\sigma}$ is the configuration vector with the occupancy variable $\sigma_i = 1$ if Li occupies the site $i$ and $\sigma_i = 0$ if the site contains a vacancy. As is common, we define $E_\text{f}(\bsym{\sigma})$ by computing the formation energy with DFT for a subset of configurations and use these values to parameterize a cluster expansion \cite{Sanchez1984,deFontaine1994} as a rapidly queryable surrogate for the formation energy. We use the \texttt{CASM} software\cite{casm}, which facilitates the construction and parameterization of cluster expansion Hamiltonians and their use in Monte Carlo simulations, to select configurations for the DFT computations and perform the statistical mechanics calculations in this work \cite{VanderVen2010,thomas2013,puchala2013}. Candidate configurations are chosen for O3-Li$_x$CoO$_2$ with Li compositions from $x=0$ to $x=1$. 

A cluster is a collection of sites on the Li sub-lattice. Given a cluster of sites $\alpha = \{i,j,\dots,k\}$, a polynomial $\phi_\alpha$ can be defined as the product of occupancy variables of those sites, i.e. $\phi_\alpha = \sigma_i\sigma_j\cdots\sigma_k$. A cluster expansion is a linear combination of the polynomials $\phi_\alpha$, leading to the following form for the formation energy:
\begin{align}
    E_\text{f}(\bsym{\sigma}) &= V_0 + \sum_\alpha V_\alpha \phi_\alpha(\bsym{\sigma}),
\end{align}
where the coefficients $V_0$ and $V_\alpha$ are optimized for a ``best fit'', and are called effective cluster interactions (ECI).

We fit a cluster expansion to the formation energy calculated using DFT for 333 configurations for the O3 host structure using the \texttt{CASM} code. A sparse regression technique that combines a genetic algorithm with weighted linear regression is used to perform the fit. In this approach, a number of candidate basis functions are created using singletons, pairs, triplets and quadruplets of lattice sites that are within a given distance from each other. The genetic algorithm is used to select a subset of these basis functions to include in the cluster expansion. The coefficients of each subset of basis functions are calculated using linear regression. We use cluster types up to quadruplets, with maximum lattice site distances of $24\;\AA$ for pairs, $8\;\AA$ for triplets and $6\;\AA$ for quadruplets, for a total of 221 candidate basis functions. During the linear regression, we apply weights, $w_i$ to bias the fit towards greater accuracy for configurations on or near the convex hull using the energy difference, $\Delta E^\text{hull}_i$ between data point $i$ and the convex hull at that configuration: 
\begin{equation}
    w_i = 15\exp\left(-\frac{\Delta E^\text{hull}_i}{0.005}\right) + 0.5
\end{equation}

Additionally, the genetic algorithm is constrained to select basis functions such that the convex hull constructed from the cluster expansion predictions for all 333 configurations consists of the same configurations as in the DFT convex hull.


\subsubsection {Symmetry-adapted order parameters}
\label{sec:order params methods}
In addition to composition, symmetry-adapted order parameters are useful for observing and tracking order-disorder transitions, as well as for identifying the various translational and rotational variants of a given ordering, such as those appearing in Figure 3a for the zig-zag ordering at composition $x \approx 1/2$. The method for identifying symmetry-adapted order parameters is laid out elsewhere; e.g., see Natarajan, et al. \cite{Natarajan2017}. In this process supercells are identified corresponding to each variant (Figure 3b). A mutually commensurate supercell is then identified that encompasses each distinct supercell and is therefore sufficient for representing all the 12 variants. For the zig-zag ordering it includes 32 Li sublattice sites (Figures 3c-3d). The sublattice compositions of this supercell form a basis that can describe each variant as a vector $\bsym{x} \in \mathbb{R}^{32}$, where each component is 1 if the corresponding sublattice sites are fully occupied by Li and 0 otherwise. 

The symmetry group, $P$, of the zig-zag ordering consists of 384 unique linear transformations between the 12 variants  each represented by a matrix in $\mathbb{R}^{32\times 32}$. Following the algorithm from Thomas and Van der Ven \cite{thomas2017}, we constructed a $P$-invariant matrix and performed its eigenvalue decomposition. It resulted in eight nonzero eigenvalues: two distinct, two repeated three times, and four repeated six times, and eight corresponding sets of eigenvectors, for a total of 32 eigenvectors. The eigenvectors formed the rows of the orthogonal transformation matrix $\bsym{Q}\in\mathbb{R}^{32\times 32}$, that maps a sublattice composition vector $\bsym{x}$ to a vector $\bsym{\eta}$ with 32 order parameters.

The subset of order parameters relevant to the zig-zag ordering was identified by operating with $\bsym{Q}$ on the vectors $\bsym{x}^{(i)}$, $i = 1,\dots, 12$ describing the variants of that ordering. Using zero indexing, components $\eta^{(i)}_7$ through $\eta^{(i)}_{31}$ of $\bsym{\eta} \in \mathbb{R}^{32}$ are zero for all variants $i = 1,\dots, 12$, and are irrelevant for describing the zig-zag ordering (Figure 3c). This defines the seven relevant order parameters, $\eta_0,\dots,\eta_6$. The first of these, $\eta_0$, is associated with one of the distinct eigenvalue and corresponds to the composition averaged over all 32 sublattice sites; i.e., $\eta_0 = x$. The other six order parameters, $\eta_1,\dots,\eta_6$ are associated with one of the degenerate eigenvalues that has six corresponding eigenvectors.


Since the  free energy density is invariant under transformations of the triangular Li sublattice that map between the ordered variants, the IDNN representation is presented with features that are symmetric functions of $\eta_0,\dots \eta_6$ under these transformations. Monomials of up to sixth order were chosen and subjected to the Reynolds operator, by summing: 
\begin{equation}
 h(\bsym{\eta}) = \sum_{\bsym{M}^{(\eta)}\in\mathcal{M}}f(\bsym{M}^{(\eta)}\bsym{\eta}),
\end{equation}
where for each $\bsym{M} \in P$ we have $\bsym{M}^\eta = \bsym{QM}$. This operation yields the P-invariant polynomial functions in Eq. (1) as IDNN features.

\subsubsection {Monte Carlo sampling}
Given $E_\text{f}(\bsym{\sigma})$, we sample within the semi-grand canonical ensemble, in which the chemical potential is specified and the corresponding composition and/or order parameters are determined through ensemble averaging. The partition function for the semi-grand canonical ensemble is the following:
\begin{align}
    \Theta &= \sum_{\bsym{\sigma}} \exp{\left(-\frac{E(\bsym{\sigma}) - M\widehat{\bsym{\eta}}(\bsym{\sigma})\cdot\widehat{\bsym{\mu}}}{k_B T}\right)} \label{eqn:part2}
\end{align}
where $\widehat{\bsym{\eta}} = \langle\eta_0,\dots,\eta_6 \rangle^\text{T}$ represents the reduced vector of order parameters with $\eta_0 = x$ and $\widehat{\bsym{\mu}} = \langle\mu_0,\dots,\mu_6\rangle$ is the corresponding vector of chemical potentials, $M$ is the number of reference supercells that tile the configuration, $k_B$ is the Boltzmann constant, and $T$ is the temperature. Each chemical potential component is the derivative of the free energy with respect to its corresponding composition or order parameter component.

The chemical potential associated with Li composition is related to the voltage as a function of composition, $V(x)$ as
\begin{equation}
    V(x) = -\frac{\mu_0^\mathrm{cathode}(x) - \mu_0^\mathrm{anode}}{e},
    \label{eqn:voltage}
\end{equation}
where $\mu_0^\mathrm{cathode}$ is the chemical potential with respect to Li for the cathode (LCO) and $\mu_0^\mathrm{anode}$ is the chemical potential of the anode. Taking the anode to be Li metal, $\mu_0^\mathrm{anode}$ is a constant equal to the Gibbs free energy of Li \cite{Aydinol1997}. It is necessary for the two chemical potentials to have a consistent reference, however. We approximate the Gibbs free energy of Li with the total energy calculated using DFT, but the chemical potential for LCO reported from Monte Carlo sampling is based on formation energies that were re-referenced to the end members, according to the linear relation in Eq. (\ref{eqn:form_energy}). Since the chemical potential is the derivative of the free energy with respect to composition, this re-referencing shifted the chemical potential of LCO by a constant $k$ equal to the derivative of Eq. (\ref{eqn:form_energy}) with respect to $x$, i.e. $k := -E^\mathrm{tot}_{\mathrm{LiCoO}_2} + E^\mathrm{tot}_{\mathrm{CoO}_2}$. Therefore, we reverse the shift in the chemical potential by subtracting $k$ from $\mu_0^\mathrm{cathode}(x)$ before computing the voltage according to Eq. (\ref{eqn:voltage}). This imposes a consistent reference for both $\mu_0^\mathrm{cathode}$ and $\mu_0^\mathrm{anode}$.

A limitation with Monte Carlo sampling within the semi-grand canonical ensemble is that it does not produce data within the unstable regions associated with phase separation. While this is sufficient for delineating phase diagrams and predicting voltage, phase field simulations require free energy information within these two-phase regions in order to consistently resolve phase interfaces. Additional Monte Carlo calculations were performed for temperatures of 260 K, 300 K, and 340 K, this time by umbrella sampling using bias potentials to sample within the unstable regions of the order-disorder transition \cite{Natarajan2017,Torrie1977,mishin2004,Sadigh2012a,Sadigh2012b}. Since the fluctuations in the chemical potential around $\eta_0 = x=1/2$ and for the locations of wells at $\eta_1,\dots,\eta_6 \approx \pm 1/2$ are important for representing the ordering in the phase field models at temperatures below $\sim$330 K, dense sampling of chemical potential data was performed in those regions for 260 K and 300 K. Additional sampling was also performed near the extreme compositions of $\eta_0 = x=\{0,1\}$ and order parameters $\eta_1,\dots,\eta_6 = \pm 1/2$ to help capture the divergent nature of the chemical potential, and in the unstable regions. The partition function used in this case, then, is the following:
\begin{align}
    \Theta &= \sum_{\bsym{\sigma}} \exp{\left(-\frac{E(\bsym{\sigma})  + \sum_{i=0}^6\phi_i(\eta_i(\bsym{\sigma}) - \kappa_i)^2}{k_B T}\right)} \label{eqn:part3}
\end{align}
where $\phi_i$ and $\kappa_i$ determine the curvature and center of the bias potential, respectively, and the inner sum is over the composition and six order parameters. Usually, $\phi_i$ can be fixed at an appropriate value while $\kappa_i$ is varied to sample across the desired composition and order parameter space within the Monte Carlo sampling routine. The ensemble average of the composition $\langle \eta_0\rangle$ and each order parameter $\langle\eta_i\rangle$, $i = 1,\dots,6$ is related to its corresponding chemical potential through the bias parameters:
\begin{align}
    \frac{1}{M}\mu_i\Big|_{\langle\bsym{\eta}\rangle} &= -2\phi_i(\langle\eta_i\rangle - \kappa_i), \qquad i=0,\ldots,6
    \label{eq:mu-eta-kappa}
\end{align}

Regions with orderings are recognized by the steepness and positive slope of the curve traced out by connecting the chemical potential data points. This indicates a strongly convex free energy well and thermodynamic preference for these structures over the corresponding composition intervals.

An ordering is also predicted at $x = 1/3$, appearing below approximately 235 K. This ordering has not been seen experimentally at room temperature, but some evidence has been reported at low temperatures \cite{ShaoHorn2003}, which is consistent with our results. The improved accuracy of the predicted order-disorder temperatures over previous first-principles work \cite{VanderVen1998} is likely due to the advanced methods of DFT$+U$ with van der Waals interactions included (\S \ref{sec:DFTmethods}). 

The Monte Carlo calculations are run with a variance in ensemble averages $\langle\eta_i\rangle=3 \times 10^{-4}$ from which the precision of the order parameter is computed as Var$(\langle\eta_i\rangle - \kappa_i)$. The precision in $\langle\mu_i\rangle$ follows using (\ref{eq:mu-eta-kappa}). The computational runtime of the smallest, average, and largest calculation are (17 s, 97 s, 64941 s) and (1 s, 166 s, and 13700 s) for standard sampling and umbrella sampling, respectively. Standard sampling has 5487 points on average per Monte Carlo simulation, and umbrella sampling has 52789.

\subsection {Integrable deep neural networks (IDNNs) for free energy representations}
\label{sec:IDNN}

The IDNN representation is obtained for the free energy density function by training on derivative data in the form of pairs of chemical potential--the label--and corresponding (symmetry-invariant functions of) composition or order parameter--the features \cite{Teichert2019,Teichert2020}. The integrability of the IDNN is built into its structure by constructing it as the derivative form of a standard fully-connected deep neural network (DNN). While it is straightforward to differentiate the equations describing the standard DNN to derive the equations for the IDNN, modern deep learning libraries make this step unnecessary. The user can simply construct a standard DNN and apply a gradient operation to create the IDNN, which is then used for training. Mathematically, a DNN can be denoted by a function $Y(\bsym{X},\bsym{W},\bsym{b})$ representing the ensemble averaged chemical potentials $\langle \mu_i\rangle$, with arguments or inputs $\bsym{X}\in \mathbb{R}^7$ representing the composition $\langle x \rangle = \langle \eta_0\rangle$ and remaining order parameters $\langle\eta_i\rangle$, $i = 1,\dots6$, weights $\bsym{W}$, and biases $\bsym{b}$. Training the DNN involves the solution of an optimization problem for the weights and biases, given the dataset $\{(\bsym{\widehat{X}}_\theta,\widehat{Y}_{\theta})\}$:
\begin{align}
    \bsym{\widehat{W}},\bsym{\widehat{b}} = \underset{\bsym{W},\bsym{b}}{\mathrm{arg\,min}}\,\mathrm{MSE}\left(Y(\bsym{X},\bsym{W},\bsym{b})\Big |_{\bsym{\widehat{X}}_\theta},\widehat{Y}_{\theta}\right)
\end{align}
The case of interest here is when the dataset $\{(\bsym{\widehat{X}}_\theta,\widehat{Y}_{\theta})\}$ is not available. Instead, we have the derivative dataset $\{(\bsym{\widehat{X}}_\theta,\widehat{y}_{\theta_k})\}$, where $\widehat{y}_{\theta_k}$ corresponds to the partial derivative of $\widehat{Y}_{\theta}$ with respect to the $k^\text{th}$ component of $\widehat{\bsym{X}}_\theta$. To use these data, the IDNN is defined as the gradient of $Y$ with respect to its inputs $\bsym{X}$, i.e. $\partial Y(\bsym{X},\bsym{W},\bsym{b})/\partial X_k$. The training is defined as follows:
\begin{align}
    \bsym{\widehat{W}},\bsym{\widehat{b}} = \underset{\bsym{W},\bsym{b}}{\mathrm{arg\,min}}\,\sum_{k=1}^n\mathrm{MSE}\left(\frac{\partial Y(\bsym{X},\bsym{W},\bsym{b})}{\partial X_k}\Big |_{\bsym{\widehat{X}}_\theta},\widehat{y}_{\theta_k}\right)
\end{align}
The resulting optimized weights $\bsym{\widehat{W}}$ and biases $\bsym{\widehat{b}}$ can be used with the function $\partial Y(\bsym{X},\bsym{\widehat{W}},\bsym{\widehat{b}})/\partial X_k$ to return a prediction of the chemical potential. Its antiderivative is exactly represented by using the same weights and biases in the function $Y(\bsym{X},\bsym{\widehat{W}},\bsym{\widehat{b}})$. For the current work, $\partial Y(\bsym{X},\bsym{\widehat{W}},\bsym{\widehat{b}})/\partial X_k$ gives the IDNN representation of the chemical potentials, and $Y(\bsym{X},\bsym{\widehat{W}},\bsym{\widehat{b}})$ is the DNN representation of the free energy.


\subsection {Sampling and active learning workflow}
\label{sec:activelearning}

In order to train an IDNN that represents the chemical potential data for composition plus the six order parameters associated with the zig-zag ordering, we must sample data in a seven-dimensional space. Rather than sampling uniformly and densely across the entire space, which would require a potentially prohibitive number of Monte Carlo simulations, we focus on the regions that are both significant physically and have features that may be difficult to capture. These regions include the energy wells related to the variants of the zig-zag ordering and the divergent behavior of the chemical potential at the boundaries of the order parameter space, including the composition end members at $\eta_0 = x=\{0,1\}$. Some general, unguided sampling of the order parameter space is also performed to capture overall trends. Thus, this first method of sampling involves the \emph{exploration} of areas known \emph{a priori} to be of interest.

Since the purpose of the sampling is to develop a surrogate model of the free energy, we combine the exploration sampling with a second method of sampling that \emph{exploits} the partially trained surrogate to identify additional areas with data that may be helpful in improving the surrogate model. For the current workflow, this exploitation consists simply of identifying a specified number of data points at which the surrogate model shows high point-wise error. The high error suggests that the landscape in this region is difficult to capture, and so more data are sampled here. The combined exploration/exploitation sampling approach forms the active learning workflow where a cycle of exploration sampling, IDNN training, and exploitation sampling is followed until a stopping criterion is met.

Before sampling within the space, it is necessary to define the boundaries of the space. The sublattice parameter space is the unit hypercube in 32 dimensions, and  $\widehat{\bsym{Q}}$ the reduced matrix obtained from $\bsym{Q}$ (in Figure 3e) by restriction to its first seven rows, transforms from the sublattice space to the order parameter space. A uniform sampling of the sublattice space, transformed through $\widehat{\bsym{Q}}$, does not produce a uniform sampling of the order parameters, however, due in part to the dimension reduction. Instead, we define bounding planes in the order parameter space. First, we restrict the space so that $\eta_7$ through $\eta_{31}$ are equal to zero, i.e., any ordering can only be a variant of the zig-zag ordering. Using this restriction, the relationship $\bsym{\eta} = \bsym{Q}\bsym{x}$, and applying the physical constraint that the sublattice compositions satisfy $0 \leq x _i \leq 1$, $i = 0,\ldots,31$, we can use the inverse of the full $\bsym{Q}$ matrix to define the following upper bounding planes:
\begin{align}
    \sum_{j=0}^6 Q^{-1}_{ij}\eta_j \leq 1, \qquad i = 0,\ldots,31
\end{align}
and lower bounding planes:
\begin{align}
    \sum_{j=0}^6 Q^{-1}_{ij}\eta_j \geq 0, \qquad i = 0,\ldots,31
\end{align}
where we have used zero indexing of vectors in $\mathbb{R}^{32}$ and $\bsym{Q}\in \mathbb{R}^{32\times 32}$.

We use the Billiard Walk \cite{Polyak2014} random sampling algorithm to sample within the bounding planes. A summary of the algorithm is as follows: Given an initial point within the bounding planes, a random trajectory and length are chosen. The trajectory is followed until reaching a bounding plane, at which point the trajectory is updated according to its reflection off the plane. The reflections continue until reaching the full length of the trajectory, which defines the next point in the Billiard Walk. This process is repeated until the desired number of globally sampled internal points is reached.

A byproduct of the Billiard Walk is a collection of quasi-random boundary points. We take a random subset of these boundary points to help capture the divergent behavior of the chemical potentials at the boundary during early iterations of the workflow. The area around the end members and energy wells are randomly sampled within a hypercube of side length 0.15. We also explicitly sample along and near the order-disorder transition paths in the order parameter space at $\eta_0 = x=0.5$.

The active learning workflow (Figure 4a,b), guided sampling with exploration, training and exploitation near the bounds, high error points, wells and unstable regions of the reduced-order $\widehat{\bsym{\eta}} \in \mathbb{R}^7$ subspace. The workflow forced a new  search for the IDNN hyperparameters on (a) the second workflow iteration, and (b) if the mean square error (MSE) calculated for the two previous IDNN models using the most recent dataset increased from one to the other. If the MSE decreased, then the workflow allowed training to continue with the previous IDNN on the most recent data (Figure 4b).

\subsection {Phase field theory and associated computational framework}
\label{sec:phasefield}

The evolution of microstructure and phase changes can be modeled using the phase field equations. The Cahn-Hilliard equation \cite{CahnHilliard1958} models the dynamics of conserved quantities, such as composition, while nonconserved order parameter fields are modeled using the Allen-Cahn equation \cite{Allen1979}. When neglecting elastic effects, the total free energy of the system with composition and $n = 6$ order parameters can be described as follows:
\begin{align}\label{eqn:free_energy}
    \Pi[x,\widehat{\bsym{\eta}}] = \int\limits_\Omega \left(f(x,\widehat{\bsym{\eta}}) + \frac{1}{2}\chi_0|\nabla x|^2 + \sum_{i=1}^6\frac{1}{2}\chi_i|\nabla\widehat{\eta}_i|^2\right)\,\mathrm{d}V
\end{align}
where $\chi_i$ are the gradient parameters, and $f(x,\widehat{\bsym{\eta}})$ is the free energy density, represented by the analytically integrated DNN in this work.

The chemical potentials $\widetilde{\mu}_i$ used in the phase field equations are given by the variational derivatives of the total free energy, such that $\widetilde{\mu}_0 := \delta\Pi/\delta x$ and $\widetilde{\mu}_i := \delta\Pi/\delta\widehat{\eta}_i$, $i = 1,\ldots,n$ (the $\widehat{\bullet}$ notation is used for clarity; however, $\widehat{\eta}_i = \eta_i$ from previous sections):
\begin{align}
    \widetilde{\mu}_0 &= \frac{\partial f}{\partial x} - \chi_0 \nabla^2 x\\
    \widetilde{\mu}_i &= \frac{\partial f}{\partial \widehat{\eta}_i} - \chi_i \nabla^2 \widehat{\eta}_i, \qquad i=1,\ldots,n
    \label{eq:continuouschempots}
\end{align}
The Cahn-Hilliard and Allen-Cahn equations, respectively, are the following:
\begin{align}
    \frac{\partial x}{\partial t} &= \nabla\cdot(\widetilde{M}\nabla\widetilde{\mu}_0) \label{eqn:CH}\\
    \frac{\partial \widehat{\eta}_i}{\partial t} &= -L\widetilde{\mu}_i, \qquad i=1,\ldots,n
    \label{eqn:AC}
\end{align}
where $\widetilde{M}$ is the mobility and $L$ is a kinetic coefficient. We substitute in the equations for the chemical potentials and write the governing equations in weak form to be solved using a mixed finite element method:
\begin{align}
    0 &= \int_\Omega \left(w_x\frac{\partial x}{\partial t} + \widetilde{M}\nabla w_x\cdot\nabla\tilde{\mu}_0\right)\mathrm{d}V - \int_{\partial\Omega}w_xj_n\mathrm{d}S\\
    0 &= \int_\Omega \left[w_{\tilde{\mu}_0}\left(\tilde{\mu}_0 - \frac{\partial f}{\partial x}\right) - \chi_0\nabla w_{\tilde{\mu}_0}\cdot\nabla x\right]\mathrm{d}V\\
    0 &= \int_\Omega \left[w_i\frac{\partial \widehat{\eta}_i}{\partial t} + L\left(w_i\frac{\partial f}{\partial \widehat{\eta}_i} + \chi_i\nabla w_i\cdot\nabla\widehat{\eta}_i\right)\right]\mathrm{d}V, \qquad i=1,\ldots,n
\end{align}
where $w_x$, $w_{\tilde{\mu}_0}$, and $w_i$ are weighting functions. For the equations written in this mixed formulation, the following Neumann boundary conditions have been applied to $x$, $\tilde{\mu}_0$, and $\widehat{\eta}_i$, $i = 1,\ldots,n$,  on $\partial\Omega$, where $\bsym{n}$ is the outward unit normal and $j_n$ is an influx:
\begin{align}
    \nabla x\cdot\bsym{n} &= 0\label{eq:dirbc-c}\\
    \widetilde{M}\nabla \tilde{\mu}_0\cdot\bsym{n} &= j_n\label{eq:neumbc-mu0}\\
    \nabla \widehat{\eta}_i\cdot\bsym{n} &= 0, \qquad i=1,\ldots,n\label{eq:dirbc-eta}
\end{align}

To compare the behavior of LCO at different temperatures, we perform phase field simulations at $260$ and $300$ K, informed by data from the Monte Carlo computations at these temperatures. The analytically integrated free energy DNNs for each of these temperatures are used to represent $f(x,\widehat{\bsym{\eta}})$ in Eq. (\ref{eq:continuouschempots}). We define the following composition-dependent surrogate function for the diffusivity $D$ at 300 K, using the predicted values from Ref. \cite{VanderVen2000}:
\begin{equation}
D = 0.01\exp(-274(1.05-x)(0.47-x)(1.-x))
\end{equation}
The diffusivity surrogate function and the predicted data appear in Figure S5, where the effective vibrational frequency $\nu^* $ is reported to be on the order of $10^{13}$ s$^{-1}$ \cite{VanderVen2000}. The mobility $\widetilde{M}$ is related to $D$ by the equation $\widetilde{M} = Dx/(k_BT)$ \cite{Jiang2016}. The value of $D$ is multiplied by 4 to approximate the diffusivity at 340 K and is divided by 4 for 260 K. This approximation was obtained from studies in which Li  diffusion in LCO was computed by a combination of DFT and kinetic Monte Carlo that accounted for the composition-dependent variation of the migration energy barrier \cite{VanderVen2000}. These computations reported the composition-dependent diffusivity at 300 and 400 K, with $D$ at 400 K being approximately 30 times the diffusivity at 300 K, from which we have drawn the approximation of the factor of 4 for every 40 K.

We obtain values of the gradient parameters $\chi_i$ in Eq. (\ref{eqn:free_energy}) such that the phase field calculation of the interface energy agrees with the DFT results given in Section \ref{sec:DFTintfc}.
Let $\gamma$ and $\beta$ be the anti-phase boundary energies of two ordered LCO variants at composition $x=0.5$ calculated by DFT and phase field theory, respectively. And similarly consider $\widehat{\gamma}$ and $\widehat{\beta}$ for the interface energies between an ordered variant at $x=0.5$ and a disordered matrix at $x>0.5$ and $\eta_i=0,i=1,\ldots,6$ calculated by DFT and phase field theory, respectively. One cannot define a disordered LCO atomic 
 structure with any degree of uniqueness, making consistent calculation of $\widehat{\gamma}$ challenging. Therefore, we use an estimation that $\widehat{\gamma} = \frac{1}{2}\gamma$ \cite{wang1998}. 

We then define an inverse phase field  problem: Given $\gamma$, find $\chi_0$ and $\chi=\chi_1=\ldots=\chi_6$ such that $\sqrt{(\beta-\gamma)^2+(\widehat{\beta}-\frac{\gamma}{2})^2} < \varepsilon$, where we set the threshold $\varepsilon = 0.1\;\text{mJ/m}$. We compute $\beta$ and $\widehat{\beta}$ on a thin rectangular domain with length and height of $0.2$ and $0.005$ m, respectively. We obtain optimal $\chi_0$ and $\chi$ by performing an exhaustive two-dimensional grid search. The inferred values are: $\chi_0 = 1\times 10^{-4}\;\text{mJ/m}$ at $260$ K and $\chi_0 = 1.88\times 10^{-4}\;\text{mJ/m}$ at $300$ K, $\chi_1,\dots,\chi_6 = 2.12\times 10^{-8}\; \text{mJ/m}$ at $260$ K and $\chi_1,\dots,\chi_6 = 4.91\times 10^{-8}\; \text{mJ/m}$ at $300$ K. At these values, the minimized function is $0.05\;\text{mJ/m}$ at $260$ K and $0.08\;\text{mJ/m}$ at $300$ K.

The simulations are performed using the finite element method with the \texttt{mechanoChemFEM} code\footnote{Code available at github.com/mechanoChem/mechanoChemFEM}, which is based on the \texttt{deal.II} \cite{dealii2017} library, and run on the Great Lakes HPC cluster at the University of Michigan. Adaptive meshing with hanging nodes and adaptive time stepping are used.

\section {Data on DFT calculations for anti-phase boundary energies}

\begin{table}[h]
\caption{Anti-phase boundary interface energy between ordered variants $\eta_1=\frac{1}{2}$ and $\eta_3=\frac{1}{2}$ (see Figure 3a) computed using large-scale DFT calculations with relaxation of ionic forces and cell stress.}
\small
\centering
\begin{tabular}{c c c c c }
\hline
Number of MC   & Number of  & $\gamma^{\prime}$\\ 
super-cells ($2N_{\textrm{cell}}$) &  atoms (electrons) & (mJ/$\textrm{m}^2$) \\ \hline\hline
2    & 448 (3904)     & 45.9       \\
4   & 896 (7808)     & 57.9 \\
6   & 1344 (11712)     & 74.4  \\\hline\hline
$\gamma$   &     & 30.9\\\hline
\end{tabular}
\label{tbl:Table1}
\end{table}
\section {Explicit nucleation}

The idea of defining explicit nucleation in phase field simulations using the nucleation rate from classical nucleation rate was proposed by Simmons et al. \cite{simmons2000,simmons2004}. The probability $P_n$ of a nucleus forming in an element within a time step $\Delta t$ is given by:
\begin{align}
    P_n = 1 - \exp(-J^* \Delta t)
\end{align}
The nucleation rate $J^*$ within an element is given by
\begin{align}
    J^*  &= ZN\beta^* \exp\left(-\frac{\Delta G^* }{k_BT}\right)\exp\left(\frac{\tau}{t}\right)
\end{align}
where $Z$ is the Zeldovich factor, $N$ is the number of atoms (Li sites in our case) in the element, $\beta^* $ is the frequency factor, $\Delta G^* $ is the activation energy to create a stable nucleus, $k_B$ is Boltzman's constant, and $T$ is the temperature. We neglect the incubation term with $\tau$. Expressions for $Z$ and $\beta^* $ are given by \cite{legoues1984}:

\begin{align}
    Z &= \frac{3v_\mathrm{o}}{4\pi^{3/2}}\left[\frac{\Delta G^* }{k_BT}\right]^{1/2}\frac{1}{{R^* }^3}\\
    \beta^*  &= \frac{4\pi{R^* }^2Dx_0}{a^4}
\end{align}
where $v_\mathrm{o}$ is the average volume per (Li) atom in the ordered phase (roughly twice the unit cell volume), $R^* $ is the critical radius, $D$ is the diffusivity (see Eq. (\ref{eq:diffusivity}) ), $x_0$ is the average composition, and $a$ is the average lattice parameter of the nucleus and matrix phases (approximated as the cube root of the unit cell volume). We can combine the three terms in the prefactor and simplify slightly, and we let $v_\mathrm{e} := Nv_o$, the volume of the element:
\begin{align}
    ZN\beta^*  &= \frac{3v_\mathrm{e}Dx_0}{R^* a^4}\left[\frac{\Delta G^*}{\pi k_BT}\right]^{1/2}
\end{align}

A similar expression of the nucleation rate, as well as an explanation of how to derive $\Delta G^* $ is presented in a review of classical nucleation theory by Sear \cite{sear2007}. The change in free energy for forming a nucleus is equal to the change in bulk energy plus the interfacial energy that is introduced. For a spherical precipitate, the total change in energy $\Delta G$ is the following
\begin{align}
    \Delta G = -\frac{4}{3}\pi R^3 \Delta G_v + 4\pi R^2\gamma
    \label{eqn:DeltaG}
\end{align}
where $R$ is the radius of the precipitate, $\Delta G_v$ is the decrease in free energy per unit volume (i.e. a positive $\Delta G_v$ signifies a drop in the free energy density; more on this term later), and $\gamma$ is the interfacial energy. $\Delta G^*$ maximizes this energy since it is the energy barrier, so we differentiate w.r.t. $R$ and set equal to zero to solve for the critical radius, $R^* $:
\begin{align}
    0 &= -4\pi {R^* }^2\Delta G_v + 8\pi R^* \gamma\\
    \implies R^*  &= \frac{2\gamma}{\Delta G_v}
\end{align}
Plugging back into Eq. (\ref{eqn:DeltaG}) gives the value for homogeneous nucleation
\begin{align}
    \Delta G^*_\mathrm{hom} &= -\frac{4}{3}\pi \left(\frac{2\gamma}{\Delta g_v}\right)^3 \Delta G_v + 4\pi \left(\frac{2\gamma}{\Delta g_v}\right)^2\gamma\\
    &= \frac{16\pi\gamma^2}{3\Delta G_v^3}
\end{align}
Strictly speaking, this is for homogeneous nucleation. Another term, $f(\theta)$, is multiplied for heterogeneous nucleation, which is a function of the angle $\theta$ between the interface and the (flat) boundary:
\begin{align}
    \Delta G^*_\mathrm{het} &= f(\theta)\Delta G^*_\mathrm{hom}\\
    f(\theta) &= \frac{1}{2} -  \frac{3}{4}\cos\theta + \frac{1}{4}\cos^3\theta
\end{align}
Note that $f(\pi/2) = 0.5$, which is the value that is used here.

\begin{figure}[h!]
\begin{minipage}[t]{0.33\textwidth}
\centering
    \includegraphics[width=\textwidth]{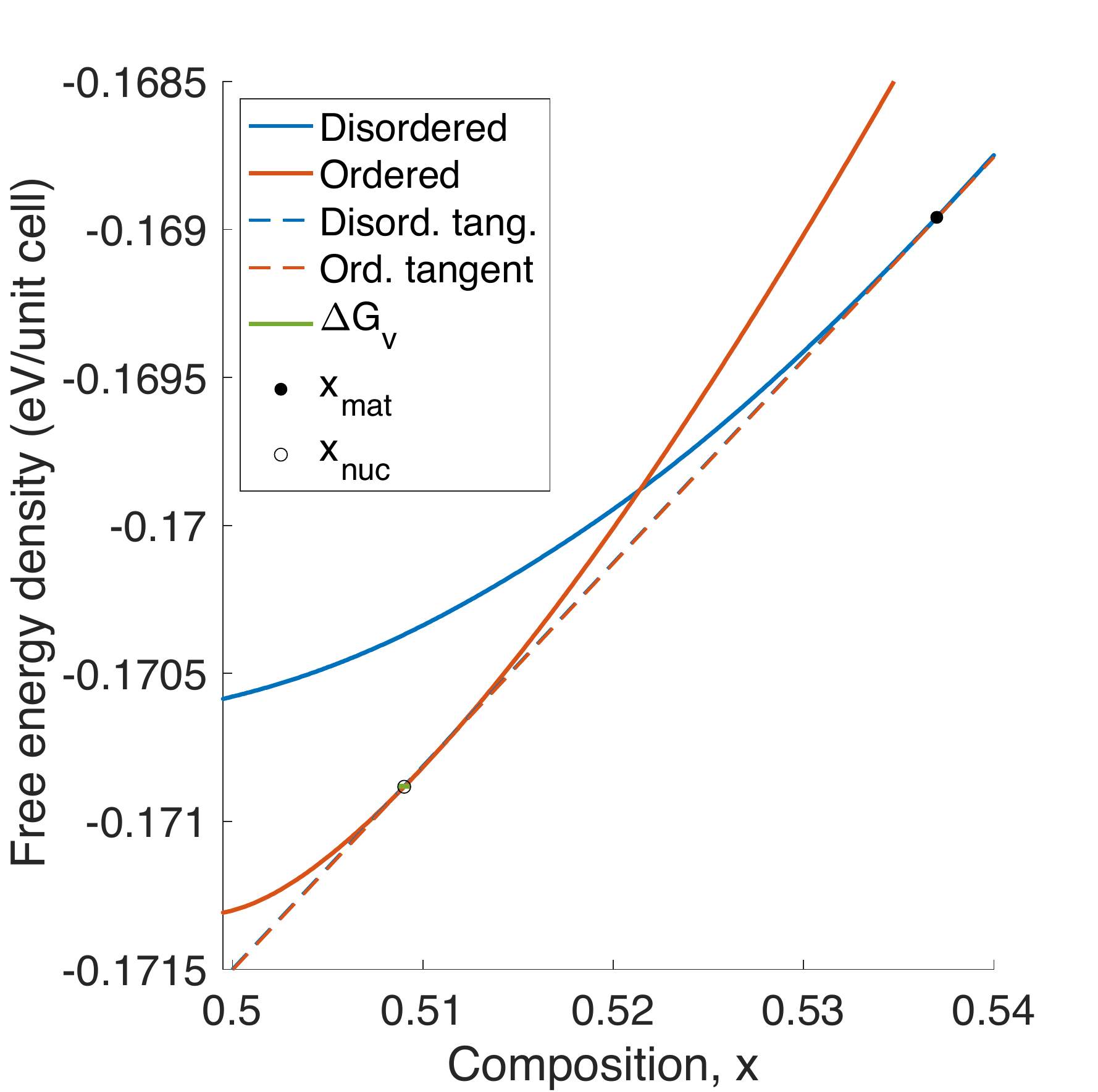}
    \subcaption{}
\end{minipage}
\begin{minipage}[t]{0.33\textwidth}
    \centering
    \includegraphics[width=\textwidth]{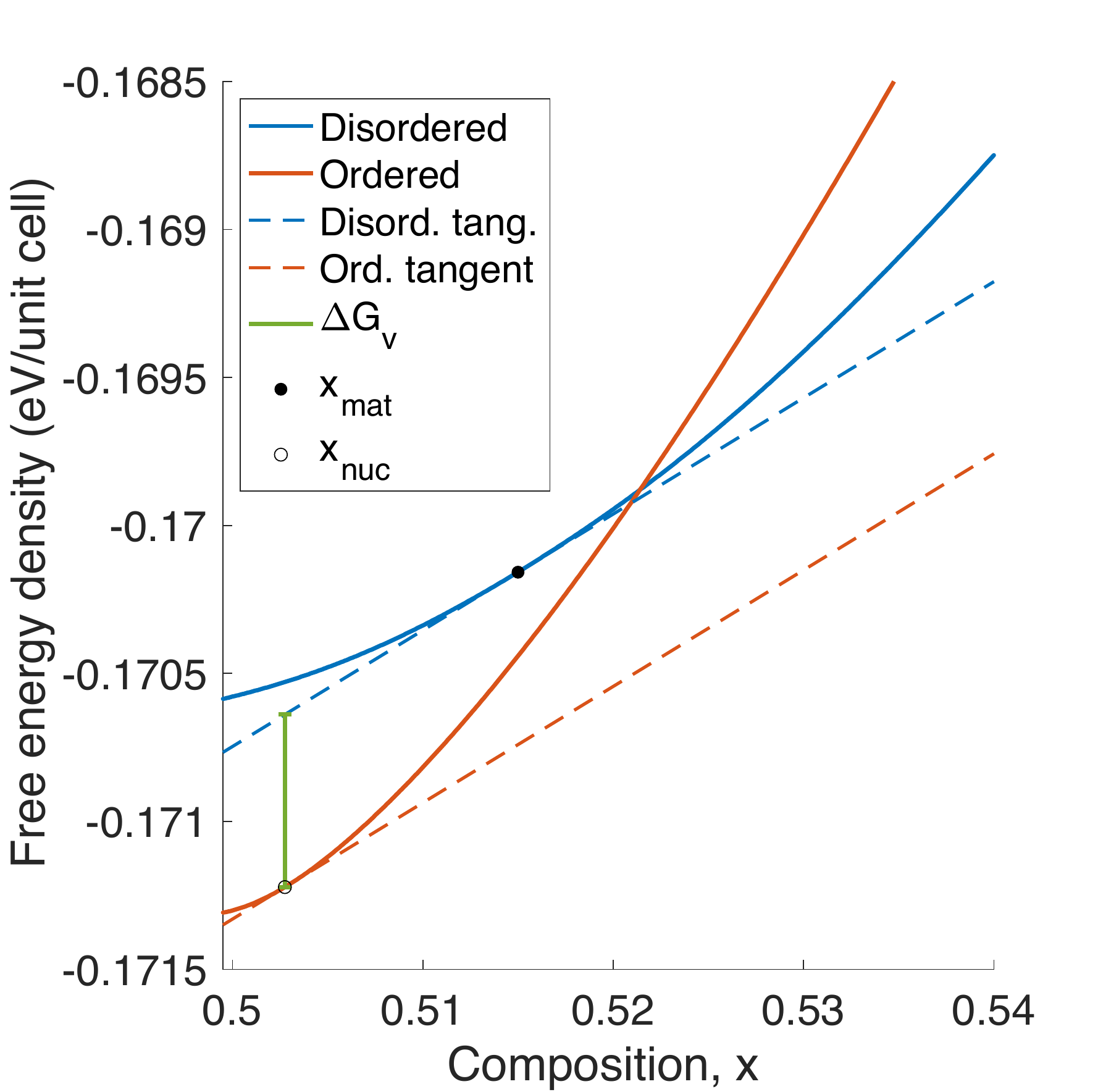}
    \subcaption{}
\end{minipage}
\begin{minipage}[t]{0.33\textwidth}
    \includegraphics[width=\textwidth]{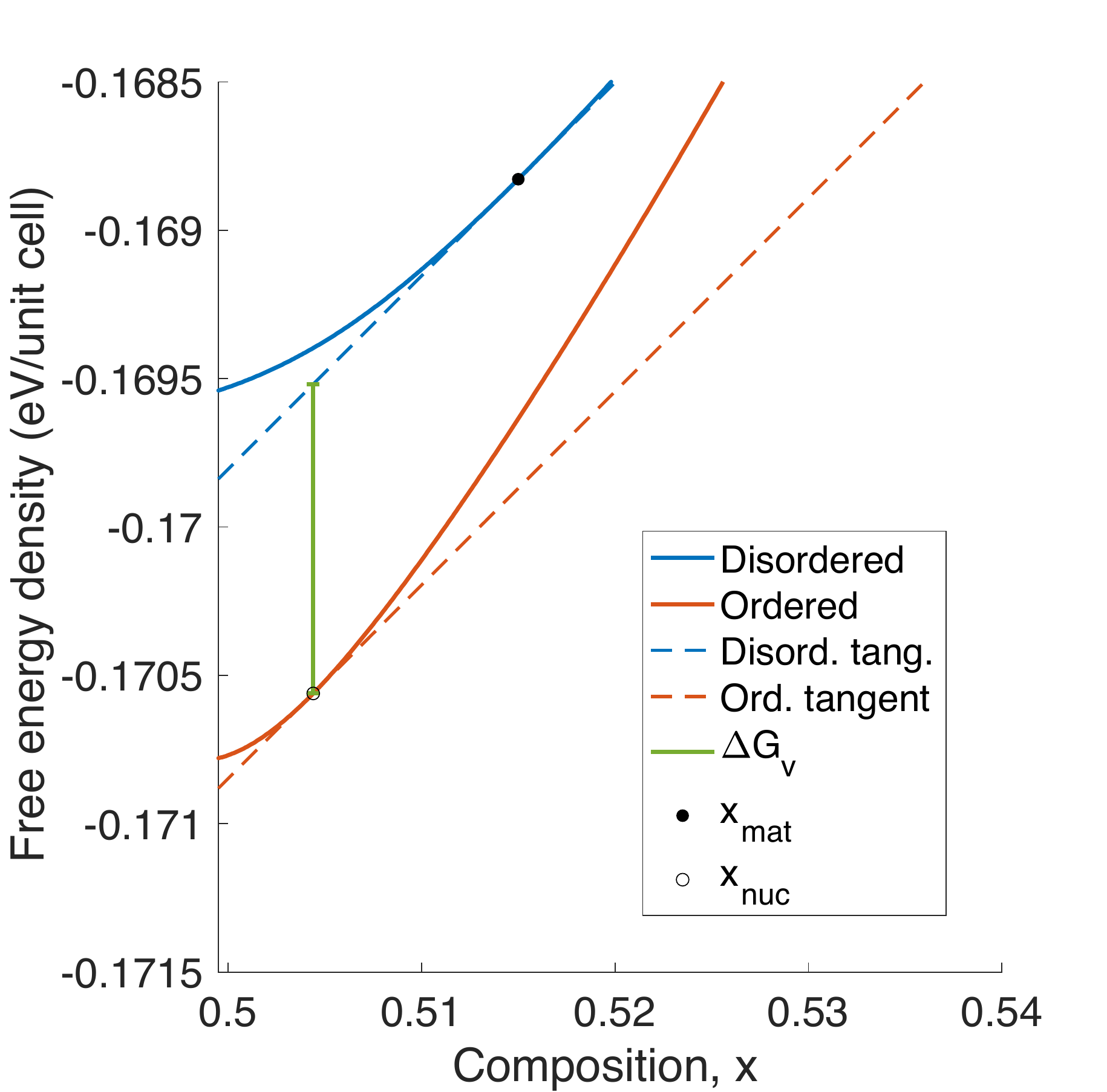}
    \subcaption{}
\end{minipage}
    \caption{(a) Demonstration of how to calculate the driving force of nucleation, $\Delta G_v$, at a composition of, $x_\text{mat}$ and $x_\text{nuc}$ of the disordered matrix and ordered nucleus, respectively. The driving force $\Delta G_v$ is zero at a composition resulting in a common tangent line. (b) $\Delta G_v$ is the vertical distance between parallel tangent lines to the ordered and disordered free energy densities, as drawn here at $300$ K. (c) $\Delta G_v$ at $260$ K is $\sim 2\times$ the value at $300$ K.}
    \label{fig:deltaGv}
\end{figure}

Poduri and Chen \cite{poduri1996} suggest calculating the driving force, $\Delta G_v$, by drawing the tangent line w.r.t. $x$ of the disordered free energy density at the composition of the potential nucleation site, currently disordered (labeled $x_\mathrm{mat}$ in Figure \ref{fig:deltaGv}a). $\Delta G_v$ is the largest difference between this disordered tangent line and the ordered free energy density. This occurs at the composition (labeled $x_\mathrm{nuc}$ in Figure \ref{fig:deltaGv}) at which the ordered tangent line is parallel to the disordered tangent line. When the disordered phase is at a composition that gives a coincident or common tangent line with the ordered phase, $\Delta G_v$ is zero (see Figure \ref{fig:deltaGv})b.

\begin{figure}[h!]
    \centering
    \includegraphics[width=0.5\textwidth]{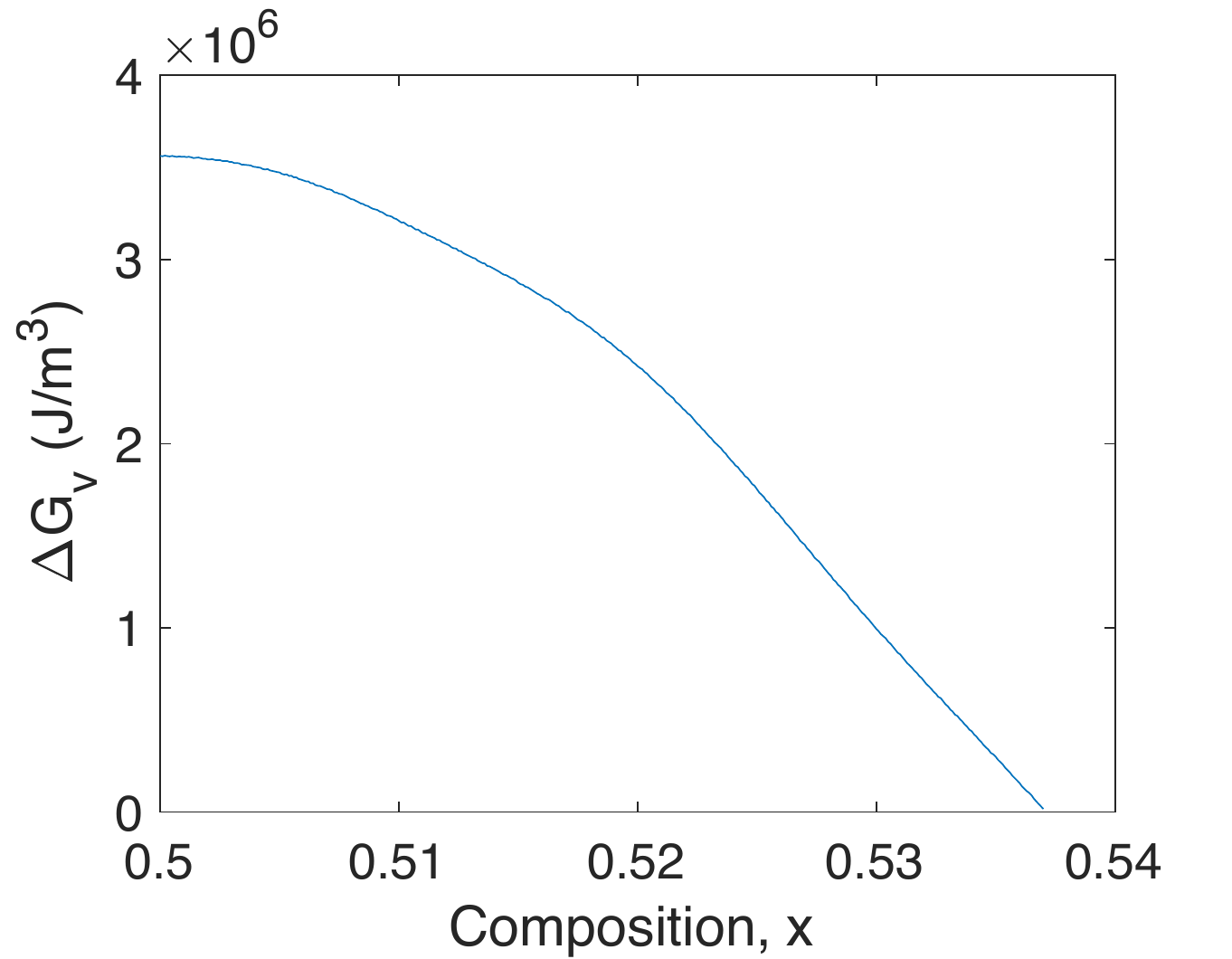}
    \caption{Plot of the driving force $\Delta G_v$ at a given composition within the disordered phase at 300 K.}
    \label{fig:deltaGv2}
\end{figure}

The free energy density DNN has units of eV/unit cell. With a unit cell volume of 32.502 \AA$^3$, the conversion factor to J/m$^3$ is $4.926\times10^9 \frac{\text{J/m}^3}{\text{eV/unit cell}}$. Figure \ref{fig:deltaGv2} shows the the driving force $\Delta G_v$ at a given composition within the disordered phase at 300 K. Within the operating range of a battery (i.e., $x \geq 0.5$), $\Delta G_v$ has a maximum of $3.563 \times 10^6$ J/m$^3$ at $x=0.5$ and becomes zero around $x = 0.537$.

\begin{figure}[h!]
\begin{minipage}[t]{0.48\textwidth}
    \centering
    \includegraphics[width=.9\textwidth]{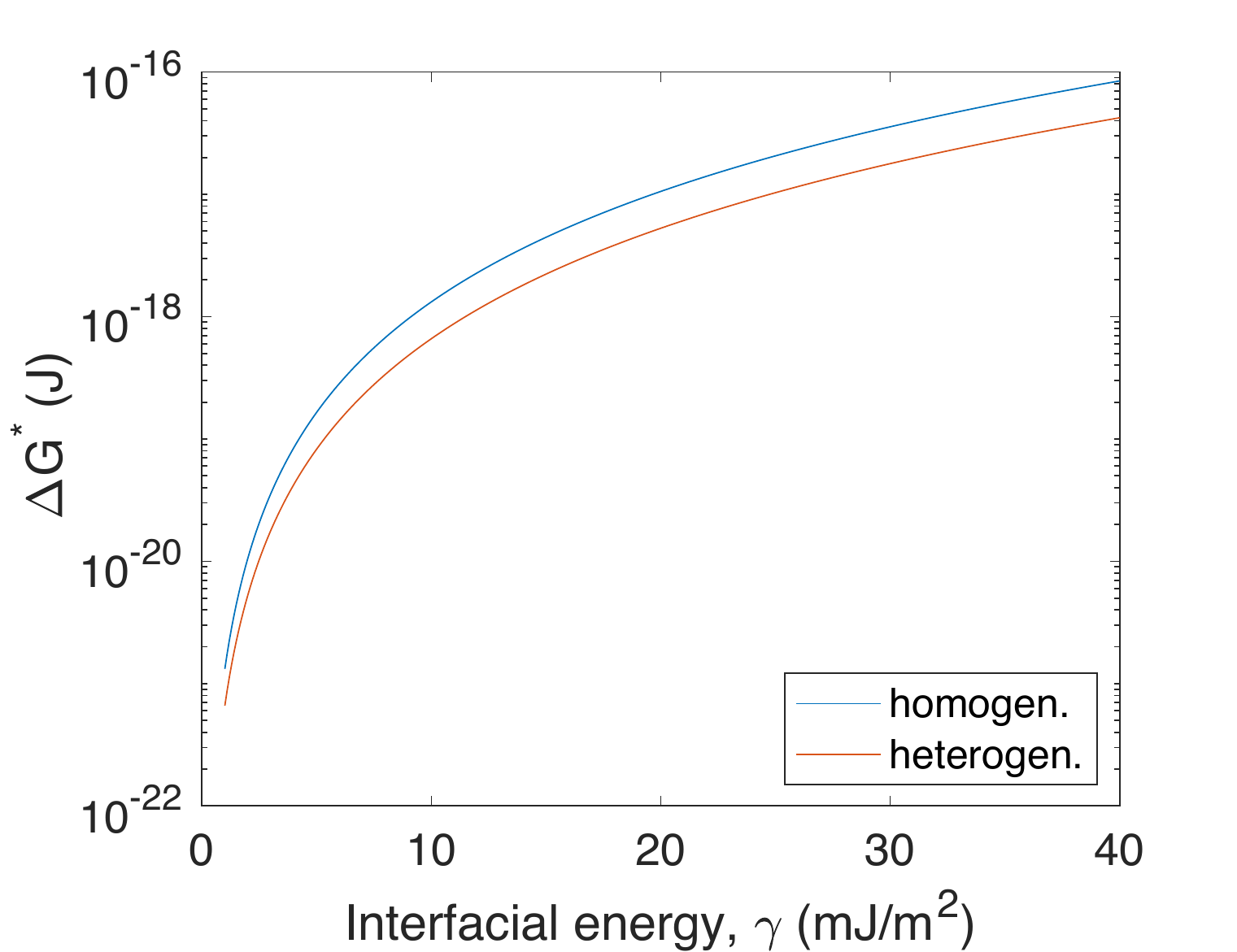}
    \subcaption{}
\end{minipage}%
\begin{minipage}[t]{0.48\textwidth}
\centering
    \includegraphics[width=.9\textwidth]{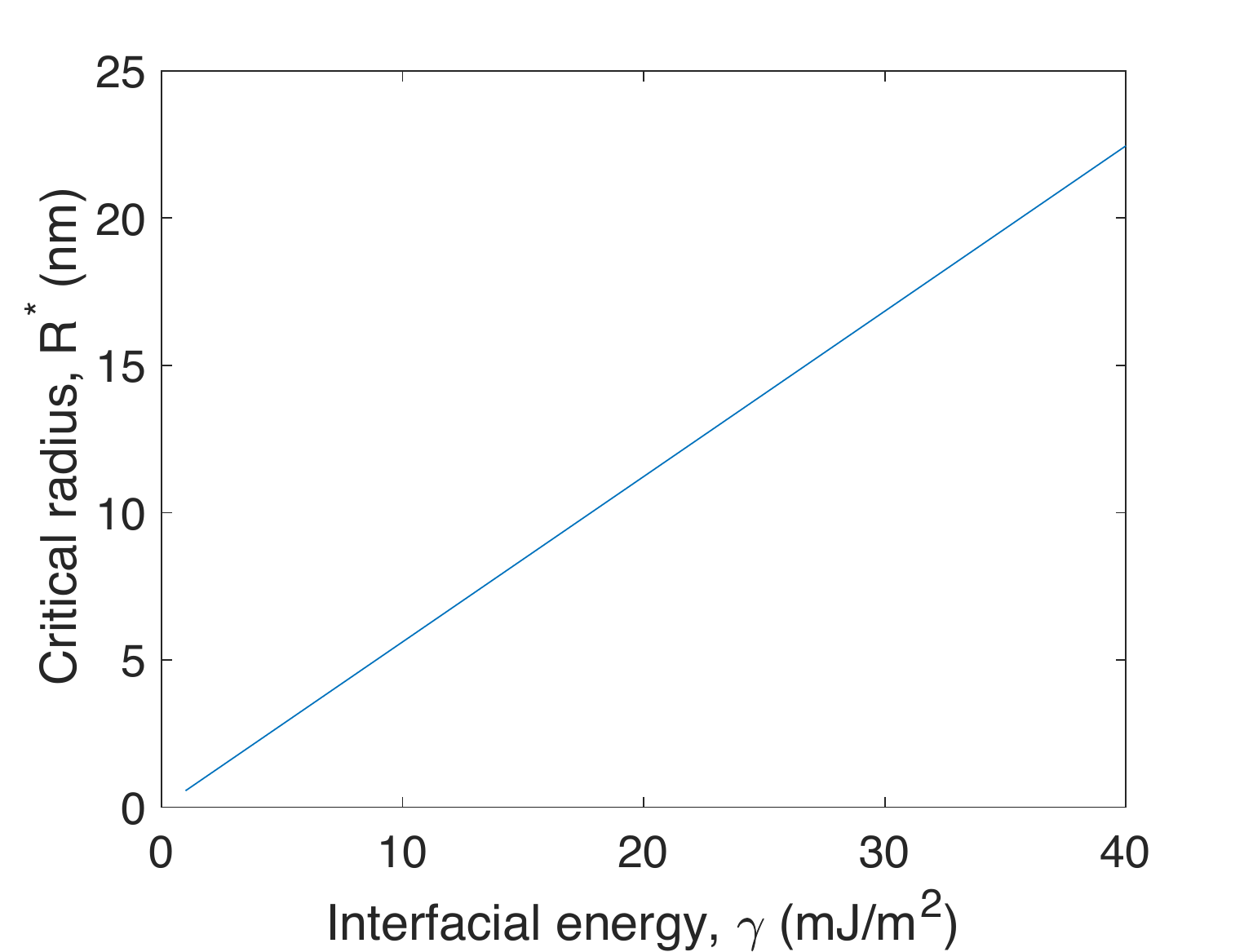}
    \subcaption{}
\end{minipage}
\begin{minipage}[t]{0.48\textwidth}
    \centering
    \includegraphics[width=.9\textwidth]{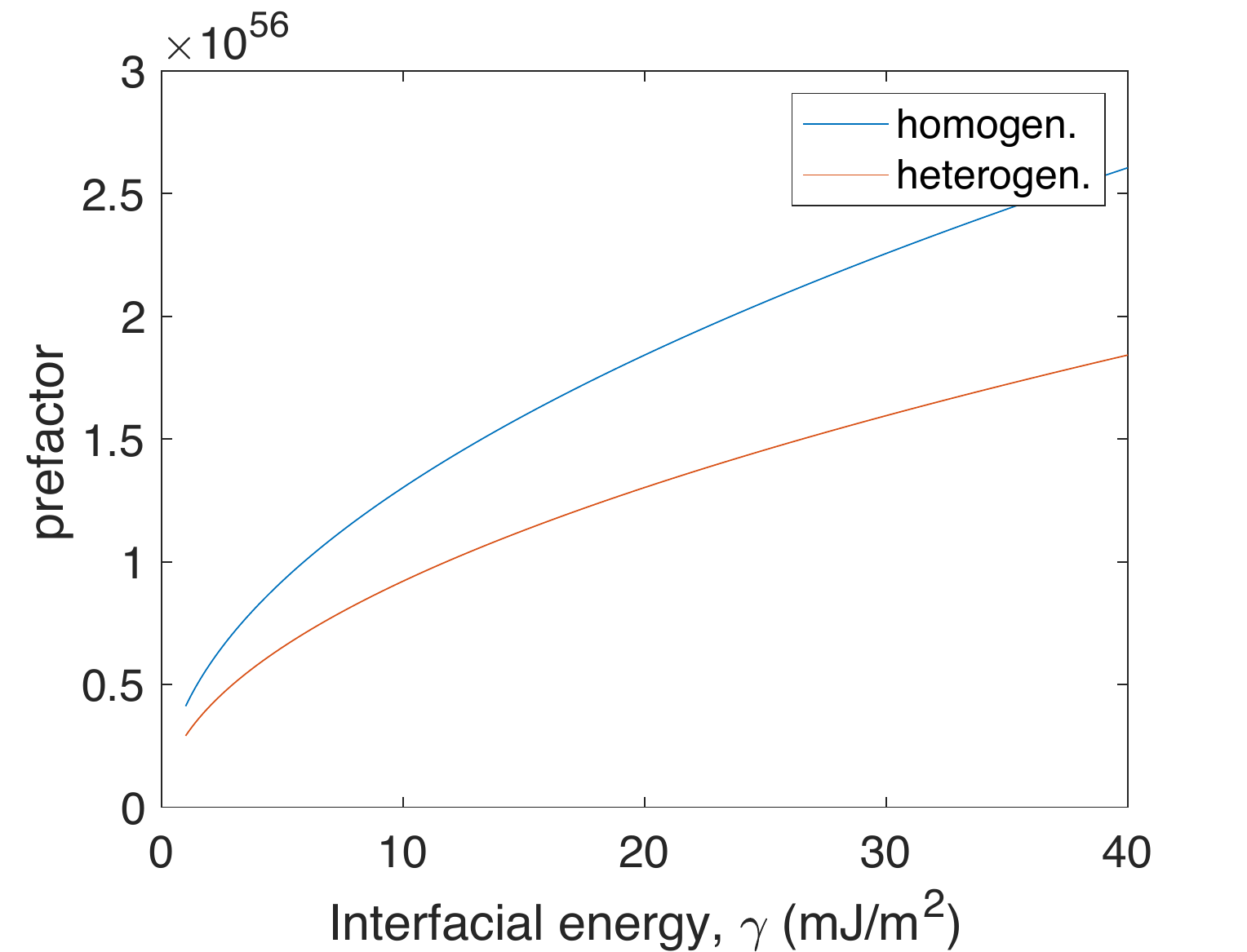}
    \subcaption{}
\end{minipage}%
\begin{minipage}[t]{0.48\textwidth}
\centering
    \includegraphics[width=.9\textwidth]{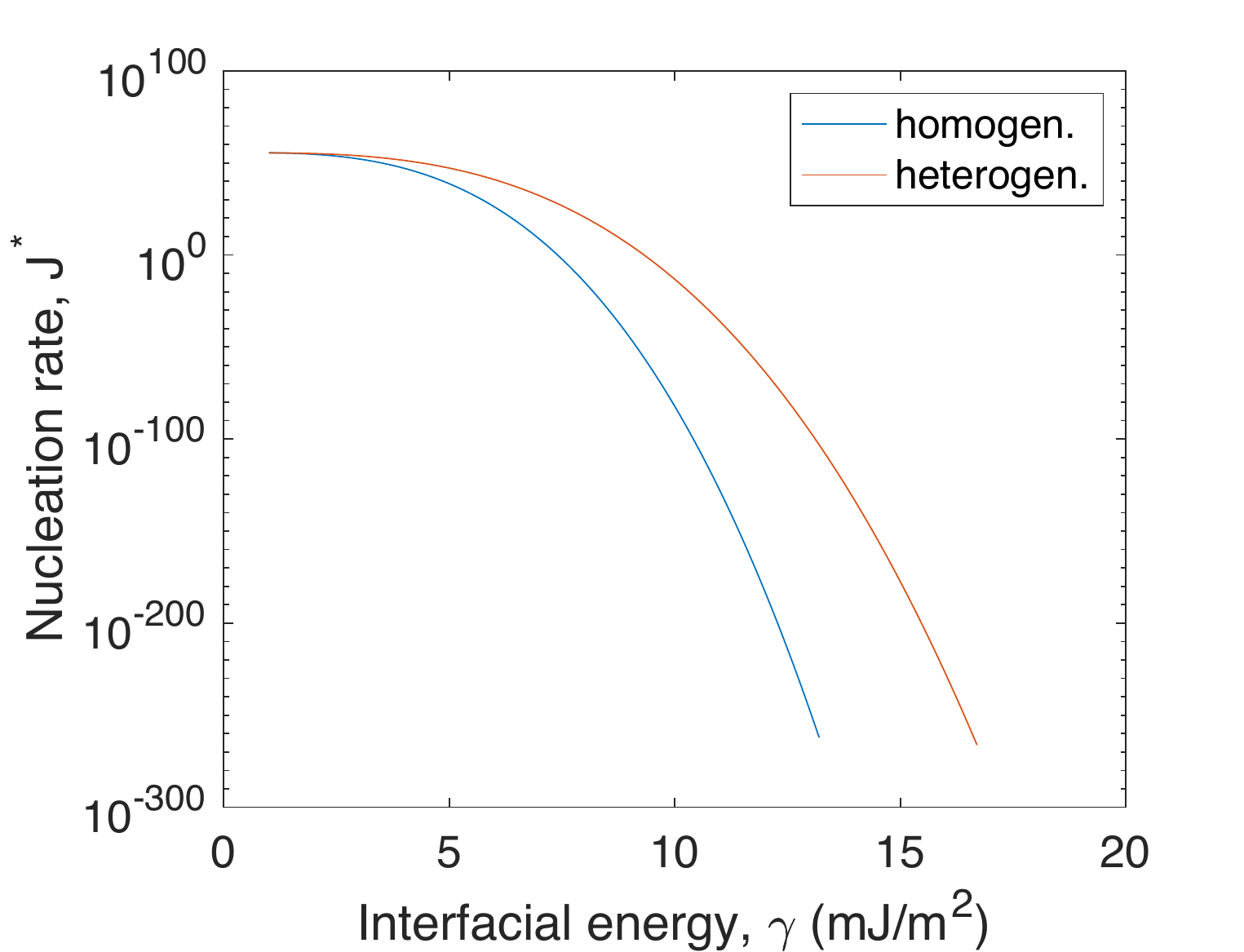}
    \subcaption{}
\end{minipage}
\begin{minipage}[t]{0.48\textwidth}
\centering
    \includegraphics[width=.9\textwidth]{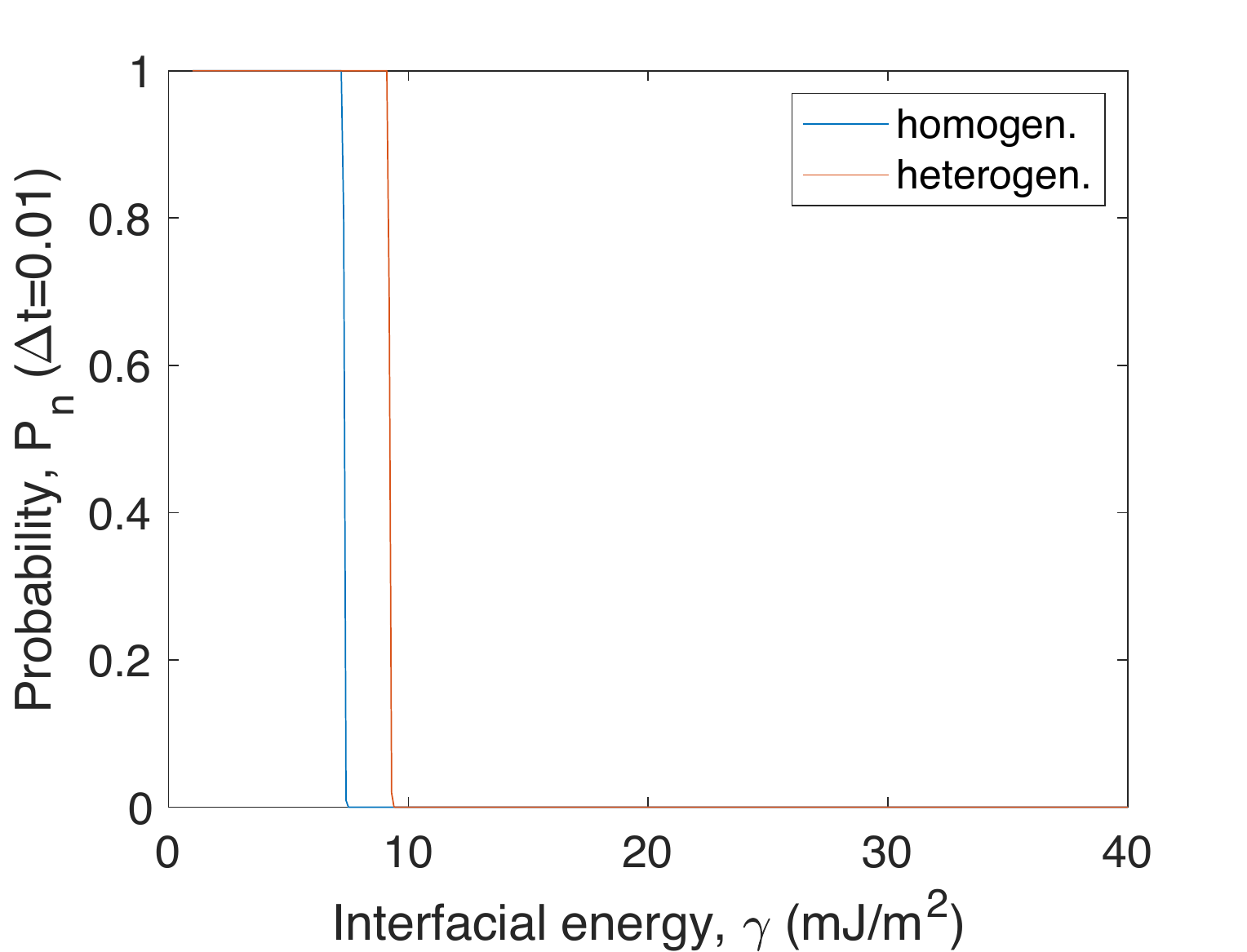}
    \subcaption{}
\end{minipage}
    \caption{(a) Critical free energy of nucleation, $\Delta G^* $, (b) critical radius, $R^* $, (c) prefactor, $ZN\beta^* $, (d) nucleation rate, $J^*$, and (e) probability of nucleation within an element over a timestep of 0.01 s, $P_n$, all as functions of the interfacial energy, $\gamma$, and using the maximum value of $\Delta G_v$.}
    \label{fig:star}
\end{figure}

To explore the effect of interfacial energy, Figure \ref{fig:star} shows $\Delta G$ and $R$ as a function of interfacial energy, using the largest driving force of $\Delta G_v = 3.563 \times 10^6$ J/m$^3$.

Under an applied voltage, $V$, the critical free energy for heterogeneous nucleation is
\begin{equation}
        \Delta G^*_\mathrm{het} = \frac{16\pi\gamma^2}{3(\Delta G_v - V\Delta x)^3}
\end{equation}
for $\Delta x = x_\text{nuc} - x_\text{mat}$. The corresponding rate of heterogeneous nucleation is then
\begin{align}
    J ^* &= ZN\beta^* \exp\left(-\frac{\Delta G^*_\mathrm{het}}{k_BT}\right)\exp\left(\frac{\tau}{t}\right)
\end{align}
Figure 5b in the main text represents the heterogeneous nucleation rate \emph{versus} $V$ and $x_\text{mat}$, accounting for the dependence of $x_\text{nuc}$ upon $x_\text{mat}$ as illustrated in the plots of Figure \ref{fig:deltaGv}.
\section {Thermodynamic representation ignoring ordering}
\begin{figure}[tb]
        \centering
\begin{minipage}[t]{0.28\textwidth}
        \centering
	\includegraphics[width=0.99\textwidth]{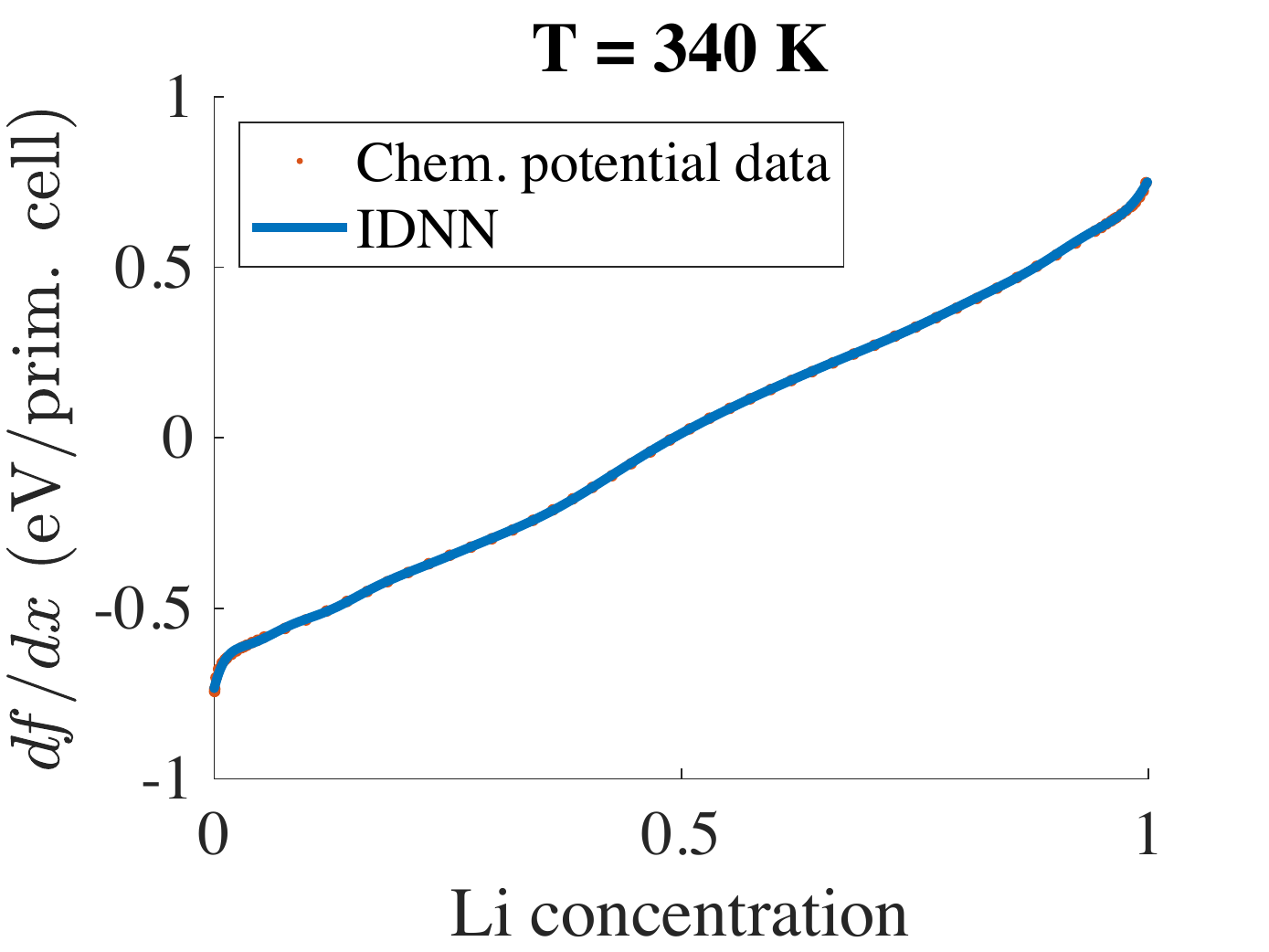}
\end{minipage}%
\begin{minipage}[t]{0.28\textwidth}
        \centering
	\includegraphics[width=0.99\textwidth]{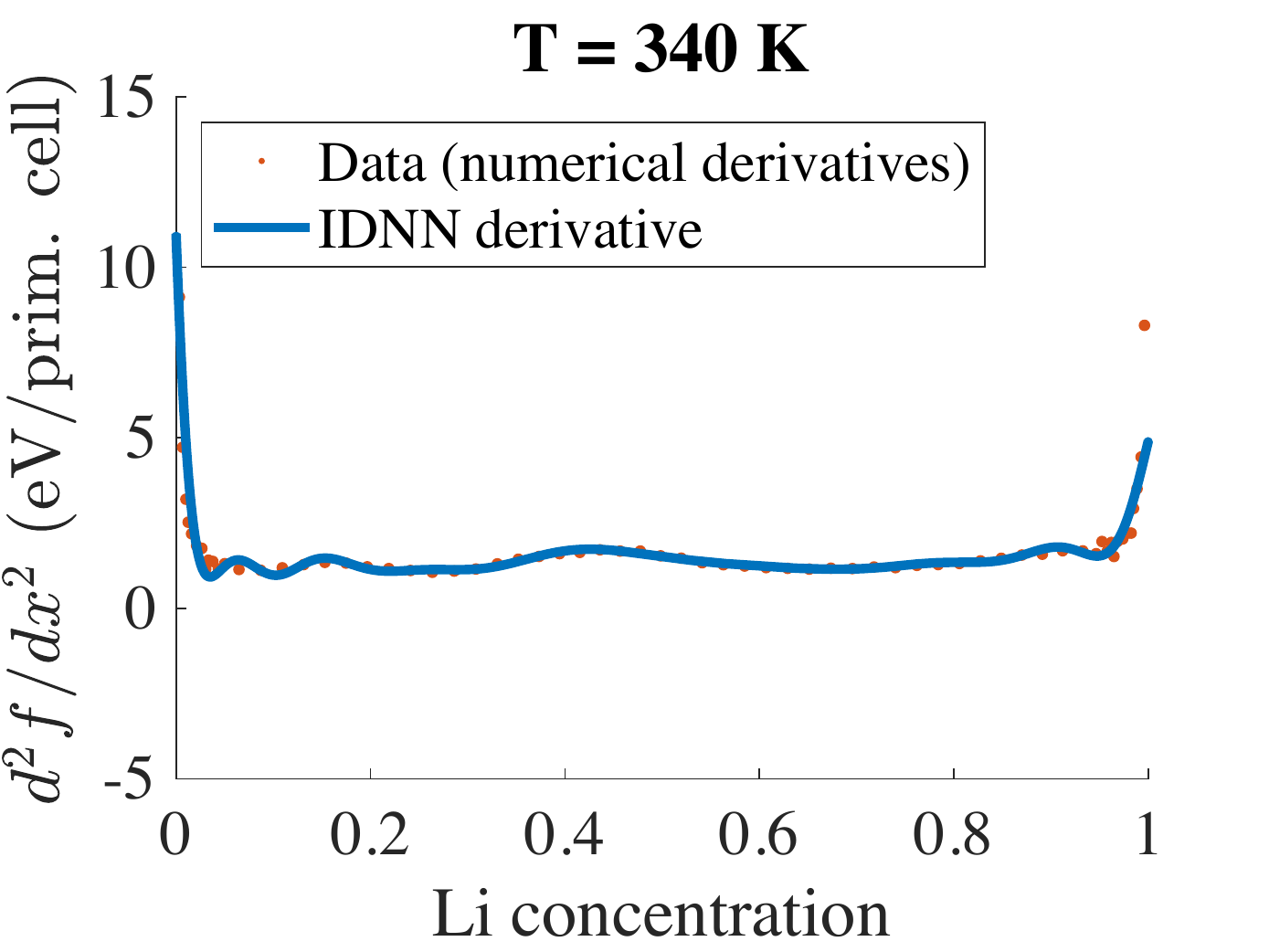}
\end{minipage}%
\begin{minipage}[t]{0.28\textwidth}
        \centering
	\includegraphics[width=0.99\textwidth]{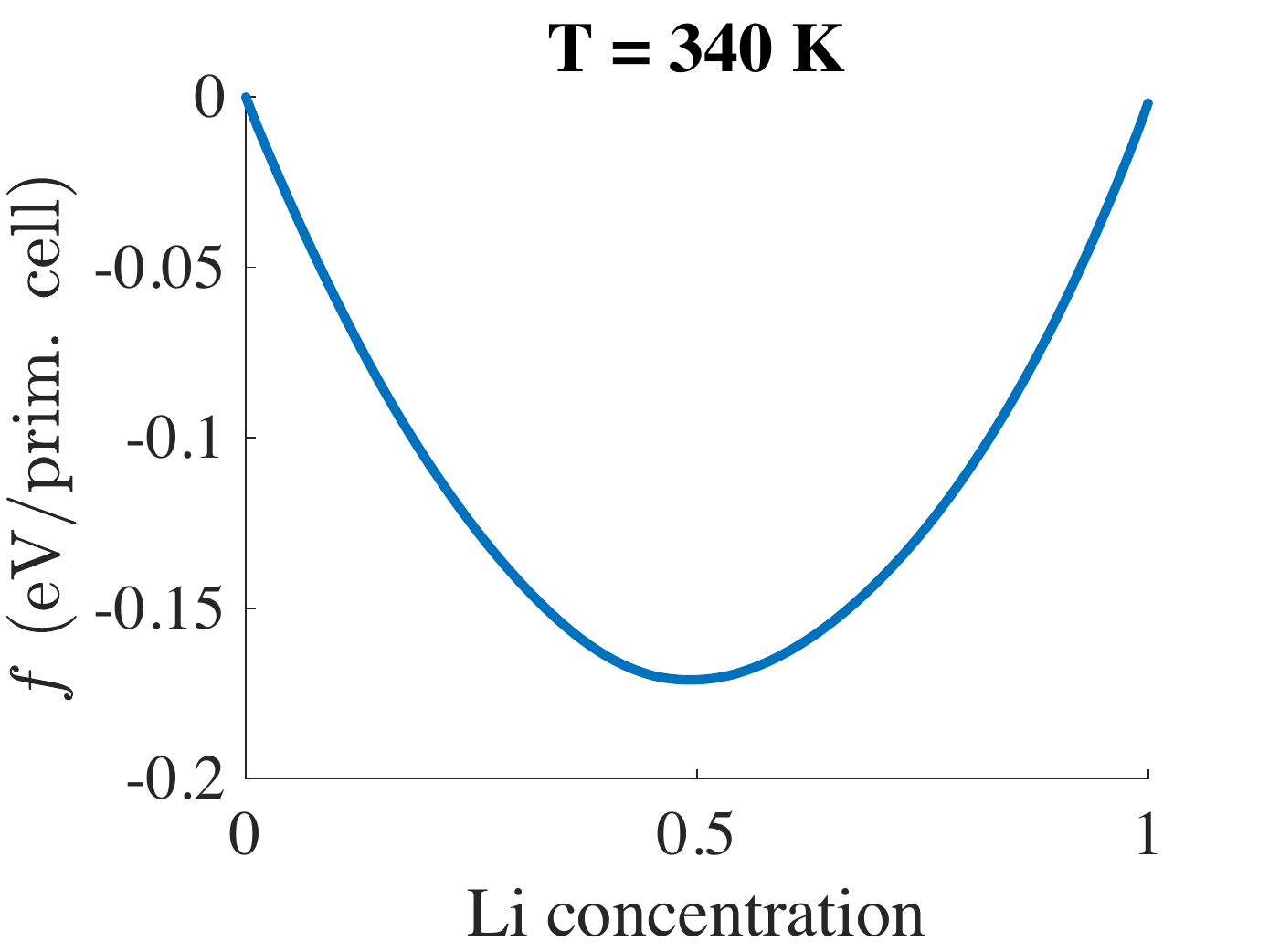}
\end{minipage}
\begin{minipage}[t]{0.28\textwidth}
        \centering
	\includegraphics[width=0.99\textwidth]{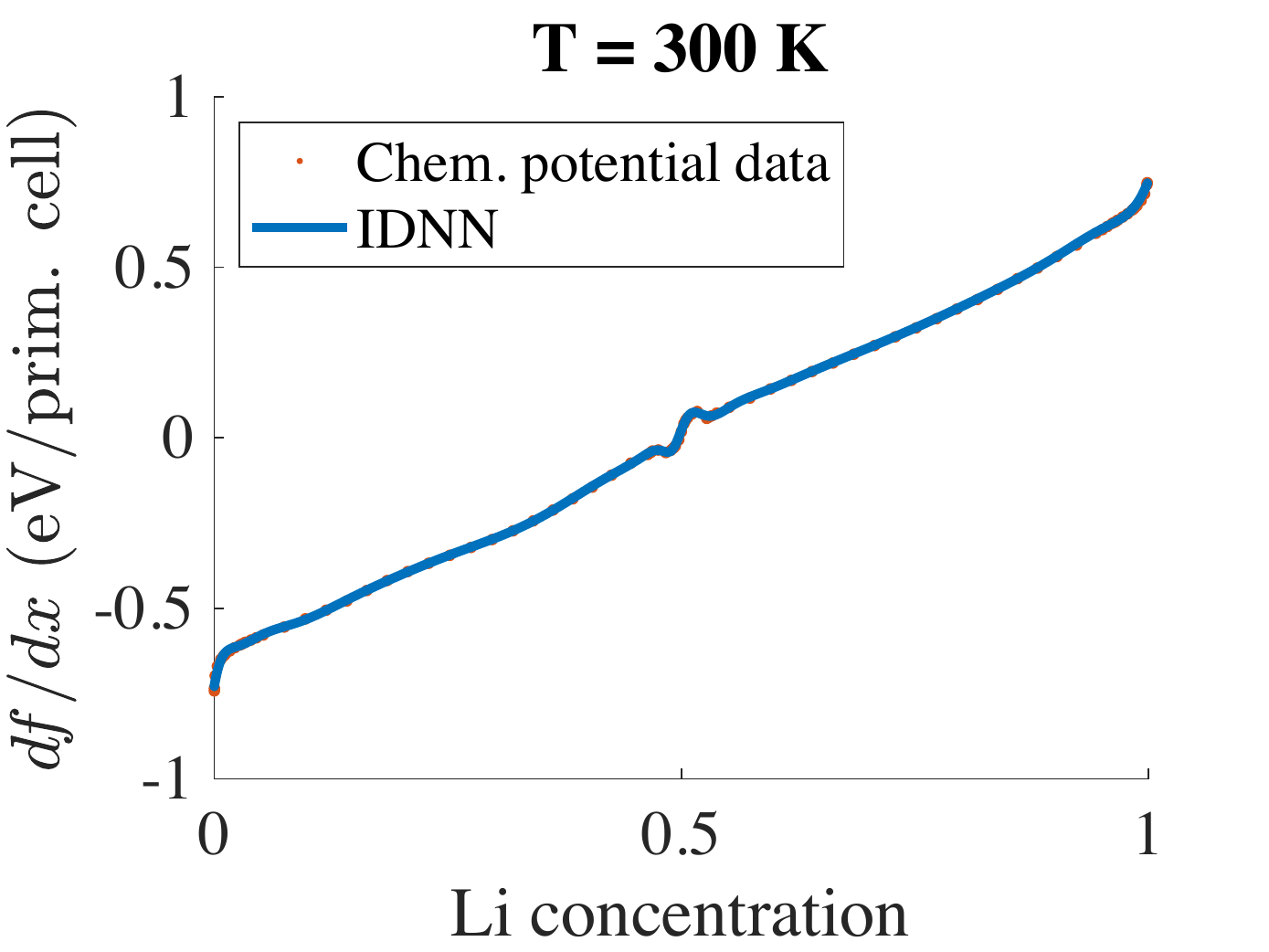}
\end{minipage}%
\begin{minipage}[t]{0.28\textwidth}
        \centering
	\includegraphics[width=0.99\textwidth]{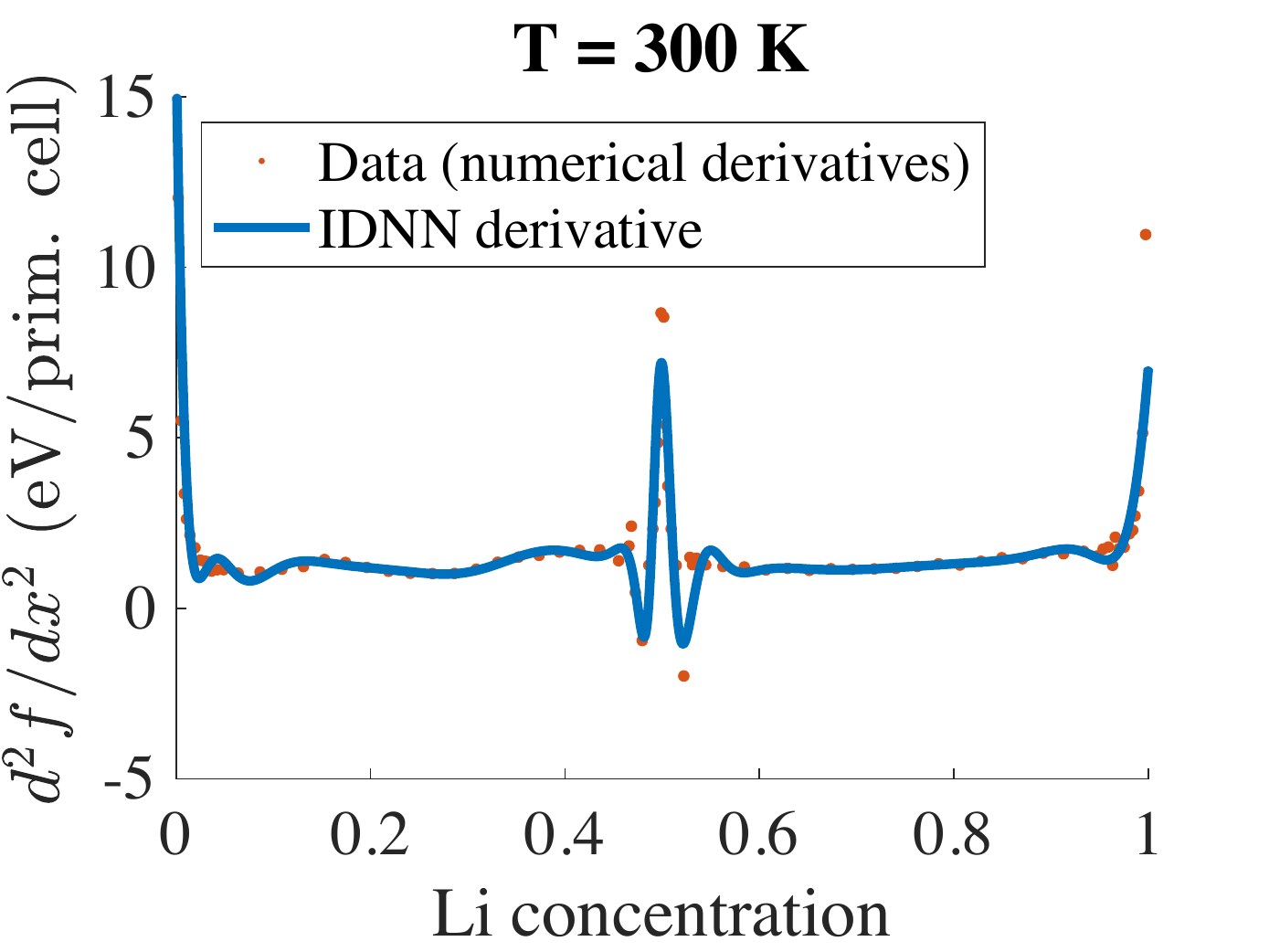}
\end{minipage}%
\begin{minipage}[t]{0.28\textwidth}
        \centering
	\includegraphics[width=0.99\textwidth]{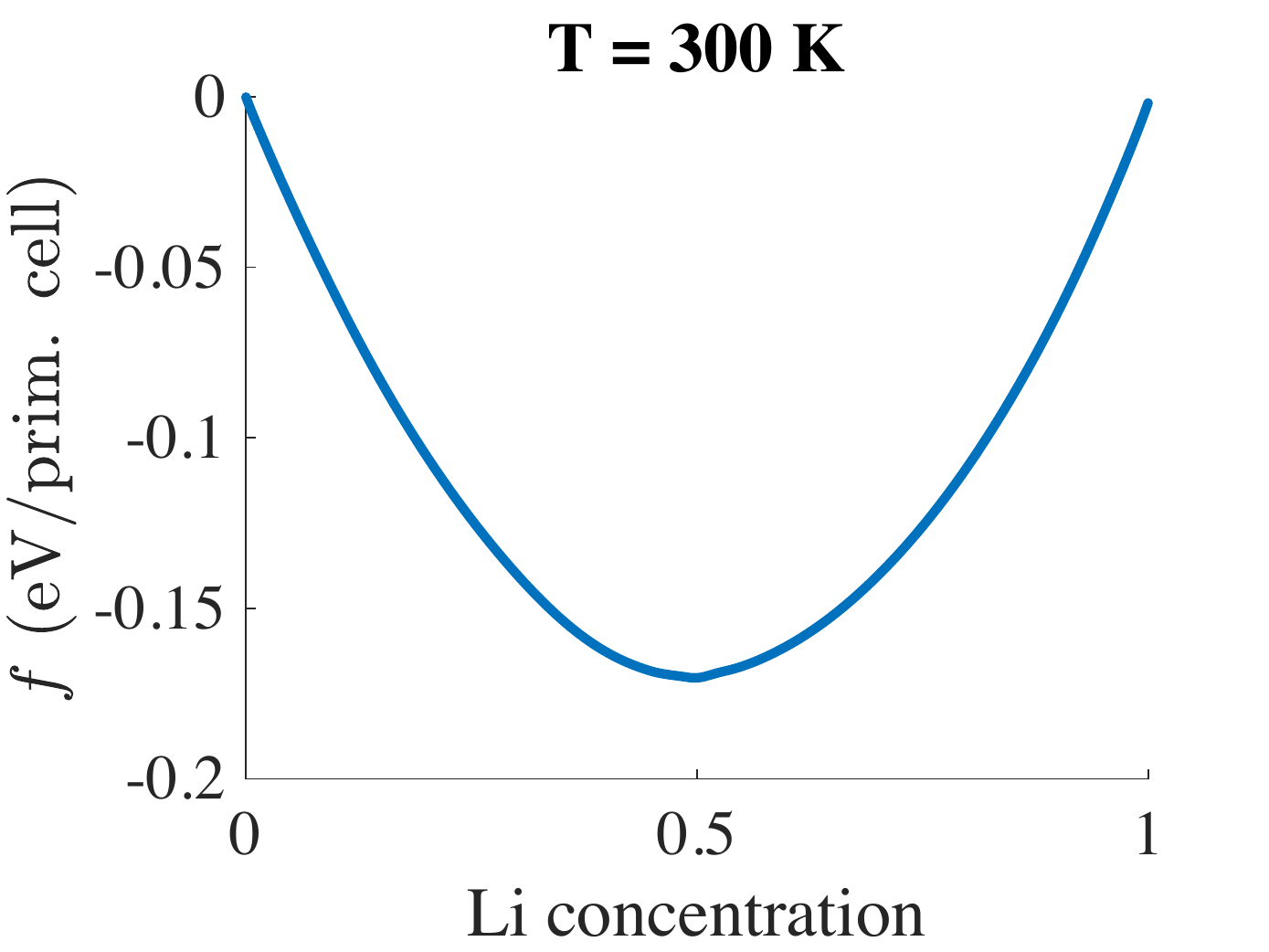}
\end{minipage}
\begin{minipage}[t]{0.28\textwidth}
        \centering
	\includegraphics[width=0.99\textwidth]{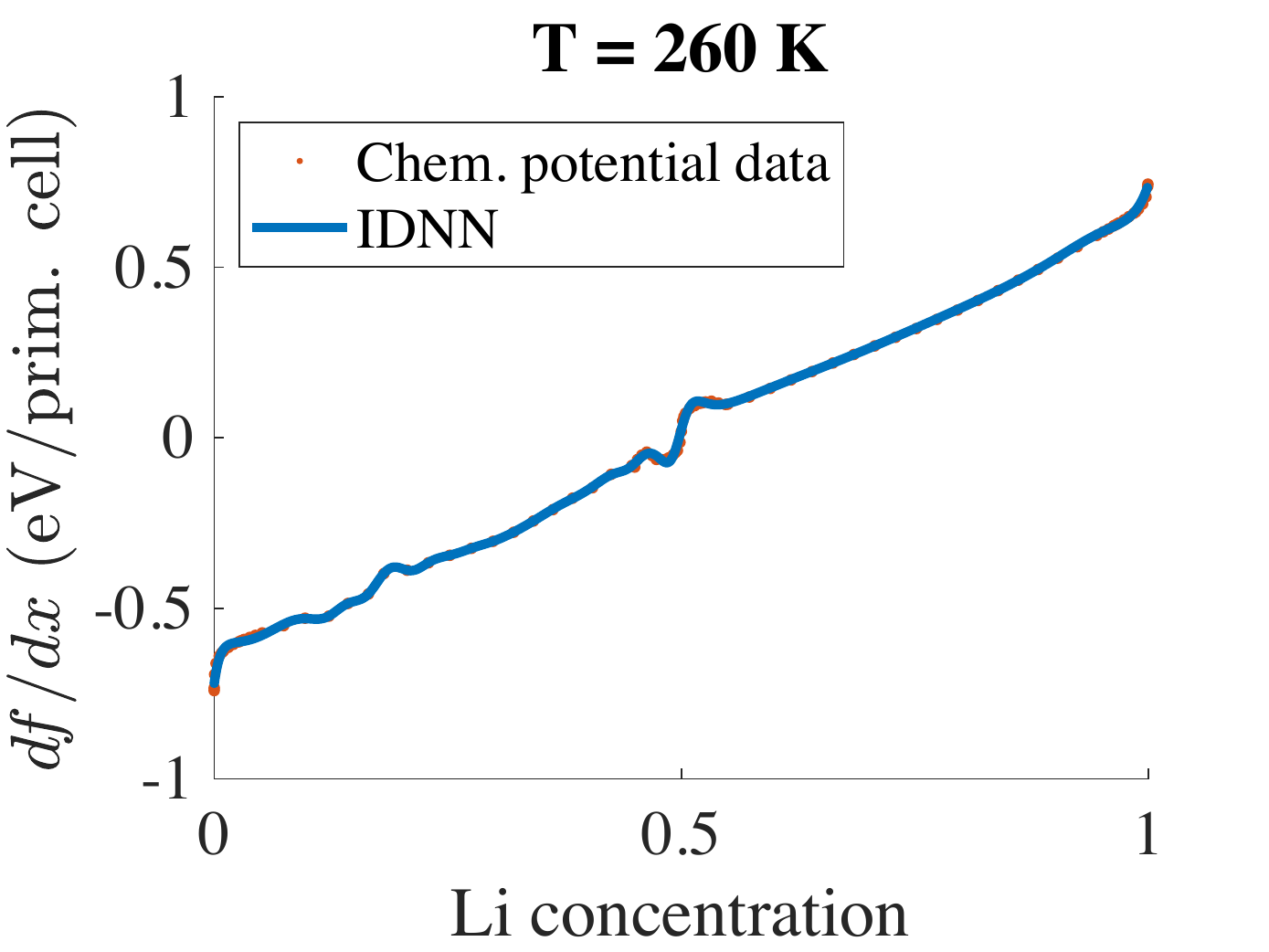}
	\subcaption{Chemical potential, $\mathrm{d}f/\mathrm{d}c$}
\end{minipage}%
\begin{minipage}[t]{0.28\textwidth}
        \centering
	\includegraphics[width=0.99\textwidth]{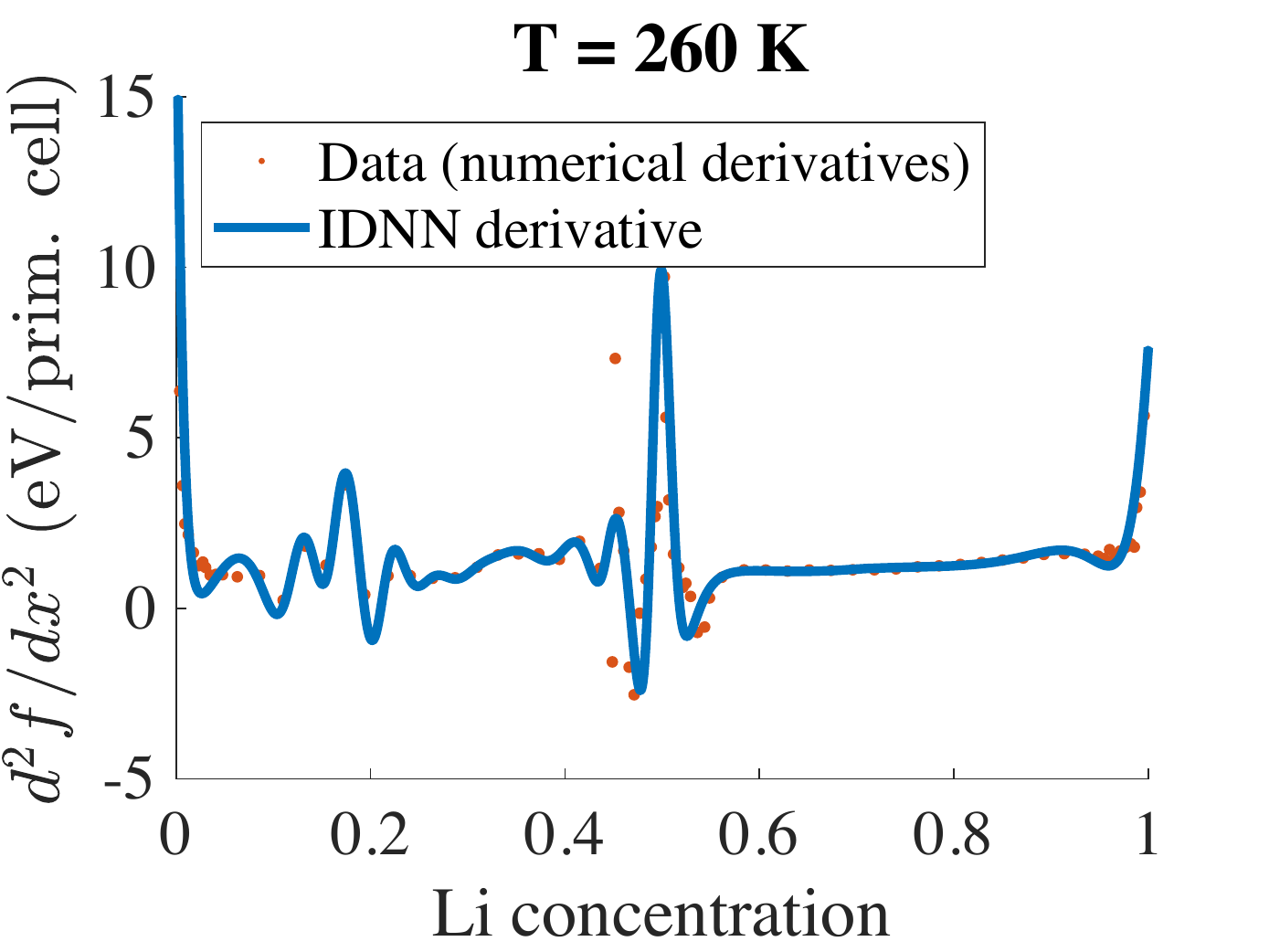}
	\subcaption{$\mathrm{d}^2f/\mathrm{d}c^2$}
\end{minipage}%
\begin{minipage}[t]{0.28\textwidth}
        \centering
	\includegraphics[width=0.99\textwidth]{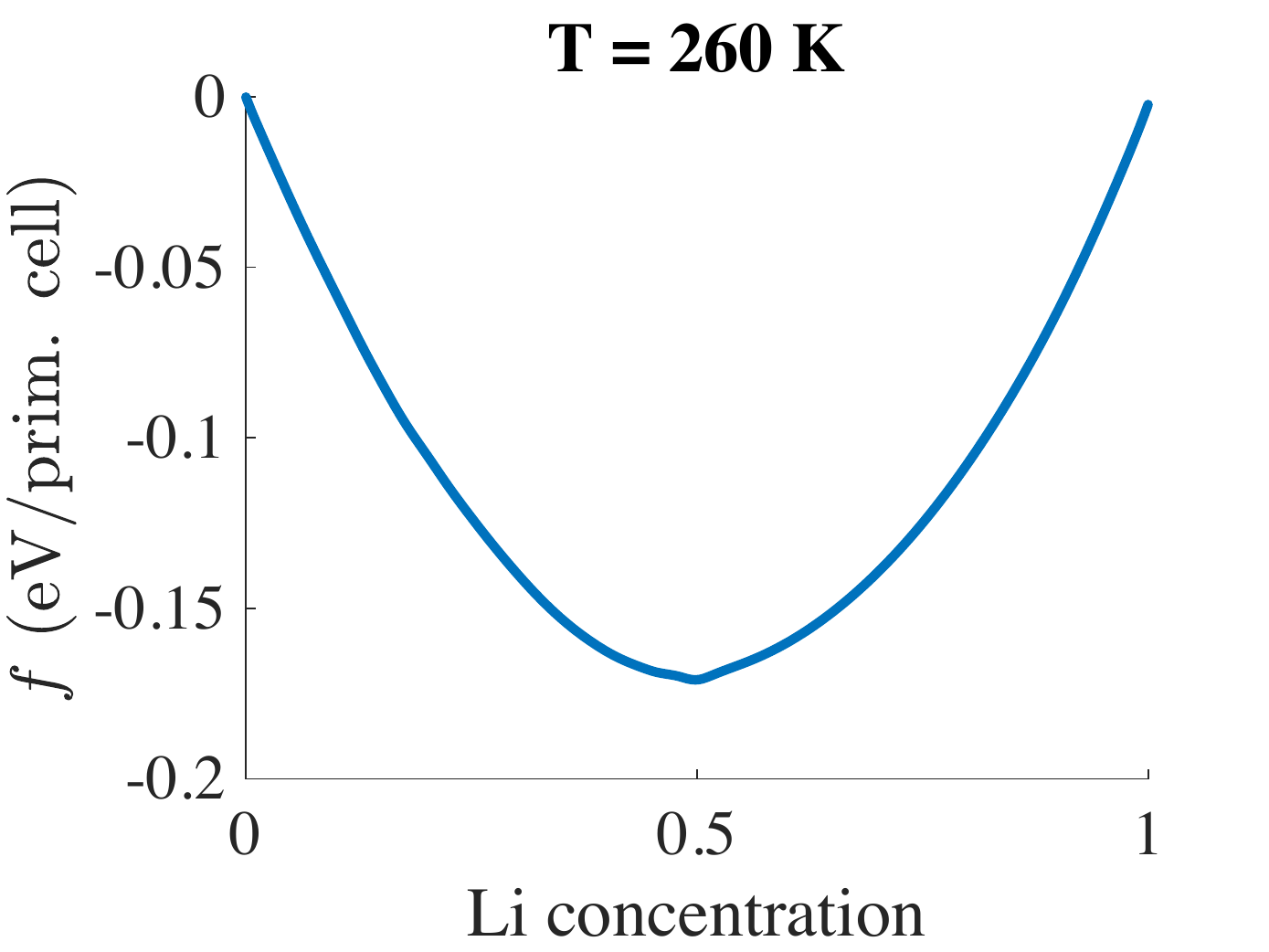}
	\subcaption{Free energy, $f(c)$}
\end{minipage}
    \begin{minipage}[t]{0.7\textwidth}
    \includegraphics[width=0.9\textwidth]{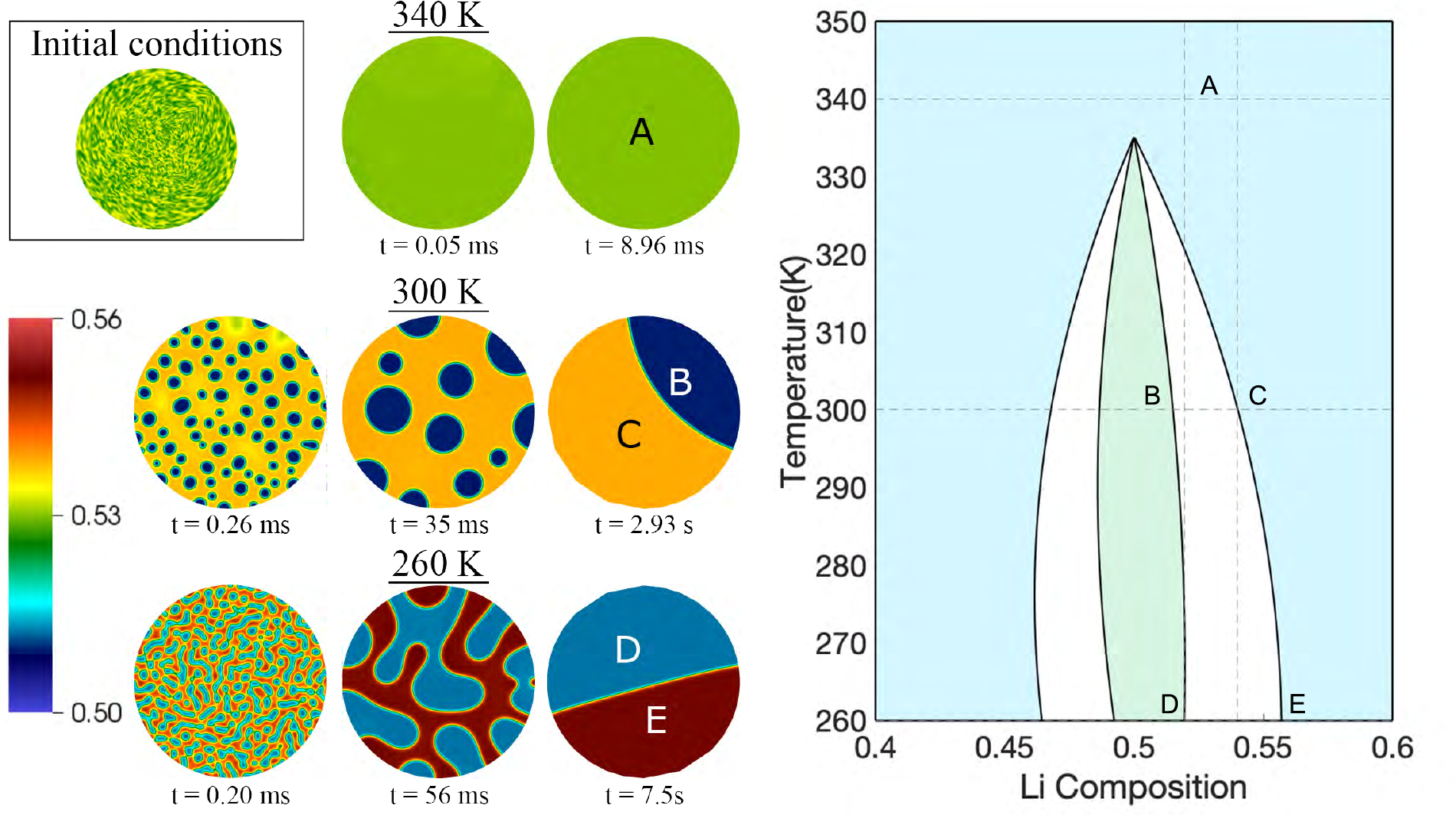}
    \subcaption{Cahn-Hilliard simulations.}
    \end{minipage}
\caption{Comparison of IDNN fits with the data for 260 K, 300 K, and 340 K: (a) the chemical potential data used for training, as sampled with the bias potentials, and the IDNN prediction, (b) numerically differentiated data and derivative of the chemical potential IDNN, and (c) the analytically integrated free energy DNN. (d) Cahn-Hilliard phase field computations of the order-disorder transition based on the free energy density function parameterized by composition, alone. Order parameters are not included; therefore, the ordered variants are not represented. The disordered regions A,C, E and ordered regions C,D are labelled.}
\label{fig:supp1}
\end{figure}

Figure \ref{fig:supp1} shows the IDNN representation obtained by training on only the free energy/chemical potential data corresponding to the composition with the composition as the only feature. These are one-dimensional representations that were obtained without consideration of ordering \cite{teichert2021li}. The corresponding phase microstructures are also shown.

\begin{figure}
    \centering
    \includegraphics[width=0.7\textwidth]{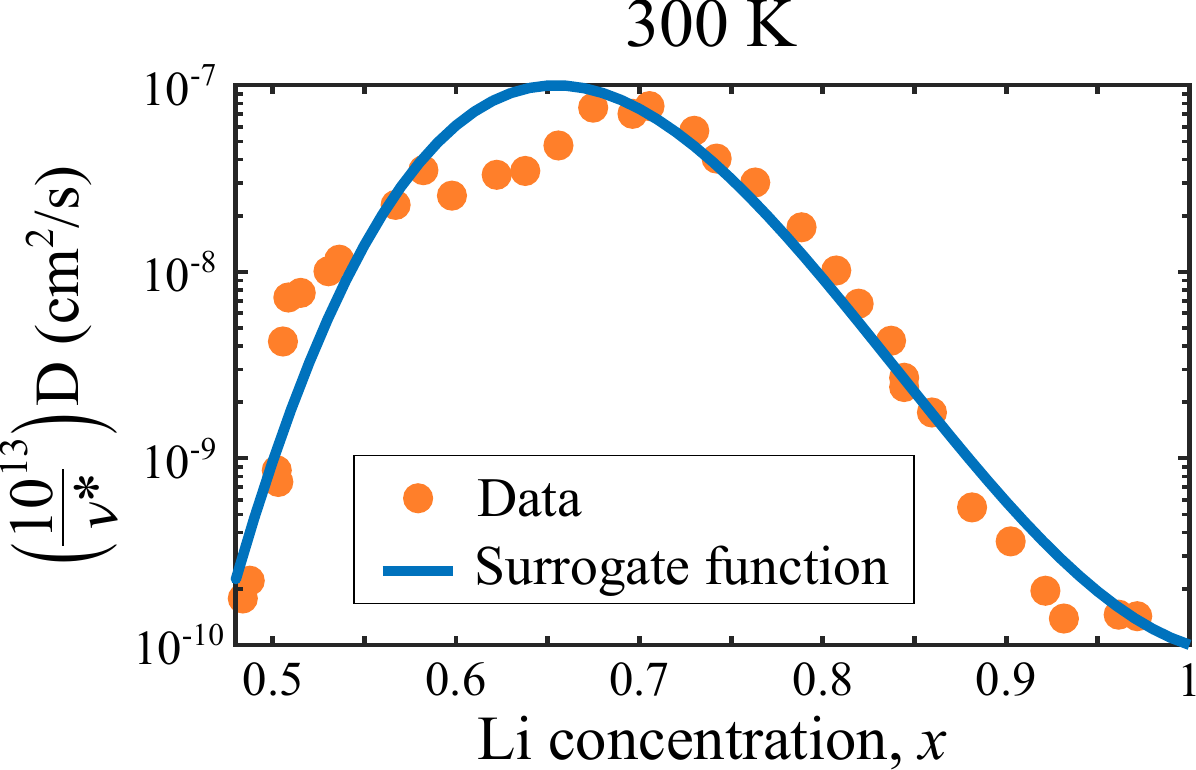}
    \caption{Li diffusivity fit to data.}
    \label{fig:supp2}
\end{figure}

Figure \ref{fig:supp2} shows the diffusivity's dependence on composition fit to data from \cite{VanderVen2000}. The fit function is:
\begin{equation}
D(x) = 0.01\exp(-274(1.05-x)(0.47-x)(1.-x))
\label{eq:diffusivity}
\end{equation}

\begin{figure}[tb]
    \centering
    \includegraphics[width=0.6\textwidth]{figures/phase_field_results_new.pdf}
        \includegraphics[width=0.7\textwidth]{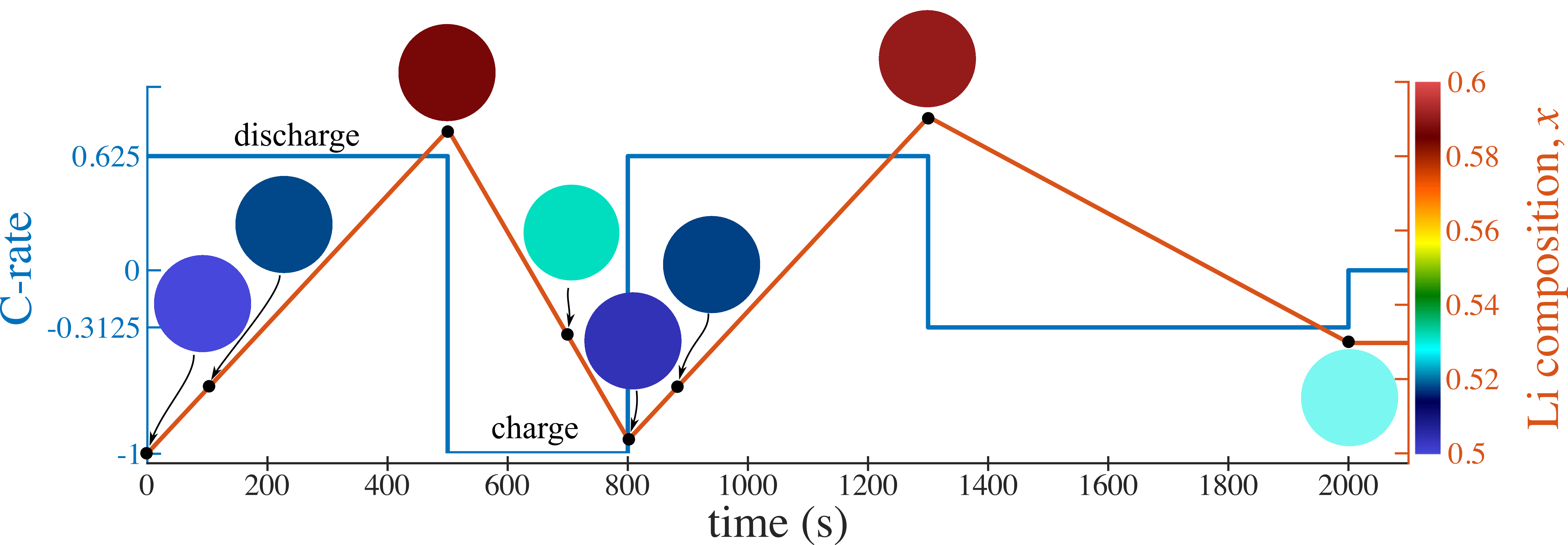}
\caption{2D phase field simulation results at 260, 300, and 340 K showing the Li composition in a 50 nm diameter particle, with initial Li composition randomly perturbed about $x = 0.53$ and no boundary flux. Compare with the results using 1 $\mu$m particles in Figure 5. Phase field results showing the Li composition field resulting from applying a cycling current density for a 50 nm diameter particle. Results are visually indistinguishable for 260, 300, and 340 K, and so only one set of results is included here. The applied current density is scaled by a factor of $1/20$ compared with the 1 $\mu$m particles to maintain the same C-rates in both particle sizes (see Figure 5). The blue line shows the applied C-rate, and the red line shows the corresponding average Li composition of the particle.}
\label{fig:Fig5_small}
\end{figure}

Results from the phase field simulations initialized with Li composition randomly perturbed about $x = 0.525$, with no boundary flux, are shown in Figure \ref{fig:Fig5_small} for the 50 nm particle. As with the larger particles, spinodal decomposition occurs at 260 and 300 K, but not 340 K. However, equilibrium is reached between 100 and 10,000 times more quickly with the smaller particles, depending on the temperature. Additionally, while the equilibrium microstructures are comparable for both sizes, the transient microstructure is much simpler for the smaller particle. 

The results from the cycling simulation for the 50 nm particles presented in Figure \ref{fig:Fig5_small} show that, unlike with the larger particles, the Li composition field is essentially the same for all three temperatures. This, again, matches the experimental findings of Choi et al. \cite{choi2006particle} which show consistent performance across the temperature range of 258 to 333 K.

\section {Symmetry functions}
The 12 symmetric variants are energetically equivalent, so it is important build that symmetry into the IDNN. We do that by making the IDNN a function of symmetry functions that encapsulate the symmetry of the system. The symmetry functions themselves are functions of the order parameters. For example, consider a simpler case (e.g., a B2 system) where the ordering can be defined by a single order parameter with a positive value to represent one variant and a negative value for the second variant. The energy should be the same whether the order parameter is positive or negative, i.e., the energy function is symmetric about $\eta = 0$. To build this in, we can make the energy a function of $\eta^2$, since $\eta^2$ encapsulates the appropriate symmetry. With more order parameters, the symmetry functions become more complicated, but the basic idea is the same.

The algorithm for defining symmetry functions is described by Thomas and Van der Ven \cite{thomas2017} in Appendix C (and Section 5) of their paper. The idea is to start with a complete set of monomials, constructed of the order parameters, of a given order and applying the Reynolds operator to each of them. Note that since we are now operating on the order parameter vectors $\bsym{\eta}$ instead of the sublattice vectors $\bsym{x}$, we first convert the transformation matrices $\bsym{M}^{(x)}_i$ through $\bsym{M}^{(\eta)}_i = \bsym{Q}\bsym{M}^{(x)}_i\bsym{Q}^{-1}$. Many of them will reduce to the zero function and there will likely be duplicates among the resulting nonzero functions, but those that do survive are symmetry invariant. Also, in most cases the final functions are no longer monomials. The process begins with first order monomials, then continues increasing the order of the monomials until a sufficient number of symmetry invariant functions are obtained.

Not all individual symmetry functions reflect the same symmetries. For example, the function
\begin{align}
    f(\bsym{\eta}) = \eta_1\eta_2\eta_3\eta_4\eta_5\eta_6
\end{align}
clarifies that the free energy can be different if you change the sign on a single order parameter, but this characteristic is missed in, say, this lower order symmetry function:
\begin{align}
    f(\bsym{\eta}) = \eta_1^2 + \eta_2^2 + \eta_3^2 + \eta_4^2 + \eta_5^2 + \eta_6^2
\end{align}

The Reynold's operator is performed by applying all transformations to the monomial's input, in turn, and summing the results, as in the following:
\begin{align}
    h(\bsym{\eta}) = \sum_{\bsym{M}^{(\eta)}\in\mathcal{M}}f(\bsym{M}^{(\eta)}\bsym{\eta})
\end{align}
where $\mathcal{M}$ is the group of all transformation matrices $\bsym{M}^{(\eta)}$ acting on the order parameter space, $h(\bsym{\eta})$ is a symmetry invariant function, and $f(\bsym{\eta})$ is the initial monomial.

We can represent the monomial $f(\bsym{\eta})$ with an exponent vector $\bsym{n}$ and an associated coefficient $a^{(\bsym{n})}$:
\begin{align}
    f(\bsym{\eta}) = a^{(\bsym{n})}\prod_{m=1}^{\mathrm{len}(\bsym{n})}\eta_m^{n_m}
\end{align}

The final symmetry functions used here include up through order six and are the following:

\begin{align}
    h_1 &= \eta_0\\
    h_2 &= \frac{2}{3}\sum_{i=1}^6\eta_i^2\\
    h_3 &= \frac{8}{3}\sum_{i=1}^6\eta_i^4\\
    h_4 &= \frac{4}{3}\left[\left(\eta_1^2+ \eta_2^2\right)\left(\eta_3^2 + \eta_4^2 + \eta_5^2 + \eta_6^2\right) + \left(\eta_3^2 +\eta_6^2\right)\left(\eta_4^2 + \eta_5^2\right)\right]\\
    h_5 &= \frac{16}{3}\left(\eta_1^2\eta_2^2 + \eta_3^2\eta_6^2 + \eta_4^2\eta_5^2\right)\\
    h_6 &= \frac{32}{3}\sum_{i=1}^6\eta_i^6\\
    \begin{split}
        h_7 &= \frac{8}{3}\big[\left(\eta_1^4+ \eta_2^4\right)\left(\eta_3^2 + \eta_4^2 + \eta_5^2 + \eta_6^2\right) + \left(\eta_3^4 +\eta_6^4\right)\left(\eta_4^2 + \eta_5^2\right) + \\
        &\phantom{= \frac{\sqrt{10}}{4}\big[}\left(\eta_1^2+ \eta_2^2\right)\left(\eta_3^4 + \eta_4^4 + \eta_5^4 + \eta_6^4\right) + \left(\eta_3^2 +\eta_6^2\right)\left(\eta_4^4 + \eta_5^4\right)\bigg]
    \end{split}\\
    \begin{split}
        h_8 &= \frac{16}{3}\big[\eta_1^2\eta_2^2(\eta_3^2 + \eta_4^2 + \eta_5^2 + \eta_6^2) + \eta_3^2\eta_6^2(\eta_1^2 + \eta_2^2 + \eta_4^2 + \eta_5^2) +\\
        &\phantom{=\frac{\sqrt{30}}{2}\big(}\eta_4^2\eta_5^2(\eta_1^2 + \eta_2^2 + \eta_3^2 + \eta_6^2)\big]
    \end{split}\\
    h_9 &= \frac{32}{3}\left(\eta_1^4\eta_2^2 + \eta_1^2\eta_2^4 + \eta_3^4\eta_6^2 + \eta_3^2\eta_6^4 + \eta_4^4\eta_5^2 + \eta_4^2\eta_5^4\right)\\
    h_{10} &= 8(\eta_1^2 + \eta_2^2)(\eta_3^2 + \eta_6^2)(\eta_4^2 + \eta_5^2)\\
    \begin{split}
    h_{11} &= \frac{64}{5}\big[\eta_1\eta_2(\eta_3^2 - \eta_6^2)(\eta_4^2 - \eta_5^2) + \\
    &\phantom{= \frac{64}{5}\big[}\eta_3\eta_6(\eta_1^2 - \eta_2^2)(\eta_4^2 - \eta_5^2) + \\
    &\phantom{= \frac{64}{5}\big[}\eta_4\eta_5(\eta_1^2 - \eta_2^2)(\eta_3^2 - \eta_6^2)\big]
    \end{split}\\
    h_{12} &= 64\sqrt{5}\eta_1\eta_2\eta_3\eta_4\eta_5\eta_6
\end{align}

\section {Monte Carlo Precision}
Each Monte Carlo run finds the ensemble average for $\bsym{\eta}$ and $\bsym{\mu}$ for a given $\bsym{\phi}$, $\bsym{\kappa}$, and temperature. The Monte Carlo calculations used in the 7D IDNN have a precision of 0.0003 for the order parameters with a confidence of 0.95. When the Monte Carlo simulation converges, $\bsym{\mu}$ is then computed using the ensemble average. We found the precision of the order parameter does not affect the determined relationship between $\bsym{\eta}$ and $\bsym{\mu}$ and therefore does not affect the accuracy of the data used to train the IDNN. However, it does effect what data is used to train the IDNN, as the determined $\bsym{\eta}$ given the inputs is less accurate. We found that a higher precision is needed at 300K and 340K to have a consistent relationship between $\bsym{\kappa}$ and $\bsym{\eta}$. When precise sampling is needed, higher precision on the order parameters would be necessary. For our current active learning method, there is randomness in both the global and local sampling, such that additional randomness in the $\bsym{\eta}$ used to find $\bsym{\mu}$ is not an issue.

\section {Lowest Free Energy Surface}

The Free Energy Surface is plotted in the $\eta_0 -\eta_1$ space with the lowest free energy curve for each $\eta_0$ (with $\eta_2$-$\eta_6$ held at 0) plotted in red. This free energy curve is found by predicting the free energy using the 7D IDNN for each point in the $\eta_0$ and positive $\eta_1$ space and determining the lowest free energy at each $\eta_0$. As the order parameters are symmetrically invariant, this holds for negative $\eta_1$ and all other order parameters. For 260K and 300K (with $\eta_2$-$\eta_6$ held at 0) there is a discontinuity in the derivative of the lowest free energy corresponding to $\eta_1$ jumping from 0 to (.33-.39). At 340K the lowest energy state is always disordered. 

\begin{figure}
    \centering
    \includegraphics[width=1\textwidth]{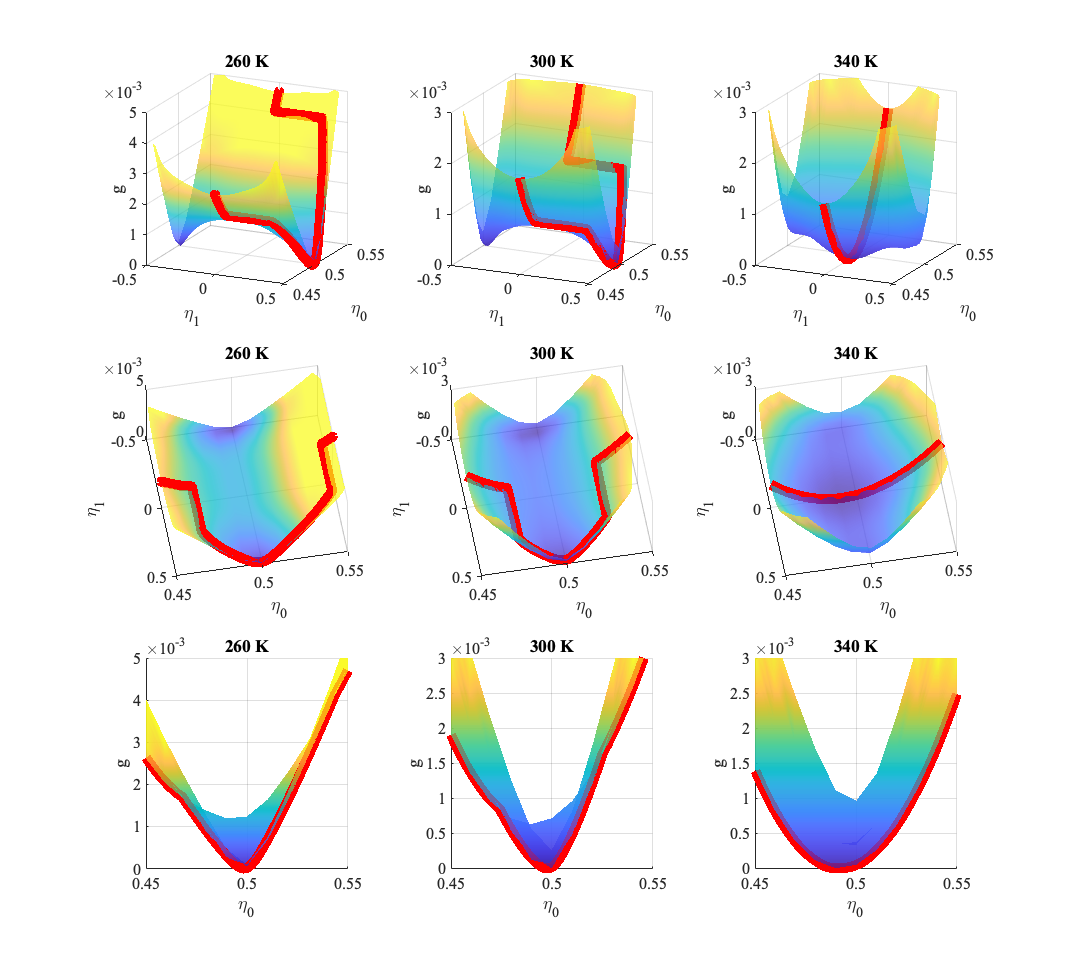}
    \caption{The free energy density surfaces plotted in the $\eta_0-\eta_1$ subspace at different orientations. The lowest free energy for each $\eta_0$ in the $\eta_1$ space as determined by the 7D IDNN plotted in red.}
    \label{fig:free energy surface}
\end{figure}

The lowest free energy curves in 7D are plotted in Figure \ref{fig:free energy curve}. These curves are found by evaluating data points over the complete $\bsym{\eta}$ space, under the conditions that $\eta_1-\eta_6$ are positive and in decreasing order (as each order parameter is symmetrically invariant) and that they satisfy the conditions outlined in the Equations 15 and 16 in the main section. For each $\eta_0$, we then determine the $\eta's$ that correspond to the lowest free energy. For 300K along the lowest free energy curve, only one order parameter is nonzero. This curve is shown in Figures \ref{fig:free energy curve} and \ref{fig:free energy surface}. For 260K, there are two distinctive non-disordered regions. In one region, both $\eta_1$ and $\eta_2$ are nonzero and jump from 0 to $\sim.18$. In the second region $\eta_2$ goes to 0 and $\eta_1$ goes to $\sim.37$. The points at which these discontinuities occur are shown by the red points in Figure \ref{fig:free energy}. In addition, the $\eta_0 -\eta_1$ curve for 260K from Figure  \ref{fig:free energy surface} is shown as the green curve in Figure \ref{fig:free energy curve}, and the discontinuities are shown by the green points in \ref{fig:free energy}.

\section {Phase Diagram}
\begin{figure}
    \centering
\begin{minipage}[t]{.9\textwidth}
        \includegraphics[width=1\textwidth]{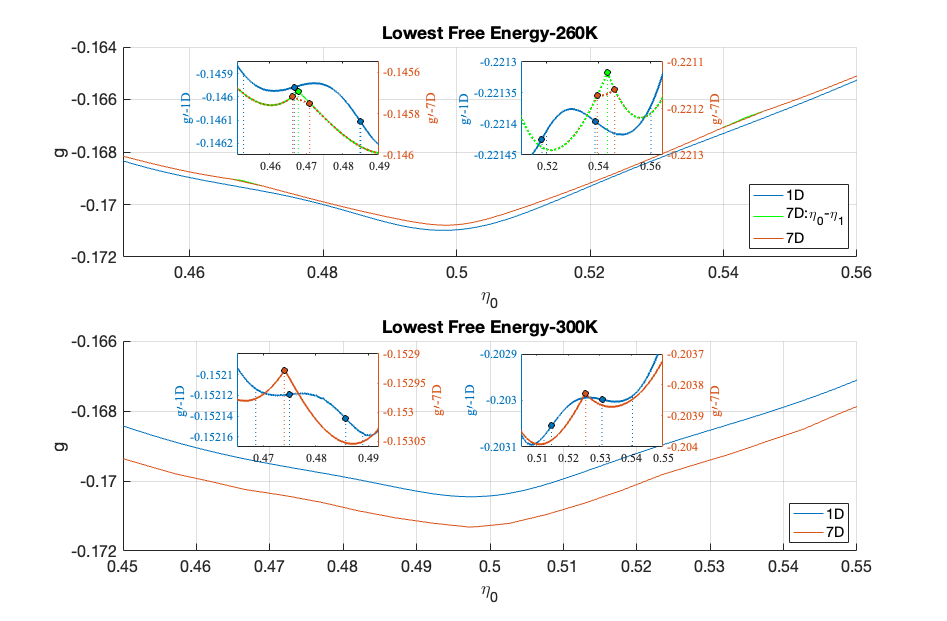}
        \subcaption{}
        \label{fig:free energy curve}
    \end{minipage}
    \hfill
   \begin{minipage}[t]{0.5\textwidth}
        \includegraphics[width=1\textwidth]{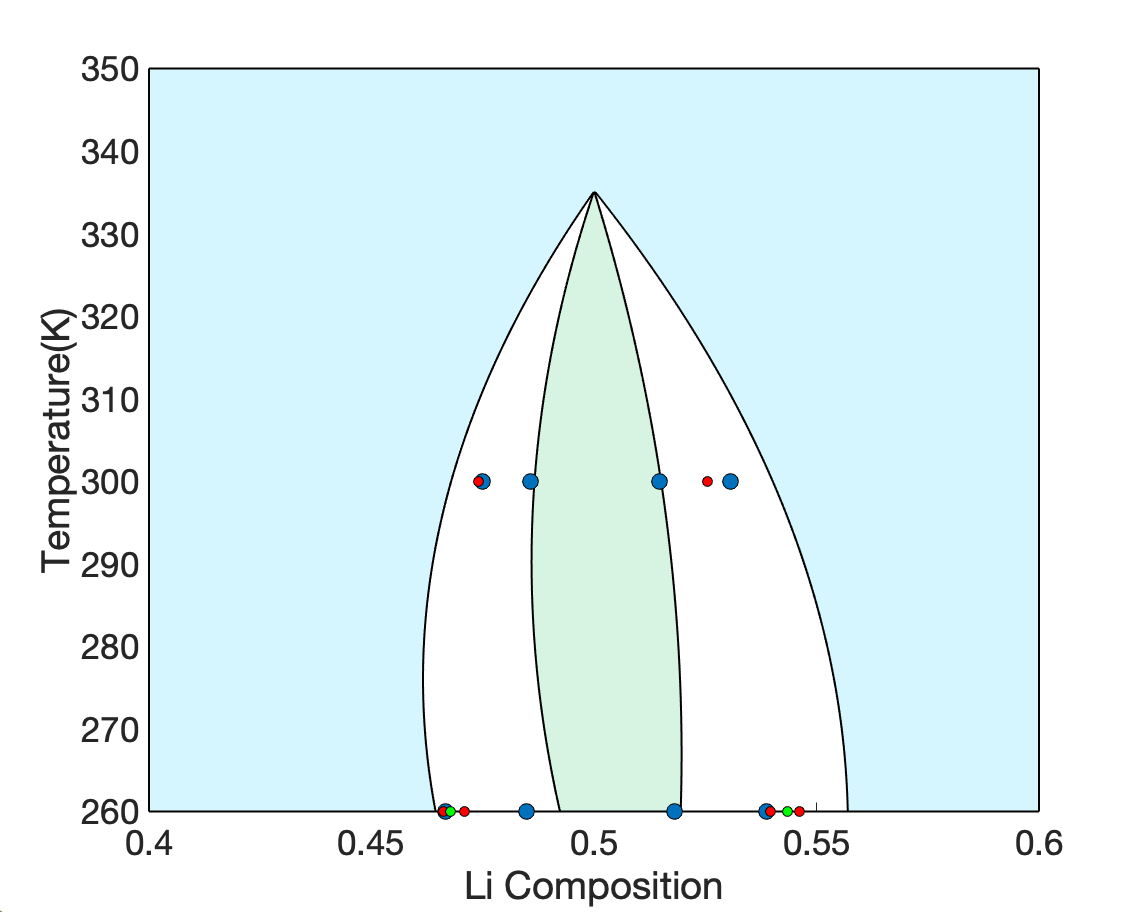}
        \subcaption{}
        \label{fig:phase diagram}
    \end{minipage}
    
    \caption{(a) The lowest free energy curves as determined by the IDNN. The curve for the 1D IDNN is shown in blue. The lowest free energy in the complete $\eta$ space is shown by the 7D curve in red. For 260K we additionally plotted the lowest free energy curve in the $\eta_0-\eta_1$ space shown in green. The points used to create the phase diagram in (b) are shown by the blue diamonds. These points occur at (0.453 0.485 0.520  0.560) for T260 and (0.469 0.489 0.512 0.540) for T300. For both (a) and (b) the points where the second derivative of the 1D IDNN is zero is shown by the blue circles. These points occur at (0.4667 0.4848 0.5181 0.5387) for T260 and (0.475 0.4857 0.5147 0.5307) for T300. The discontinuities in the free energy of the 7D IDNN are represented by the green and red points. For T260 the discontinuities in the full $\eta$ space(red points) are found at (0.466 0.471 0.540 0.546). In the $\eta_0-\eta_1$ space the green points are found at (0.468 0.544). For T300 the discontinuities are found at (0.474 0.525) The boxes in (a) highlighting the convexity of the two-phase region have a linear term added to the chemical potential to exaggerate the convex region, such that $g' = g + l \eta_0$. For T260: l is 0.05, -0.1002 for the left and the right box respectively. Similarly, for T300: l is 0.037 and -0.065. }
    \label{fig:free energy}
\end{figure}

Our Phase Diagram (Figure \ref{fig:phase diagram}) is constructed using the free energy predicted by the 1D IDNN trained on the results of 1D Monte Carlo with umbrella sampling. A constructed set of tangent lines on the free energy curve, at compositions of Li greater than 0.5, and less than 0.5, are used to find the boundaries to the region of phase instability. The points used in the phase diagram are shown in blue diamonds in Figure \ref{fig:free energy curve}. 

To determine the region of spinodal decomposition we can also use the second derivative of free energy to determine the boundary. A characteristic of the  region is that the free energy curve is convex, while otherwise the free energy curve is concave. For the 1D Monte Carlo the 2nd derivative of free energy in the two-phase region is negative and the boundary to the region can be found when $\frac{\partial^2 g}{\partial x^2}=0$. These points are plotted as blue circles in the Figure \ref{fig:free energy curve}.  

For the 7D IDNN, the convexity in the free energy is found in regions with all negative eigenvalues. However, we have found that this exists only outside the boundary regions for $\eta_1-\eta_6$ as described in section 4.4 (Equations 15 and 16) of the main text. We expect this accounts for the difference between the 1D and the 7D curve. This means we cannot find the boundaries of the spinodal decomposition region using eigenvalues. Instead, we look at the $\eta's$ corresponding to the lowest free energy curve. 

\begin{figure}
    \centering
    \includegraphics[width=1\textwidth]{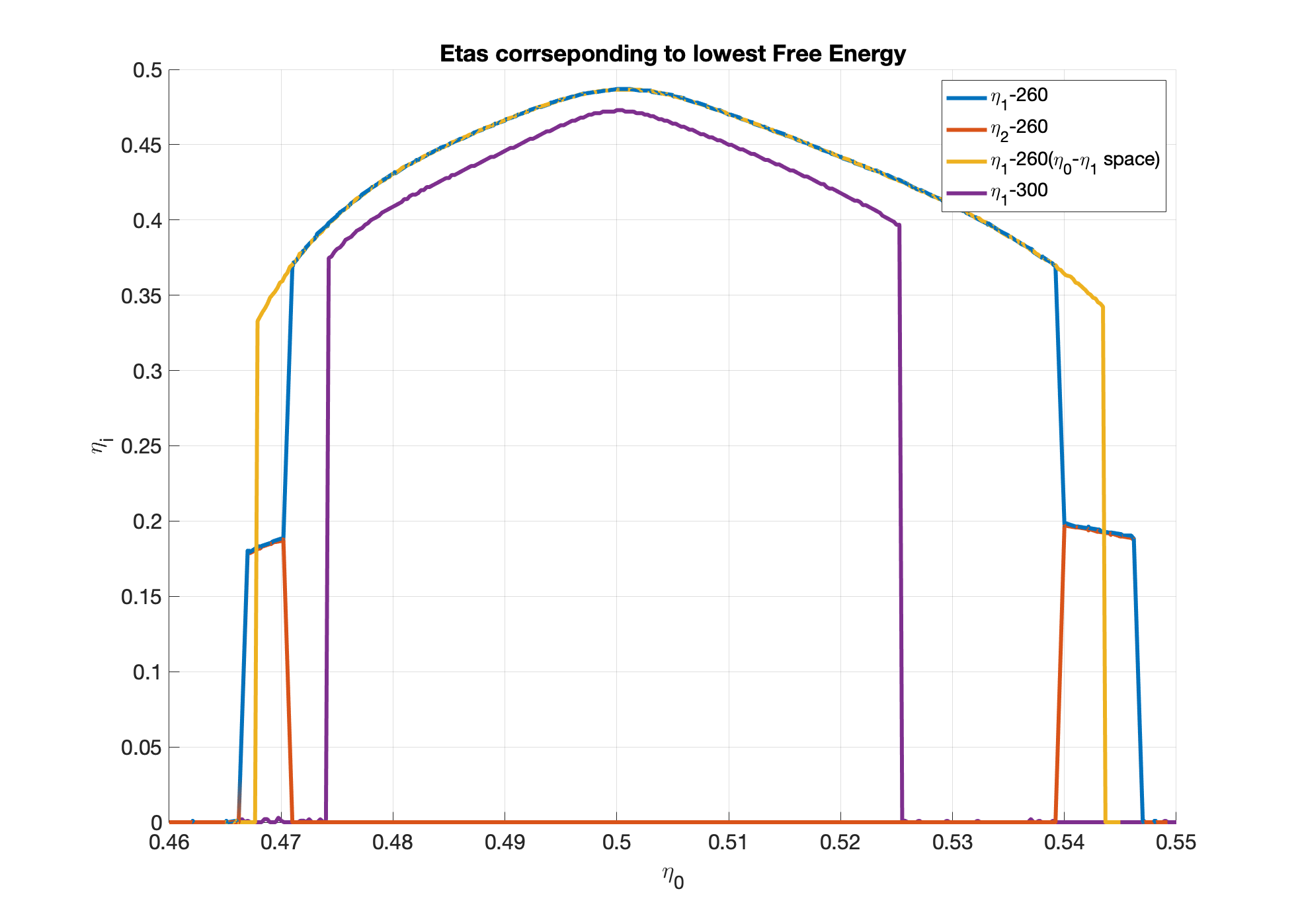}
    \caption{This figure shows the $\eta$'s corresponding to the lowest free energy curve. The red and blue curves show how $\eta_1$ and $\eta_2$ vary corresponding to the lowest free energy at 260K. The yellow curve shows how $\eta_1$ varies given $\eta_2-\eta_6$ are zero for 260K. The purple curve shows how $\eta_1$ varies for 300K given $\eta_2-\eta_6$ are zero. }
    \label{fig:free energy etas}
\end{figure}

There is a discontinuity in the $\mu_0$ space, where the $\eta_1$ jumps. These jumps are shown in Figure \ref{fig:free energy etas}.  These discontiunites are plotted by the green and red points in \ref{fig:free energy}. Since we do not see a zero or negative $\frac{\partial^2 g}{\partial x^2}$ we use the discontinuities to approximate the transition between disordered and two phase. From Figure \ref{fig:free energy} we can see that the discontinuity occurs at a similar $\eta_0$ to the 1D disorder-spinodal region boundary. The ordered to spinodal decomposition boundary is more difficult to find in 7D and additional sampling might be needed to properly capture the boundary in the IDNN.  We do expect the 1D and lowest free energy from the 7D IDNN to show the same trends. This discrepancy could be caused by inaccuracies in the Monte Carlo calculations, sub optimal sampling, or not using an optimal machine learning model.

\clearpage
\bibliographystyle{unsrt}
\bibliography{references}